\documentclass[twocolumn]{aastex61}
\usepackage{amsmath}
\received{}
\revised{}
\accepted{}
\submitjournal{ApJ}

\shorttitle{ALMA Observations of COMs in G10.6-0.4}
\shortauthors{Law et al.}
\begin{document}

\title{Sub-arcsecond Imaging of the Complex Organic Chemistry in Massive Star-forming Region G10.6-0.4}

\correspondingauthor{Charles J. Law}
\email{charles.law@cfa.harvard.edu}

\author[0000-0003-1413-1776]{Charles J.\ Law}
\affiliation{Center for Astrophysics $|$ Harvard \& Smithsonian, 60 Garden St., Cambridge, MA 02138, USA}

\author[0000-0003-2384-6589]{Qizhou Zhang}
\affiliation{Center for Astrophysics $|$ Harvard \& Smithsonian, 60 Garden St., Cambridge, MA 02138, USA}

\author[0000-0001-8798-1347]{Karin I. {\"O}berg}
\affiliation{Center for Astrophysics $|$ Harvard \& Smithsonian, 60 Garden St., Cambridge, MA 02138, USA}

\author[0000-0003-1480-4643]{Roberto Galv{\'a}n-Madrid}
\affiliation{Instituto de Radioastronom{\'i}a y Astrof{\'i}sica, Universidad Nacional Auton{\'o}ma de M{\'e}xico, PO Box 3-72, 58090 Morelia, Michoacan, Mexico}

\author{Eric Keto}
\affiliation{Center for Astrophysics $|$ Harvard \& Smithsonian, 60 Garden St., Cambridge, MA 02138, USA}

\author[0000-0003-2300-2626]{Hauyu Baobab Liu}
\affiliation{Academia Sinica Institute of Astronomy and Astrophysics, 11F of AS/NTU Astronomy-Mathematics Building, No.1, Sec. 4, Roosevelt Rd, Taipei 10617, Taiwan}

\author[0000-0002-3412-4306]{Paul T. P. Ho}
\affiliation{Academia Sinica Institute of Astronomy and Astrophysics, 11F of AS/NTU Astronomy-Mathematics Building, No.1, Sec. 4, Roosevelt Rd, Taipei 10617, Taiwan}



\begin{abstract}
Massive star-forming regions exhibit an extremely rich and diverse chemistry, which in principle provides a wealth of molecular probes, as well as laboratories for interstellar prebiotic chemistry. Since the chemical structure of these sources displays substantial spatial variation among species on small scales (${\lesssim}10^4$~au), high angular resolution observations are needed to connect chemical structures to local environments and inform astrochemical models of massive star formation. To address this, we present ALMA 1.3~mm observations toward OB~cluster-forming region G10.6-0.4 (hereafter ``G10.6") at a resolution of 0.14$^{\prime\prime}$~(700~au). We find highly-structured emission from complex organic molecules (COMs) throughout the central 20,000~au, including two hot molecular cores and several shells or filaments. We present spatially-resolved rotational temperature and column density maps for a large sample of COMs and warm gas tracers. These maps reveal a range of gas substructure in both O- and N-bearing species. We identify several spatial correlations that can be explained by existing models of COM formation, including NH$_2$CHO/HNCO and CH$_3$OCHO/CH$_3$OCH$_3$, but also observe unexpected distributions and correlations which suggest that our current understanding of COM formation is far from complete. Importantly, complex chemistry is observed throughout G10.6, rather than being confined to hot cores. The COM composition appears to be different in the cores compared to the more extended structures, which illustrates the importance of high spatial resolution observations of molecular gas in elucidating the physical and chemical processes associated with massive star formation.
\end{abstract}

\keywords{astrochemistry --- ISM: individual objects (G10.6-0.4) --- ISM: lines and bands --- ISM: molecules --- stars: formation --- stars: massive}



\section{Introduction} \label{sec:intro}

Although high-mass stars (${>}8~M_{\odot}$) play a dominant role in cosmic evolution, chemical enrichment of the interstellar medium, galaxy formation and evolution, and galactic star formation, a detailed understanding of their formation remains elusive due to their short evolutionary timescales, intrinsic rarity, large distances, and heavy obscuration \citep{Zinnecker07, Tan14, Motte18}. A further complication arises due to the complex, multi-component process of forming massive protostars, which substantially alter their surroundings, as they become energetic enough to ionize their natal molecular clouds. This first produces hypercompact (HC) \ion{H}{2} regions of sizes ${\sim}0.01$~pc \citep{Kurtz00}, which then expand to form ultracompact (UC) \ion{H}{2} regions of sizes ${\sim}0.1$~pc \citep{Churchwell02, Hoare07}, and ultimately with time, compact \ion{H}{2} regions, before all of the material surrounding the newly-formed stars are destroyed or dispersed by powerful stellar winds. During this process, these regions possess an extremely rich and diverse chemistry, as the chemical composition of molecular gas involved in star formation is greatly influenced by the physical changes that occur during the star formation process \citep{vanDishoeck98}.

In particular, the elevated temperatures and densities generated from collapsing material leads to the production and destruction of various molecular species and results in a high degree of chemical complexity \citep{Schilke01, Bisschop07, Belloche13}, which in turn leads to the emergence of hot molecular cores. These hot cores, which exhibit intense emission in numerous complex organic molecules (COMs), are often associated with hot and dense regions near massive young stellar objects (MYSOs). However, the complex chemical and physical processes driving these regions are not fully understood, despite decades of observational \citep{Blake87, Gibb00} and modeling efforts \citep[][]{Charnley92, Charnley95, Charnley95MNRAS, Garrod06, Garrod08, Laas11, Choudhury15}. For instance, the presence of externally-heated hot cores located in the vicinity of UC \ion{H}{2} regions \citep{Wyrowski99, DeBuizer03, Mookerjea07} has challenged the notion that hot cores are simply a stage in the evolutionary sequence of massive protostars. In reality, hot cores may also originate as chemical manifestations of UC \ion{H}{2} regions on their environments, rather than simply being precursors to UC \ion{H}{2} regions, and be driven, in part, by COM production from ice photochemistry due to abundant UV and X-ray photons \citep{Oberg16_chem_rev}.

Numerous studies at low spatial resolution confirmed the presence of a COM-rich chemistry in MYSOs \citep[e.g.,][]{Turner91, Schilke97, Tercero10}, while those conducted at higher resolution revealed that this chemistry is highly-structured on small scales ($\lesssim 10^4$~au) \citep[e.g.,][]{Beuther05, Mookerjea07, Qin10, Guzman18, Moscadelli18, Bogelund19, Gieser19, Molet19}. Such observations, which directly resolve the spatial differences between the emission of distinct species allow us to unambiguously discern the regions traced by different molecular species and families of molecules, providing further insight into their formation processes and chemistry. As a result, they are crucial for constraining a variety of outstanding questions related to the chemical evolution of MYSOs: How frequently and in what way are hot cores associated with UC \ion{H}{2} regions \citep[e.g.,][]{Mookerjea07, Lindberg17}? What processes and evolutionary stages are responsible for COM production in massive star-forming regions \citep[e.g.,][]{Garrod17, ElAbd19}? Are there several distinct COM chemistries that can be used to trace different aspects of star formation \citep[e.g.,][]{Fontani07, Suzuki18, Tercero18}? Are O- and N-bearing COMs systematically spatially offset from one another \citep[e.g.,][]{Blake87, Wyrowski99, Qin10, Jimenz_Serra12}?

To address these questions, we investigated the small-scale spatial structure of a representative set of COMs in the well-known massive star-forming region G10.6-0.4 (hereafter ``G10.6"). G10.6 is a ${\sim}10^6~L_{\odot}$ OB cluster-forming region \citep{Casoli86, Lin16} at a distance of 4.95~kpc \citep{Sanna14}. Unresolved observations with the Submillimeter Array (SMA) revealed that the dense gas surrounding its central proto-OB-cluster is particularly COM- and hydrocarbon-rich \citep{Jiang15, Wong18}. Hints of chemical substructure were seen in ${\sim}1^{\prime \prime}$ resolution SMA observations of a sample of shock tracer molecules, especially in E$_{\rm{u}} \approx 100$--$150$~K lines of SO$_2$ and OCS \citep{Liu_10ApJ_722_262L}. These observations suggested a rich, spatially-variable complex chemistry taking place in the central $3^{\prime \prime}$ (${\sim}0.09$~pc).

In this paper, we present a detailed study of the complex chemistry toward the central 20,000~au of G10.6 using ALMA Band 6 observations at $0.14^{\prime \prime}$. In Section \ref{sec:obs}, we describe the observations and data reduction. In Section \ref{sec:data_analysis}, we discuss the methods for extracting spatially-resolved gas physical conditions and show a detailed example for CH$_3$OH. We present rotational temperature and column density maps, as well as spatial correlations for all molecules in our sample in Section \ref{sec:results}. We discuss the COM chemistry of G10.6 and compare it with other well-studied regions in Section \ref{sec:discussion} and summarize our conclusions in Section \ref{sec:conclusions}.

\section{Observations} \label{sec:obs}

\subsection{Observational Details} \label{sec:obs_details}

G10.6 was observed on 9-10 September 2016 and 19 July 2017 in Band 6 as part of the ALMA Cycle 5 project 2015.1.00106.S. The first execution block included 36 antennas with projected baseline lengths between 15 and 3144~m (11--2462~k$\lambda$). The second execution block included 41 antennas with projected baseline lengths between 15 and 3697~m (11--2896 k$\lambda$). The on-source integration times were 67~min and 66~min, respectively, for a total on-source integration time of 2.2~hr. Precipitable water vapor was 0.50~mm and 0.46~mm with system temperatures ranging between 54--75~K and 50--70~K for each execution block, respectively. The correlator setup was identical for both execution blocks and included spectral windows centered at 217.9, 220.0, 232.0, and 233.9~GHz with a uniform resolution of 976.563 kHz (${\sim}1.3$~km~s$^{-1}$). Each spectral window had a bandwidth of 1.875~GHz for a total bandwidth of 7.5~GHz. The spectral setup was motivated by the primary science goal of studying the kinematics of the H$30\alpha$ hydrogen radio recombination line, which will be described in a future paper. The auxiliary molecular line data were obtained gratuitously, and for this reason, the dataset was not necessarily optimized for maximal coverage of COM lines.

The phase center of both observations was located at R.A. (J2000)$=$18$^{\rm{h}}$10$^{\rm{m}}$28$^{\rm{s}}$.683, Dec. (J2000)$=$ $-$19$^{\circ}$55$^{\prime}$49$^{\prime \prime}$.07. The full-width at half-maximum (FWHM) of the ALMA primary beam was 25.1$^{\prime \prime}$, and as a result, primary beam corrections were not necessary. For both executions, the quasar J1924-2914 was used for bandpass calibration and J1832-2039 was used for phase calibration, while J1733-1304 served as the flux calibrator. We assume a 10\% flux calibration uncertainty.

\subsection{Data Reduction, Continuum Subtraction, and Imaging}  \label{sec:data_redux}
Initial data calibration was performed by ALMA / NAASC staff. The reduction of the first execution used CASA version 4.7.0, while the reduction of the second execution and all subsequent imaging used CASA version 4.7.2 \citep{McMullin07}. The data cubes were imaged using the \texttt{tclean} task with a Briggs robust weighting parameter of 0.5 to achieve a compromise between resolution and sensitivity to extended emission. The resulting images have a square pixel size of $0.018^{\prime \prime}$ and an overall size of ${\sim}43^{\prime \prime}$, which is approximately twice the primary beam FWHM at 1.3~mm. We then trimmed these images to the innermost 9$^{\prime \prime}$ by 7$^{\prime \prime}$ region, which is known to contain the densest gas and hence the most complex and varied chemistry \citep[e.g.,][]{Liu_10ApJ_722_262L}.

Due to the high sensitivity and dynamic range of the observations, no sufficiently ``line-free" channels exist across the entirety of the map, especially toward regions containing the densest gas. As a result, we were unable to subtract the continuum emission in the Fourier domain and instead performed the continuum subtraction in the image plane. We followed the procedures of \citet{Jorgensen16} to define the continuum in a homogeneous and automated way based on the density distribution of channel intensities in each pixel. In brief, the flux density distribution of each pixel was first fit with a symmetric Gaussian. Then, the centroid of this fit was used to define a fitting range for a skewed Gaussian fit and the new centroid value (no longer necessarily symmetric) was recorded as the continuum level for that particular pixel. For pixels with only continuum contribution and no line emission, the distribution is a symmetric Gaussian centered at the continuum level with a width corresponding to the rms noise $\sigma$, while pixels with both continuum and line emission are modified by an exponential tail toward higher values. \citet{Jorgensen16} report an accuracy of $2\sigma$ using this method, and as this study focuses on strong lines (${>}20\sigma$), the continuum subtraction contribution to our total error budget is minimal. The accuracy of our continuum subtraction was confirmed via visual inspection, and we did not find it necessary to implement more sophisticated continuum subtraction methods \citep[e.g.,][]{Sanchez18_statcont, Molet19}.

\begin{deluxetable}{cccccc}
%\tablenum{4}
\tablecaption{Overview of Spectral Windows\label{tab:Table1}}
\tablehead{[-.3cm]
\colhead{Frequency Range} & \colhead{Beam Size} & \colhead{P. A.} & \multicolumn{2}{c}{rms} \\[-0.2cm]
\colhead{[GHz]} & \colhead{[$^{\prime \prime} \times ^{\prime \prime}$]} & \colhead{[$^{\circ}$]} & \colhead{[mJy~beam$^{-1}$]} & \colhead{[K]} \\[-0.55cm]}
\startdata
216.948 -- 218.823 & $0.15\times0.13$ & 118 & 0.60 & 0.79 \\
219.053 -- 220.928 & $0.15\times0.13$ & 110 & 0.65 & 0.84 \\
231.054 -- 232.929 & $0.14\times0.12$ & 115 & 0.63 & 0.85 \\
232.949 -- 234.824 & $0.14\times0.12$ & 120 & 0.65 & 0.87 \\
\enddata
\end{deluxetable}

We derived the continuum rms for each spectral window as the median value of all pixels without significant continuum emission. To do so, we excluded the central 0.06~pc region of G10.6, which possesses highly-structured continuum emission, and point source-like emission associated with an independent UC \ion{H}{2} region located to the west. The derived continuum rms was then used to derive rms per spectral bin. Overall, we found a typical continuum rms of ${\sim}0.60$--$0.65$ mJy~beam$^{-1}$ (${\sim}0.79$--$0.87$~K) with a synthesized beam size of $0.14^{\prime \prime}$, which corresponds to an approximate physical scale of 700~au at the distance of G10.6. A summary of beam sizes and rms values per spectral window is shown in Table \ref{tab:Table1}. Fully reduced spectral cubes are publicly available and included in the supporting materials of this paper.

\subsection{Overview of G10.6} \label{sec:obs_overview}

G10.6 is a well-suited target for a study of complex chemistry at high spatial resolution. Several \ion{H}{2} and UC \ion{H}{2} regions are present across the central ${\sim}10$~pc area, which suggests that massive star formation is simultaneously occurring over the entire molecular cloud \citep{Ho86, Sollins05}. G10.6 exhibits a centrally concentrated distribution of gas \citep{Lin16} with several approximately radially-aligned, 5--10~pc dense gas filaments, which connect to a central ${\sim}1$~pc-scale massive molecular clump \citep{Liu_12ApJ_745_61L, Liu17}. Higher angular resolution observations of dense molecular gas tracers revealed that some of these large-scale filaments are converging to a rotating, ${\sim}0.6$~pc-scale massive molecular envelope \citep{Omodaka92, Ho94, Klaassen09, Liu10ApJ_725_2190L, Liu_10ApJ_722_262L,  Liu11}. The massive proto-OB stars and associated UC \ion{H}{2} region \citep{Ho86ApJ_305_714H, Sollins05_detailed} are deeply embedded in this rotationally-flattened \citep{Guilloteau88, Keto87, Keto88} envelope, where low density gas around the rotational axis has either been photoionized or dispersed by expanding ionized gas \citep{Liu10ApJ_725_2190L, Liu11}. High excitation (E$_{\rm{u}}~{\gtrsim}~100$~K) molecular lines trace a ${\sim}0.1$~pc ``hot toroid" at the center of this massive molecular envelope \citep{Liu10ApJ_725_2190L, Beltran11}, which hosts the luminous proto-OB cluster. VLA observations of NH$_3$ (3,3) satellite hyperfine line absorption at $0.1^{\prime \prime}$ showed that the dense gas comprising the hot toroid is extremely clumpy \citep{Sollins05}. Overall, the geometric configuration of the central ${\sim}1$~pc-scale massive molecular clump resembles that of a scaled-up version of a low-mass star-forming core and disk system viewed edge-on. For a more detailed, multi-scale view of G10.6 and its environment, see \citet{Liu17}.

\subsection{Observed Small-Scale Structure of G10.6}

\begin{figure*}
\centering
\includegraphics[width=\linewidth]{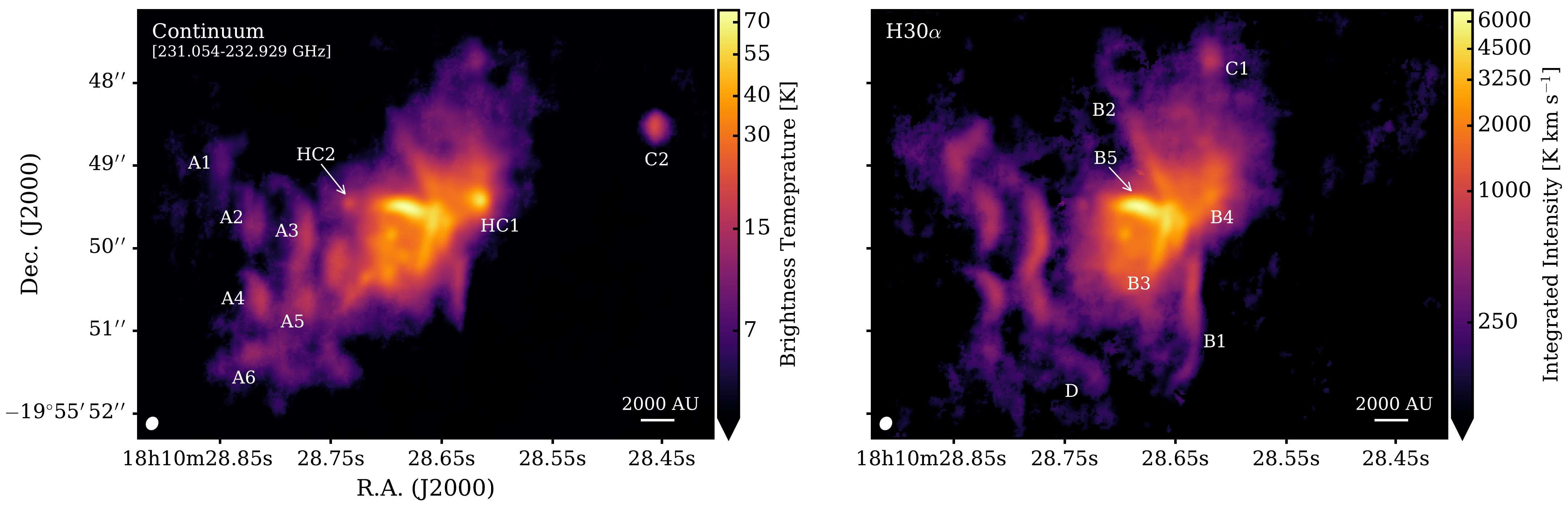}
\caption{Overview of 1.3~mm continuum (\textit{left}) and H30$\alpha$ line emission (\textit{right}) in G10.6. Only S/N~${\geq}5\sigma$ continuum emission is shown. Prominent morphological features are labeled as A1--A6, B1--B6, C1--C2, D, and HC1--HC2. A log10 stretch is applied to the color scales of both panels to increase the visibility of substructures throughout G10.6. The synthesized beam is shown in the lower left of each panel.}
\label{fig:Fig1}
\end{figure*}

To better contextualize COM emission in the central region of G10.6, it is beneficial to first understand the broader physical conditions and gas structures as revealed by the continuum emission and distribution of ionized gas as shown in Figure \ref{fig:Fig1}. For consistency, we adopt the same nomenclature used in \citet{Sollins05_detailed} and denote prominent features as A1--A6, B1--B5, C1--C2, and D. Features labeled as A correspond to localized arcs of ionized gas, and those marked as B are associated with a central cavity evacuated by a wide-angle outflow, as traced by strongly-emitted ionized gas. In particular, B1, B3, and B4 define the southwest edges of this outflow, while B2 and B5 mark the northeast edges. B5 is also the location of an elongated ridge containing the brightest H30$\alpha$ line emission and thus we also refer to it as the ``ionized ridge". Features labeled as C correspond to nearby but distinct \ion{H}{2} regions. C1  (``C" in \citet{Sollins05_detailed}) indicates a separate HC \ion{H}{2} region  located to the northwest, while C2 marks an independent UC \ion{H}{2} region to west of the main \ion{H}{2} region of G10.6. Feature D illustrates the presence of diffuse continuum and H30$\alpha$ line emission, which is present around the entire \ion{H}{2} region except toward the evacuated outflow cavity. We also introduce the HC1--HC2 markings to indicate the locations of two newly-discovered hot molecular cores, which are characterized by compact continuum emission. HC1 is located along B4, making it coincident with the northwest outflow edge, while HC2 lies to the west of B5 and toward the periphery of the northeast edge of the ionized cavity.

Although more spatially-extended, the ionized gas largely traces the continuum emission, as expected, since the majority of the millimeter continuum originates from free-free emission \citep{Liu_10ApJ_722_262L}. However, we see a local maximum in continuum emission toward both HC1 and C2, but do not observe similar peaks in the H30$\alpha$ line, which indicates contributions from thermal dust emission. A similar but less pronounced effect is seen toward HC2.  Interestingly, in the continuum map, we see diffuse material behind and between the ionized arcs traced by the H30$\alpha$ line emission, which implies that these structures are comprised of dusty neutral material whose surfaces are ionized. In fact, the H30$\alpha$ line map is more morphologically similar to the VLA 1.3~cm continuum presented in \citet{Sollins05_detailed} than to the 230~GHz continuum.

\section{Data Analysis} \label{sec:data_analysis}

\subsection{Molecule Sample}
The ALMA observations of G10.6 cover a wide bandwidth, which provides coverage of numerous rotational transitions of many molecular species. We selected a chemically representative set of COMs commonly observed toward massive star-forming regions. These molecules are: CH$_3$OH, CH$_3$OCH$_3$, CH$_3$OCHO, CH$_3$CHO, NH$_2$CHO, CH$_3$CH$_2$CN, CH$_2$CHCN, CH$_3$CN, and HNCO. We also included the isotopologues $^{13}$CH$_3$CN, CH$_3^{13}$CN, and $^{13}$CH$_3$OH, which not only provide valuable information about isotopic abundances in G10.6, but also trace denser and more optically thin gas relative to their main species. Each species had a sufficient number of lines to robustly derive rotational and column densities via a population diagram method over a wide spatial area in G10.6. Table \ref{tab:tab2} summarizes the number of identified lines, range of upper state energies, and Einstein A coefficients covered by these lines. An example of the analysis process is shown for CH$_3$OH in the following subsections.

\begin{deluxetable*}{ccccccccccc}
\tablecaption{Summary of Targeted Species\label{tab:tab2}}
\tablehead{[-.3cm]
\colhead{Species} & \colhead{Name} &  N$_{\rm{lines}}$\tablenotemark{a} & \colhead{E$_{\rm{u}}$} & \colhead{A$_{\rm{ul}}$} & \colhead{Catalog} & \multicolumn{2}{c}{Fitting Threshold\tablenotemark{b}}  \\[-0.05cm]
\cline{7-8} & & & & & & \\[-.45cm]
& & & [K] & $\times 10^{-5}$ [s$^{-1}$] & & \colhead{N$_{\rm{lines}}$}  & \colhead{$\Delta$E$_{\rm{u}}$ [K]} \\[-.55cm]}
\startdata
CH$_3$OH        & Methanol       & 10 & 46 -- 802  & 2.04 -- 4.69     & CDMS & 3  & 51   \\
$^{13}$CH$_3$OH &                & 5  & 48 -- 594  & 1.30 -- 5.27     & CDMS & 3  & 206 \\ 
CH$_3$OCH$_3$   & Dimethyl Ether & 7  & 81 -- 673  & 1.44 -- 9.14     & CDMS & 3  & 57 \\
CH$_3$OCHO      & Methyl Formate & 31 & 100 -- 282 & 3.42 -- 18.39    & JPL  & 5  & 100 \\ 
CH$_3$CHO       & Acetaldehyde   & 11 & 82 -- 183  & 28.84 -- 44.45   & JPL  & 4  & 50 \\
NH$_2$CHO       & Formamide      & 10 & 61 -- 175  & 74.75 -- 147.23  & CDMS & 9  & 114 \\
CH$_3$CN        & Methyl Cyanide & 11 & 69 -- 931  & 10.10 -- 63.71   & JPL  & 7  & 257 \\
$^{13}$CH$_3$CN &                & 10 & 78 -- 942  & 30.45 -- 107.96  & JPL  & 8  & 350 \\
CH$_3^{13}$CN   &                & 8  & 69 -- 419  & 60.71 -- 92.31   & JPL  & 6  & 179 \\
CH$_3$CH$_2$CN  & Ethyl Cyanide  & 31 & 33 -- 837  & 3.04 -- 105.49   & CDMS & 18 & 172 \\
CH$_2$CHCN      & Vinyl Cyanide  & 12 & 135 -- 435 & 216.94 -- 321.65 & CDMS & 5  & 25 \\
HNCO            & Isocyanic Acid & 10 & 58 -- 1050 & 6.68 -- 24.02    & CDMS & 3  & 170 \\
\enddata
\tablenotetext{a}{Combined multiplets are counted as one line and we only consider unblended lines.}
\tablenotetext{b}{This empirical threshold corresponds to the minimum number of unblended lines and range in E$_{\rm{u}}$ values that we required to be present in each pixel for a fit to be attempted. Such cutoffs were employed to ensure that all reported rotational temperature and column density determinations were robust.}
\end{deluxetable*}

\begin{figure*}
\centering
\includegraphics[width=\linewidth]{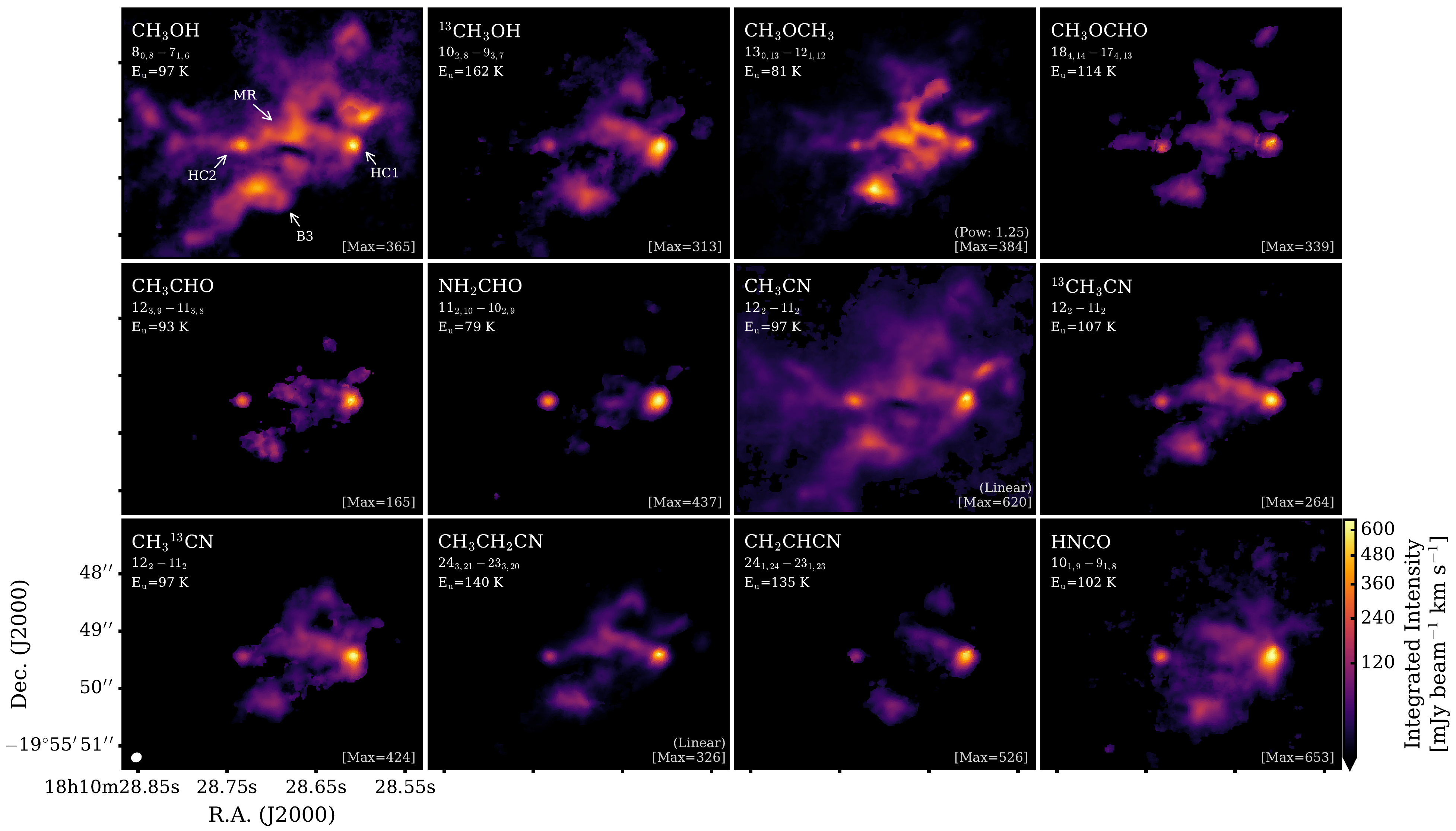}
\caption{Integrated intensity maps for transitions with E$_{\rm{u}} \approx 100$~K for all molecules in our sample. A square root stretch, unless otherwise indicated, is applied to each panel to better illustrate line substructure. The peak integrated line intensity is shown in the lower right of each panel.}
\label{fig:Fig_int_sum}
\end{figure*}

To illustrate the chemical and physical diversity across G10.6,  Figure \ref{fig:Fig_int_sum} shows integrated intensity maps for representative lines with comparable excitation conditions (E$_{\rm{u}}\approx100$~K) from each COM in our sample. We also extracted spectra from representative positions labeled in Figures \ref{fig:Fig1} and \ref{fig:Fig_int_sum}. Spectra covering half of the observed bandwidth are shown in Figure \ref{fig:Fig2}, while the other half are shown in Figure A\ref{fig:FigA1} in Appendix \ref{appendix:Extra_Spec_Appendix}. Key lines are labeled, but we did not attempt a detailed identification of all lines within the spectral coverage and instead focus on those originating from the selected COMs. A high degree of variation in line strength, width, and richness is evident. Some COMs, e.g., NH$_2$CHO and CH$_2$CHCN, are seen primarily toward HC1, while other COMs, such as CH$_3$CN and CH$_3$OH, are observed in almost all positions, albeit at varying intensity levels.

\begin{figure*}
\centering
\includegraphics[width=\linewidth]{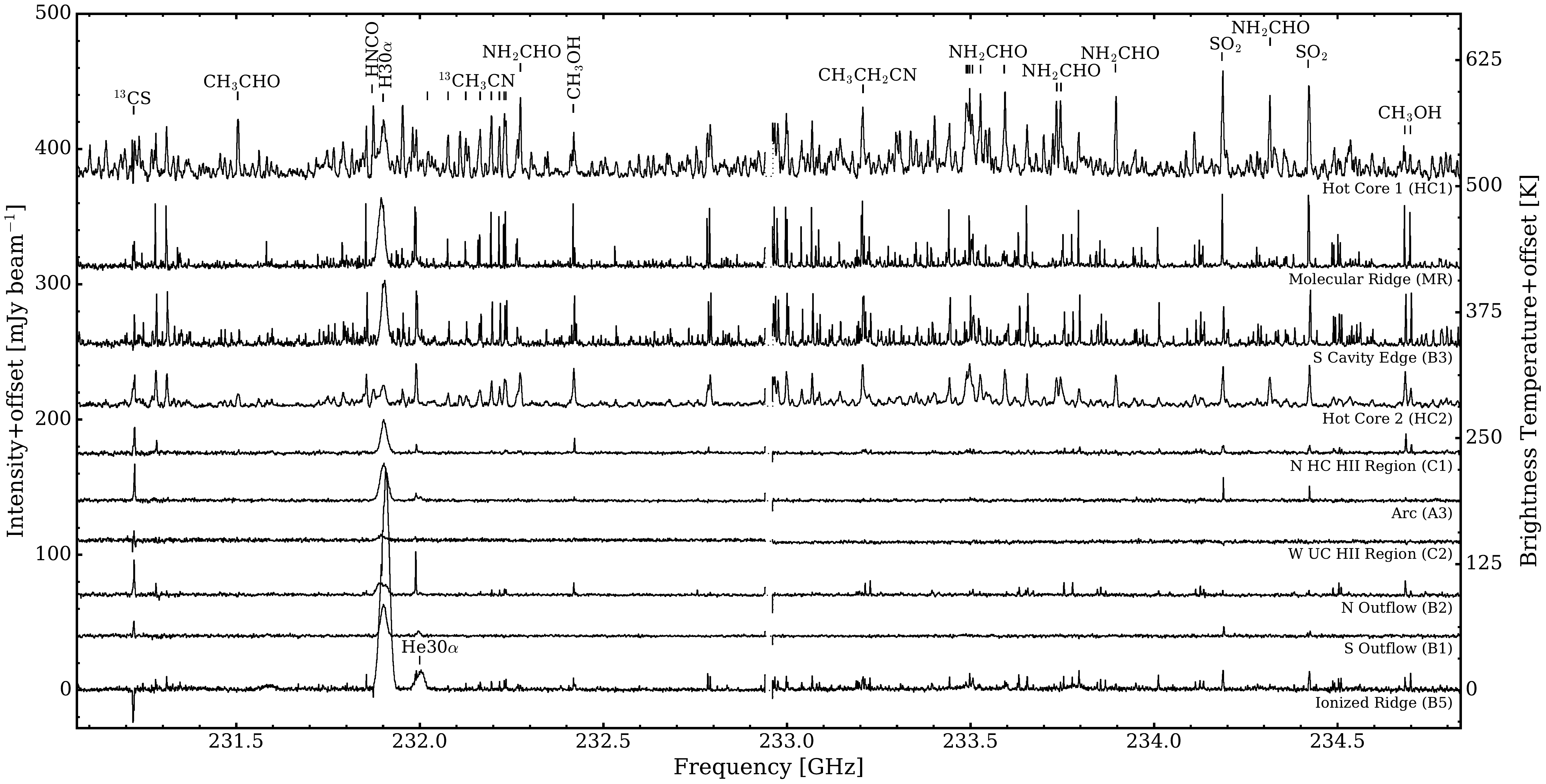}
\caption{Spectra extracted from single pixels toward ten representative positions across the central 20,000~au of G10.6. Dashed lines represent gaps in observational coverage between spectral windows.}
\label{fig:Fig2}
\end{figure*}

\subsection{Fitting Process}
For each species in the selected COM sample, we identified and fit all unblended transitions with single Gaussian profiles. All lines belonging to a single species were fit consistently. Blended lines were removed via visual inspection, while line assignments were checked by verifying that the best-fit model for each species did not predict lines at frequencies, covered by our observations, where no emission is detected.

\begin{figure*}
\centering
\includegraphics[width=\linewidth]{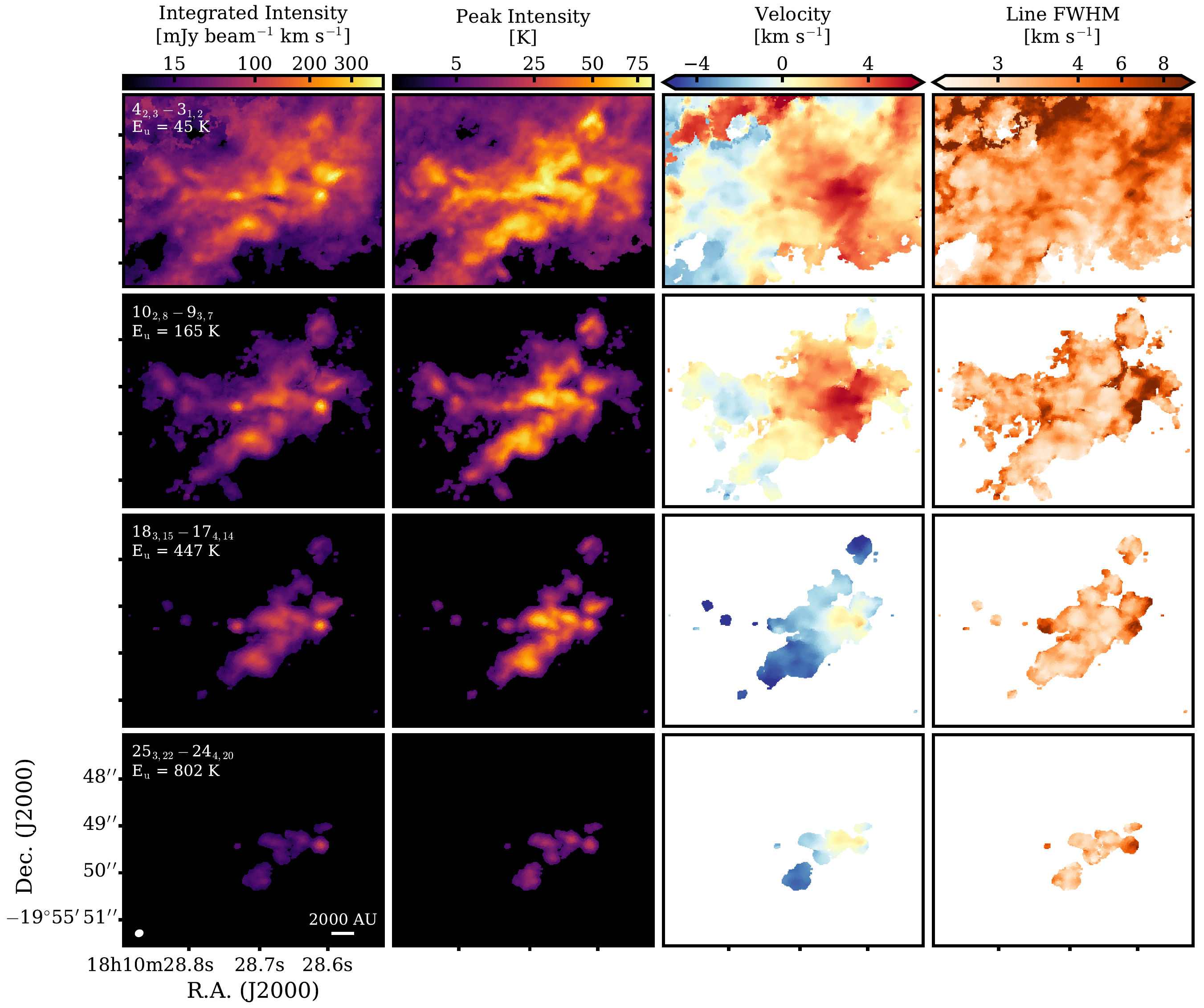}
\caption{Integrated intensity, peak intensity, systemic velocity, and line FWHM (\textit{columns from left to right}) are shown for four CH$_3$OH transitions spanning 45--802~K in E$_{\rm{u}}$ toward G10.6. Peak line intensity maps are included, as it is often easier to identify spatial trends across wide dynamic ranges. Each column shares the same color scale. To highlight gas substructure, a square root stretch is applied to the integrated and peak intensity maps, while an arcsinh stretch is used for the line FWHM map.}
\label{fig:Fig4}
\end{figure*}

In order to verify the spatial consistency of the spectral fits, we also visually inspected integrated intensity, systemic velocity, and line FWHM maps for each transition included in our analysis. In all cases, we found a reasonably coherent and smoothly-varying morphology throughout G10.6 (see Appendix \ref{appendix:COM_Morph_Kine_Appendix}). We also confirmed that, in each case, the distribution of velocity and line FWHM values were not sharply peaked toward either the minimum or maximum limits designated in the fitting code. A representative set of maps are shown in Figure \ref{fig:Fig4} for CH$_3$OH.

We then constructed maps of rotational temperature and column density for each COM species using a rotational diagram analysis \citep[e.g.,][]{Goldsmith99} on a pixel-by-pixel basis. To ensure secure determinations of rotational temperature and column density across a large spatial extent in G10.6, we restricted our fittings to those pixels with a sufficient number of lines with a wide coverage of E$_{\rm{u}}$, as shown in Table \ref{tab:tab2}.

\begin{figure}
\centering
\includegraphics[width=\linewidth]{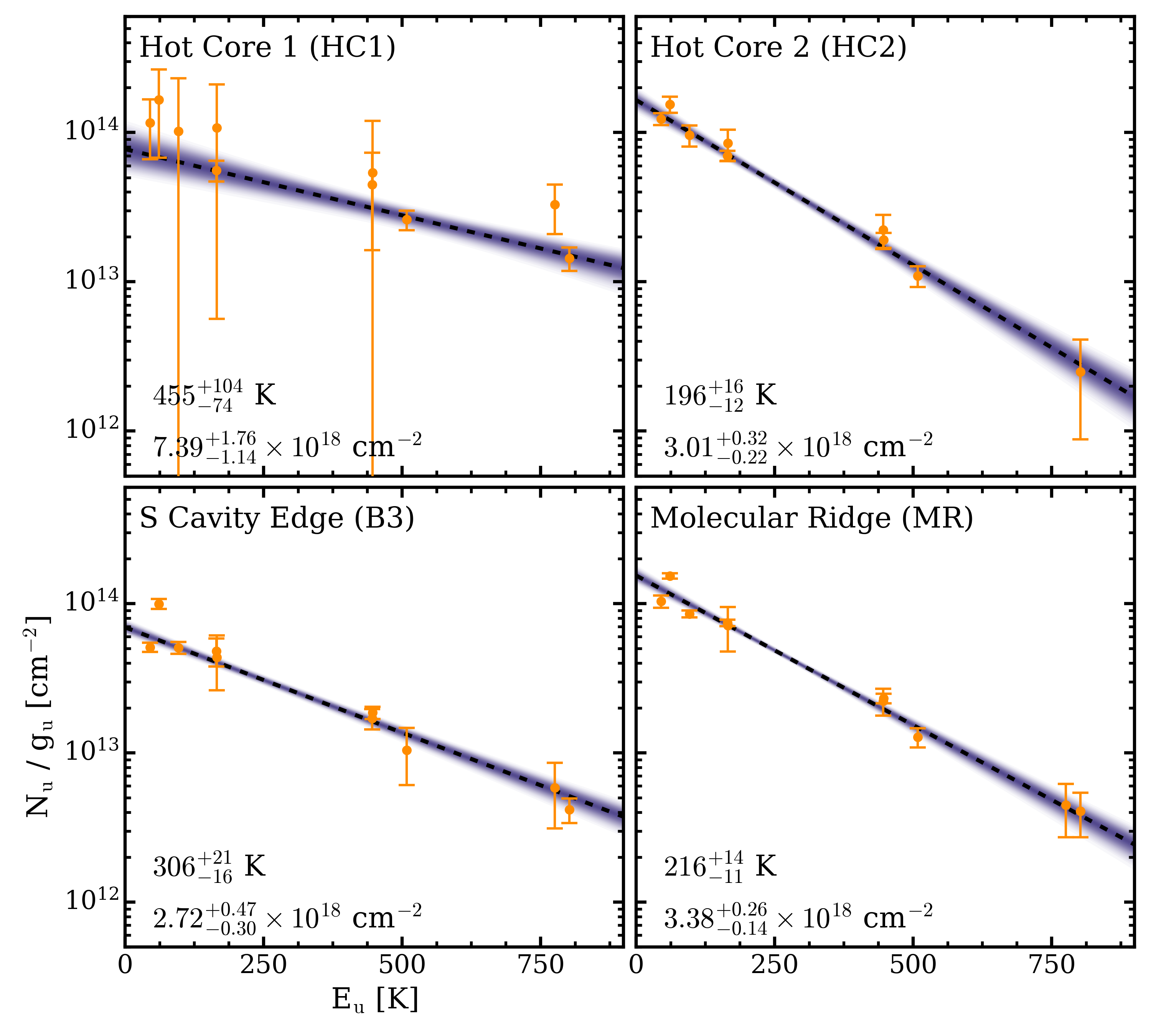}
\caption{CH$_3$OH rotational diagrams toward positions of interest in G10.6. Transitions are shown in orange and random draws from the fit posteriors are plotted in purple. The fit is indicated by a dashed, black line and the derived values are shown in the lower left corners.}
\label{fig:Fig5}
\end{figure}

We used the Markov Chain Monte Carlo (MCMC) code \texttt{emcee} \citep{Foreman13} to generate posterior probability distributions of column densities and rotational temperatures consistent with the observed data. Random draws from these posteriors are plotted in purple for several positions toward G10.6 in Figure \ref{fig:Fig5} with $\tau$-corrected values of N$_{\rm{u}}/$g$_{\rm{u}}$ plotted against E$_{\rm{u}}$. The derived parameters and uncertainties are listed as the 50th, 16th, and 84th percentiles from the marginalized posterior distributions, respectively.

In general, we initially explored a parameter space from $10^{14}$~cm$^{-2} \leq $ N$_{\rm{col}} \leq 10^{20}$~cm$^{-2}$ and 10~K $\leq$ T$_{\rm{rot}} \leq 800$~K, based on physical gas conditions expected in massive star-forming regions \citep{Bisschop07, Suzuki18, Molet19}. After this first model run, we ran a second iteration with narrower, species-specific grids and then performed quality checks to ensure that the fitted T$_{\rm{rot}}$ and N$_{\rm{col}}$ values were not sharply peaked at the extremes of our priors. For each line and species, we also visually inspected the fitted $\tau$ values across G10.6 and ensured that $\tau$ was spatially well-behaved and that $\tau \not \gg 1$. Optically-thick lines were excluded from the rotational diagram analysis. A detailed discussion of the line fitting process and rotational temperature and column density determination is presented in Appendix \ref{appendix:fit_details_appendix}.

\begin{figure*}
\centering
\includegraphics[width=\linewidth]{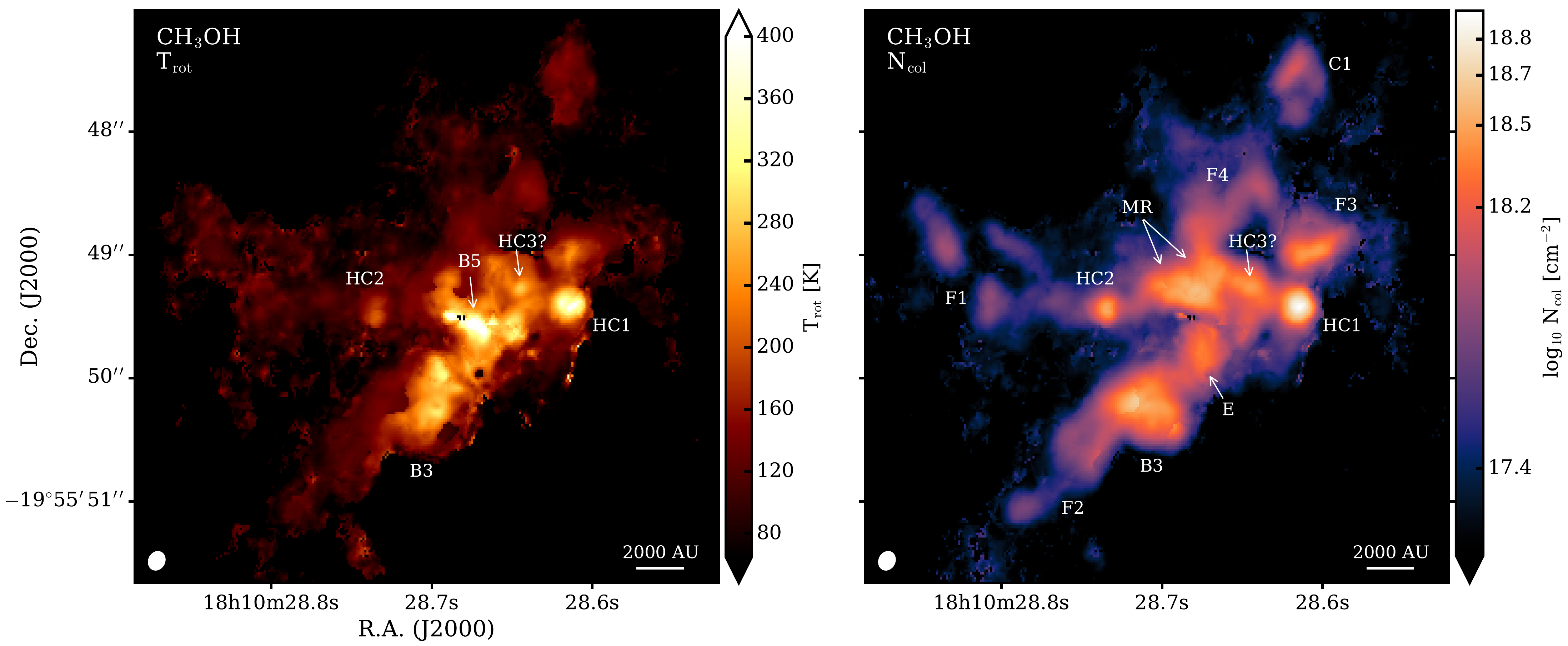}
\caption{Rotational temperature (\textit{left}) and column density (\textit{right}) maps for CH$_3$OH. Prominent gas substructures are labeled as HC1-3, MR, E, and F1-4, in addition to the nomenclature from \citet{Sollins05_detailed} used in Figure \ref{fig:Fig1}. An arcsinh stretch is applied to the color scale of the column density map to increase the visibility of substructures throughout G10.6. The synthesized beam is shown in the lower left of each panel.}
\label{fig:Fig7}
\end{figure*}

The resultant rotational temperature and column density maps derived for CH$_3$OH are shown in Figure \ref{fig:Fig7}. As a result of this MCMC fitting process, we also derived formal temperature and column density uncertainties, which were found to be low toward the central region of G10.6, but became large toward the edges of the map. This is not unexpected as we detect fewer high E$_{\rm{u}}$ lines toward the edges of the map and the lines that we do detect decrease in intensity relative to those toward the central region of G10.6, which increases the Gaussian fitting errors. A similar trend was found in all of the molecules in our sample and in no cases did we observe large spatial regions of unexpectedly high uncertainties. In addition to the initial quality cuts shown in Table \ref{tab:tab2}, we also excluded pixels with particularly high uncertainties (${>}200\%$) and those with highly discontinuous and outlier values. Thus, a lack of model result for a specific pixel does not necessarily imply the absence of molecular emission, as is clearly seen when comparing integrated line intensities (Figure \ref{fig:Fig_int_sum}) and column density maps (Figure \ref{fig:Fig10}) for several species.

Column densities and rotational temperature maps were derived using the method above with no modifications for six out of nine species in our sample. However, three molecules warranted special treatment: CH$_3$OCHO, CH$_3$CH$_2$CN, and CH$_3$CN. We ware unable to derive independent rotational temperatures for CH$_3$OCHO, which exhibited unusually large scatter in its rotational diagrams, and instead adopted those of the chemically-similar molecule CH$_3$OH \citep[e.g.,][]{Garrod06, Garrod08} to derive column density. A two component rotational diagram was required to fit the majority of CH$_3$CH$_2$CN emission and regions of optically thick CH$_3$CN emission required the use of optically thin isotope CH$_3^{13}$CN. These latter two cases are discussed in more detail in the following two subsections.

\subsection{Two CH$_3$CH$_2$CN Temperature Components}

All line data were further analyzed to explore whether the data were better fit by two distinct temperature components rather than a single component. A two-component fit was only favored for CH$_3$CH$_2$CN, in which we clearly identified two separate temperature components throughout most of G10.6. Wherever two components were needed, most of the column density is carried by the colder component, which we therefore designate as the primary component. To not overfit the data, we restricted this two-component analysis to instances where the secondary component had a rotational temperature that was $1.5\times$ in excess of that of the primary component. Otherwise, we fit that pixel with a single rotational temperature. We defined the total column density of CH$_3$CH$_2$CN as the sum of both components and used the total column density for all subsequent analysis. Previous indications of multiple CH$_3$CH$_2$CN temperature components were seen toward Orion KL by \citet{Daly13}, who also observed a similar transition between different components at E$_{\rm{u}}\approx$150--200~K.

\begin{figure*}
\centering
\includegraphics[width=0.49\linewidth]{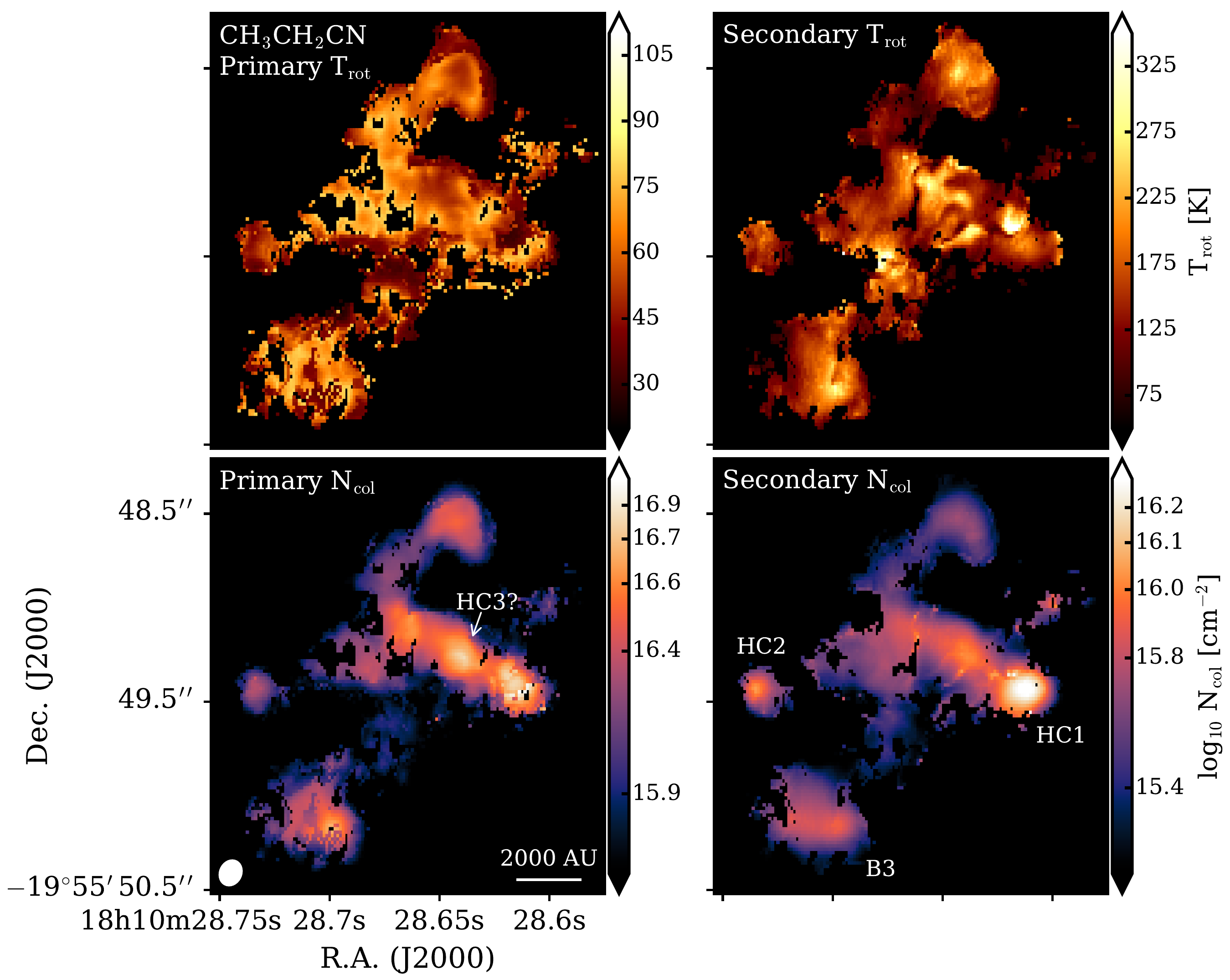}
\includegraphics[width=0.49\linewidth]{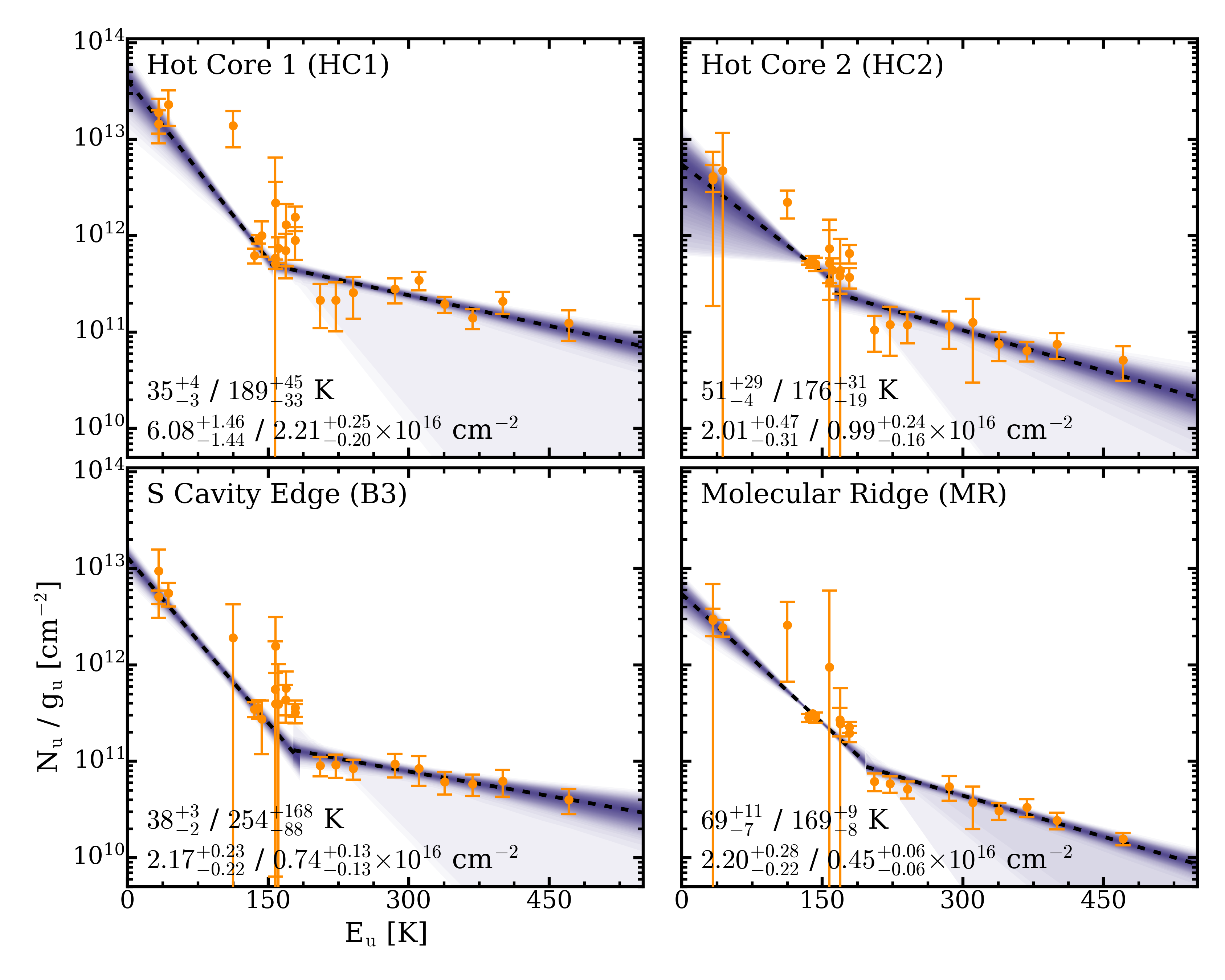}
\caption{Two-component rotational diagram analysis for CH$_3$CH$_2$CN. \textit{Left four panels}: Primary and secondary rotational temperature and column density maps. \textit{Right four panels}: Representative rotational diagrams showing two-component fits toward regions of interest in G10.6.}
\label{fig:CH3CH2CN_2_comp}
\end{figure*}

\subsection{Optically Thick CH$_3$CN and Isotopologues}

Initial efforts to fit the central regions of CH$_3$CN revealed optically thick gas, as indicated by poor fittings and unphysical temperatures (T$_{\rm{rot}} \gtrsim 1000$~K). High line optical depths ($\tau \sim 0.5$--$1.0$) were also observed in the 13--12 K-ladder of the $^{13}$CH$_3$CN isotopologue with the K=0--3 lines having $\tau \gtrsim 1$. However, the 12--11 K-ladder of CH$_3^{13}$CN was found to be optically thin ($\tau < 0.1$)\footnote{The observed optical depth differences between $^{13}$CH$_3$CN 13--12 and CH$_3^{13}$CN 12--11 are somewhat surprising, given their similar A$_{\rm{ul}}$ and E$_{\rm{u}}$ values, and may be due to self-absorption which is not resolved in frequency, line blending, or different $^{12}$C/$^{13}$C ratios for non-equivalent carbon atoms.}, allowing for secure rotational temperature and column density determinations. This approach requires a $^{12}$C/$^{13}$C scaling factor, however, which is not known for CH$_3$CN in G10.6. 

To derive this factor, we determined the $^{12}$C/$^{13}$C ratio from the column density ratios of CH$_3$OH/$^{13}$CH$_3$OH. A detailed discussion of this process is provided Appendix \ref{appendix:Iso_CH3OH_Appendix}. Overall, we find a median column density ratio of 22.8, which is the value we adopt. This is a factor of two lower than the expected ratio of 43 \citep{Milam05} at the distance of G10.6 (D$_{\rm{GC}} = 3.9$~kpc), but similar low values of $^{12}$C/$^{13}$C have been derived in COMs at core scales in star-forming regions, e.g., hot corinos \citep{Jorgensen16} and hot cores \citep{Sanna14_G023, Beltran18, Bogelund19}. However, the carbon isotopic ratio in G10.6 needs to be revisited with more comprehensive isotopologue data and as a result, the reported CH$_3$CN column densities in high density regions should only be considered accurate within a factor of 2.

\section{Results} \label{sec:results}

\subsection{CH$_3$OH Substructure} \label{sec:ch3oh_substrc}

COMs exhibit complex emission morphologies and substantial gas substructures across G10.6. As shown in Figure \ref{fig:Fig7}, this is well-illustrated by the rotational temperature and column density maps of CH$_3$OH, the most spatially-extended map in our sample. When appropriate, we refer to COM features, namely B3-5, C1, and HC1-2, using the nomenclature introduced above. For instance, hot cores HC1 and HC2 are clearly identified by their compact, high column densities and elevated temperatures. The south cavity edge (B3) contains a spatially-extended region of hot, high density gas, while the northeast cavity edge (B5) exhibits the hottest gas (${\gtrsim}400$~K) anywhere in G10.6 and possesses a relative deficit of molecular gas. Since B5 is also the location of an ionized ridge of material containing the brightest H30$\alpha$ line emission in G10.6, the hot, low column density gas associated with B5 may be the result of ionized gas heating and ultimately destroying some of surrounding molecular gas. The C1 region, which appeared compact and symmetric in continuum and ionized gas, is instead highly-structured in molecular gas and based on its arrow-shaped morphology, appears to be a cometary HC \ion{H}{2} region \citep[e.g.,][]{Wood89}.

Newly-identified gas substructures which have no obvious correspondence to the \cite{Sollins05_detailed} features, are labeled as E, F1-F4, and MR. Feature E denotes a clump of gas within the central cavity of G10.6 and is bounded by the cavity edges B3 and B5. Four separate filamentary structures are marked as F. F1 is a tri-pronged structure on the eastern edge of G10.6, while F2 is comprised of a single narrow filament extending beyond B3. The F3 filament lies directly to the north of HC1 and may be related to hot core activity or may be simply a further extension of the ``V"-shaped F4 filament located to the north of the MR. An elongated and somewhat irregular ``Z"-shaped structure is seen immediately to the north of B5. As it is readily identified by a band of high density gas, we thus refer to this structure as the ``molecular ridge," denoting it as MR. The shape of the MR suggests that it is interacting with the ionized gas in B5 and its relative molecular richness (e.g., Figure \ref{fig:Fig2}) and high column density may be attributed to external heating from B5. The western portion of the MR appears to be marginally resolved into a core-like structure, which is tentatively labeled as HC3.

\subsection{Rotational Temperatures and Column Densities} \label{sec:rot_temp_and_col_dense}

\begin{figure*}
\centering
\includegraphics[width=\linewidth]{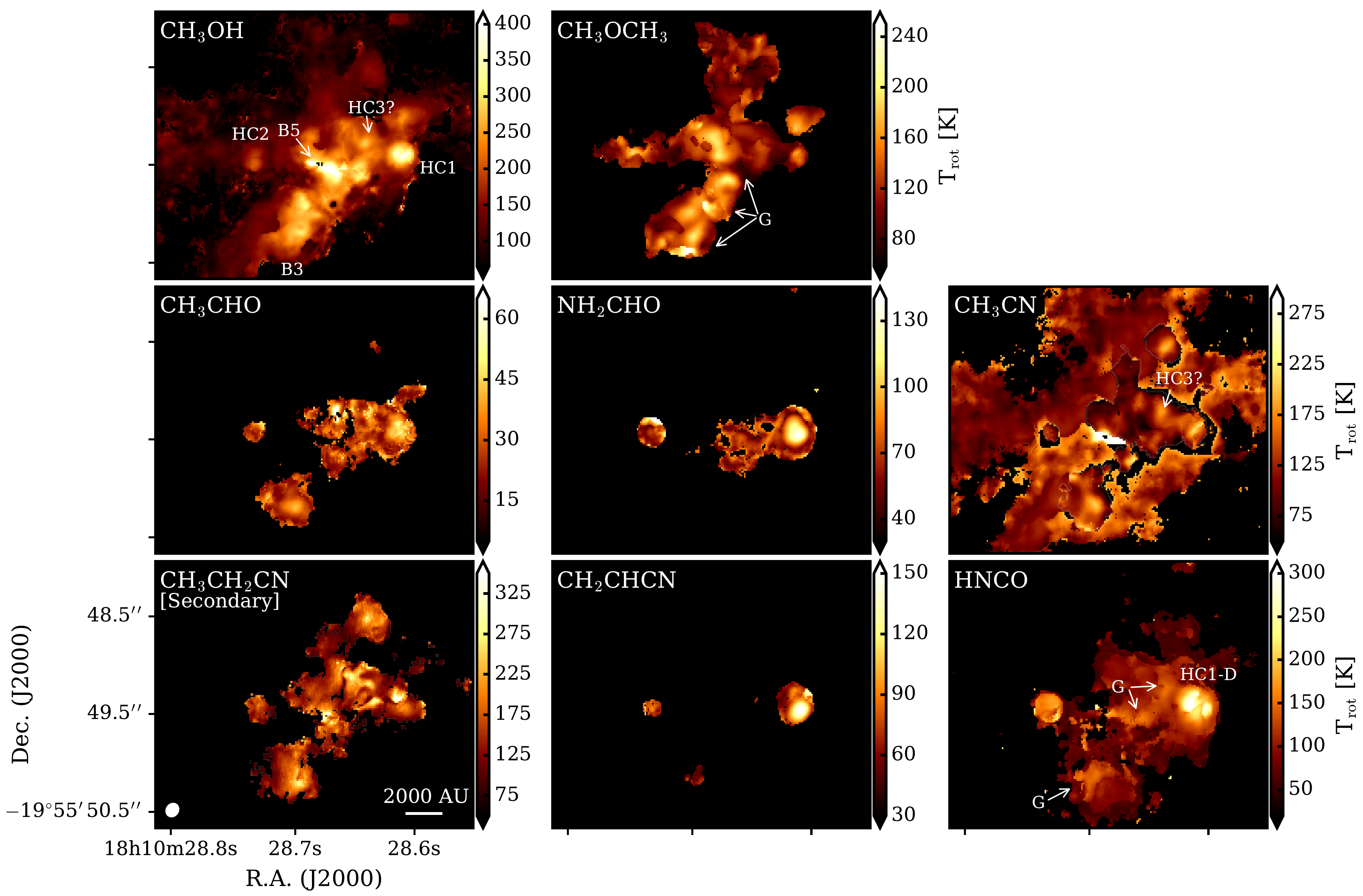}
\caption{Rotational temperature maps for all species in our sample. The secondary temperature component of CH$_3$CH$_2$CN is shown. The white contours in CH$_3$CN indicate the regions where rotational temperatures were derived using the CH$_3^{13}$CN isotopologue, as detailed in the text.}
\label{fig:Fig9}
\end{figure*}

\begin{figure*}
\centering
\includegraphics[width=\linewidth]{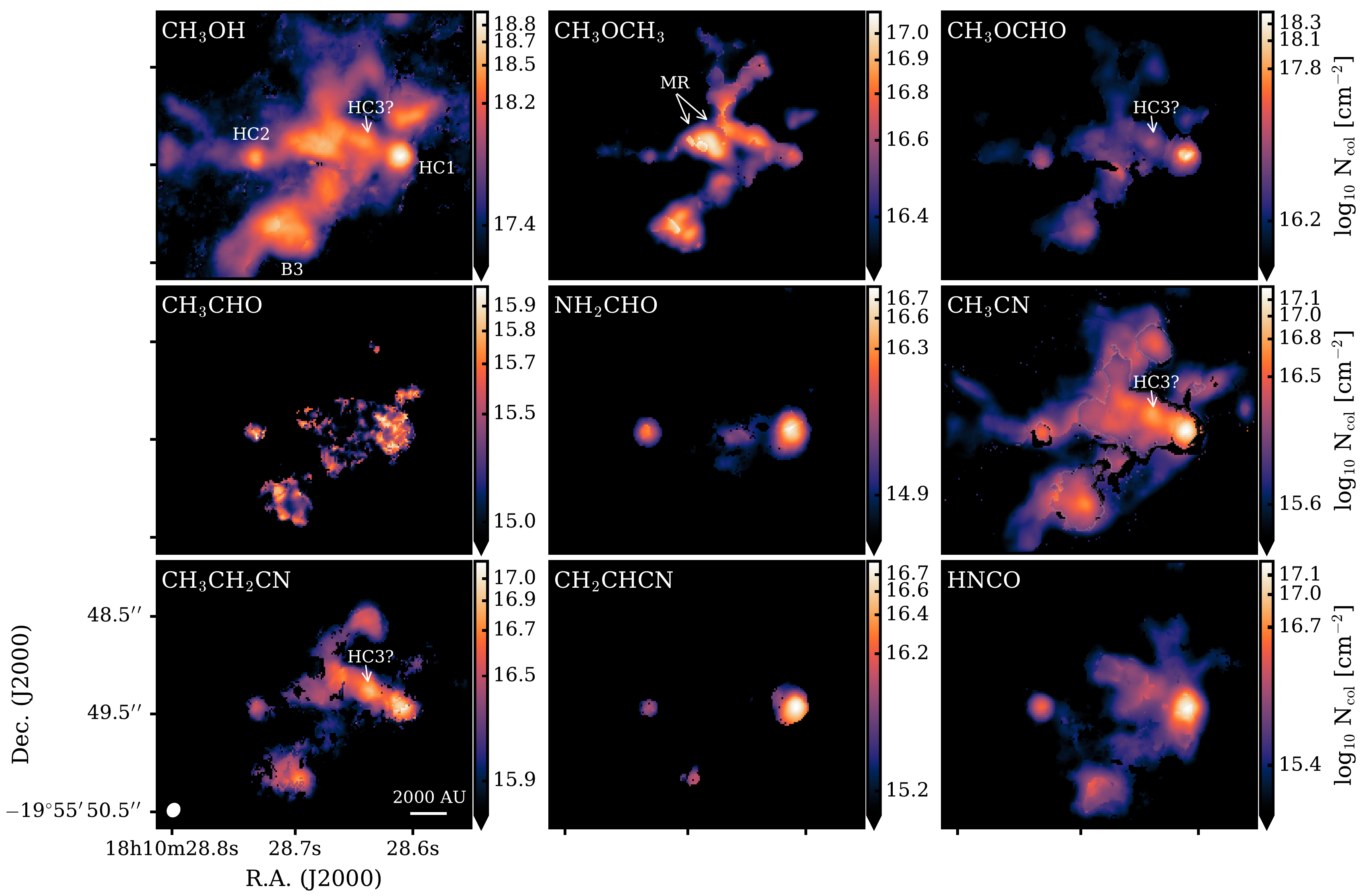}
\caption{Column density maps for all species in our sample. The patchy nature of the CH$_3$CHO is due to higher fitting uncertainties, rather than true small-scale variations. The total column density, primary plus secondary components, is shown for CH$_3$CH$_2$CN. The white contours in CH$_3$CN indicate the regions where column densities were derived using the CH$_3^{13}$CN isotopologue, as detailed in the text. An arcsinh stretch is applied to the color scales of all panels to increase the visibility of substructures throughout G10.6.}
\label{fig:Fig10}
\end{figure*}

Spatially-resolved rotational temperature and column density maps for all molecules in our sample are shown in Figures \ref{fig:Fig9} and \ref{fig:Fig10}, respectively. In terms of spatial coverage, NH$_2$CHO and CH$_2$CHCN are the most limited, CH$_3$OH and CH$_3$CN are the most extended, and the remaining species have intermediate spatial distributions. HC1 is prominent in all species, exhibiting elevated rotational temperatures and high column densities. HC2 is also seen in all column density maps, but is less enhanced in rotational temperature (except for HNCO). The MR and B3 also consistently exhibit high column densities in the majority of COMs. Filamentary features are seen in most molecules, although with varying extents and column densities. For instance, CH$_3$OCH$_3$ and CH$_3$OCHO show narrow features, e.g., F1 and F4, while CH$_3$CN is the only species with extended substructure comparable to that of CH$_3$OH. Interestingly, HC3 is consistently marginally resolved (e.g., CH$_3$OCHO, CH$_3$CN, CH$_3$CH$_2$CN), indicating the potential presence of high density clumps/cores that are unresolved in our observations.

The rotational temperature maps reveal that there is no single temperature that characterizes all COMs at a specific location. Differences in temperatures span 100s of K, indicating that multiple environments are traced along the line of sight in this complex region. On average, CH$_3$CHO exhibits the lowest rotational temperatures (${\lesssim}60$~K), while NH$_2$CHO and CH$_2$CHCN have moderate temperatures (${\sim}70$--$150$~K). CH$_3$CN, CH$_3$OH, CH$_3$OCH$_3$, CH$_3$CH$_2$CN, and HNCO all possess substantially warmer temperatures (${\sim}$150--400~K).

Within the N-bearing COM family, we notice prominent differences between the spatial distributions of complex cyanides (CH$_3$CN, CH$_3$CH$_2$CN, CH$_2$CHCN) and HNCO. HNCO also displays a different temperature pattern, including a dual-peaked hot core 1 (HC1-D), a prominent HC2, and curved bands of elevated temperatures, which are marked as G in Figure \ref{fig:Fig9}. Together, these distinct column density and temperature patterns suggest that there is no single N-bearing COM chemistry and that the amount of nitrogen incorporated into different kinds of organics strongly depends on the local environment.

Among O-bearing species, CH$_3$OH, CH$_3$OCH$_3$, and CH$_3$OCHO each trace different structures in G10.6. CH$_3$OH is widely distributed throughout G10.6, as described in Section \ref{sec:ch3oh_substrc}, while CH$_3$OCHO exhibits a significant enhancement in HC1 and shares broad morphological similarities with CH$_3$CHO, notwithstanding its increased uncertainties as reflected in a less uniform column density map. CH$_3$OCH$_3$ has an inverted ``C"-shaped distribution of hot, clumpy gas (labeled as G) that connects the western edge of the MR to B3 and presents its highest column density in the MR rather than in HC1. This temperature and column density distribution is not seen in any other molecule in our sample, and may reflect the fact that CH$_3$OCH$_3$ is responding differently to strong radiation fields from nearby B5 than the other O-bearing COMs. By contrast, CH$_3$OH only shows weak enhancements along the MR, and CH$_3$OCHO shows none at all. Notably, the CH$_3$OCH$_3$ and CH$_3$OCHO maps appear strikingly different, which is somewhat surprising considering that all proposed chemical formation routes \citep[e.g.,][]{Garrod06} indicate that they are directly linked.

\begin{deluxetable*}{lccclcccc}
\tablecaption{COM Rotational Temperatures and Column Densities at Regions of Interest in G10.6 \label{tab:Table3}}
\tablehead{[-.3cm]
\colhead{} & \colhead{T$_{\rm{rot}}$} & \colhead{N$_{\rm{X}}$} & \colhead{N$_{\rm{X} / \rm{CH}_{3}\rm{OH}}$ } & \colhead{} & \colhead{T$_{\rm{rot}}$} &  \colhead{N$_{\rm{X}}$} & \colhead{N$_{\rm{X} / \rm{CH}_{3}\rm{OH}}$ } \\[-.15cm]
\colhead{} & \colhead{[K]} & \colhead{[$10^{16}$ cm$^{-2}$]} & \colhead{[\%]} & & \colhead{[K]} & \colhead{[$10^{16}$ cm$^{-2}$]} & \colhead{[\%]}   \\[-.7cm]}
\startdata
\textbf{Hot Core 1 (HC1)}	&			&		&		&	\textbf{Hot Core 2 (HC2)}	&			&		&		 \\[-.1cm]
CH$_3$OH			&	$455^{104}_{74}$	&	$739^{176}_{114}$	&	$100$	&	CH$_3$OH				&	$196^{16}_{12}$	&	$301^{32}_{22}$	&	$100$	\\ 
CH$_3$OCH$_3$			&	$147^{51}_{55}$	&	$5.51^{2.95}_{1.38}$	&	$0.75$	&	CH$_3$OCH$_3$				&	$101^{68}_{75}$	&	$3.51^{1.82}_{1.03}$	&	$1.2$	\\ 
CH$_3$OCHO			&	$\mathit{455^{104}_{74}}$	&	$259^{216}_{141}$	&	$35$	&	CH$_3$OCHO				&	$\mathit{196^{16}_{12}}$	&	$12.0^{7.5}_{7.1}$	&	$4.0$	\\ 
CH$_3$CHO			&	$49^{38}_{20}$	&	$0.397^{0.026}_{0.024}$	&	$0.054$	&	CH$_3$CHO				&	$24^{13}_{5}$	&	$0.625^{0.032}_{0.036}$	&	$0.21$	\\ 
NH$_2$CHO			&	$130^{2}_{2}$	&	$4.23^{0.36}_{0.33}$	&	$0.57$	&	NH$_2$CHO				&	$63^{7}_{4}$	&	$1.27^{0.22}_{0.18}$	&	$0.42$	\\ 
CH$_3$CN			&	$\mathit{201^{79}_{47}}$	&	$14.8^{5.02}_{2.74}$	&	$2.0$	&	CH$_3$CN				&	$\mathit{131^{42}_{17}}$	&	$3.65^{0.821}_{0.479}$	&	$1.2$	\\ 
CH$_3^{13}$CN			&	$201^{79}_{47}$	&	$0.654^{0.218}_{0.119}$	&	$0.088$	&	CH$_3^{13}$CN				&	$131^{42}_{17}$	&	$0.164^{0.036}_{0.021}$	&	$0.054$	\\ 
CH$_3$CH$_2$CN:			&	\ldots	&	$8.29^{1.48}_{1.45}$	&	$1.1$	&	CH$_3$CH$_2$CN:				&	\ldots	&	$3.00^{0.53}_{0.35}$	&	$1.0$	\\ 
\quad Primary			&	$35^{4}_{3}$	&	$6.08^{1.46}_{1.44}$	&	$0.82$	&	\quad Primary				&	$51^{29}_{4}$	&	$2.01^{0.47}_{0.31}$	&	$0.67$	\\ 
\quad Secondary			&	$189^{45}_{33}$	&	$2.21^{0.25}_{0.20}$	&	$0.30$	&	\quad Secondary				&	$176^{31}_{19}$	&	 $0.99^{0.24}_{0.16}$	&	$0.33$	\\ 
CH$_2$CHCN			&	$139^{41}_{15}$	&	$6.19^{0.73}_{0.55}$	&	$0.84$	&	CH$_2$CHCN				&	$84^{41}_{15}$	&	$0.747^{0.058}_{0.053}$	&	$0.25$	\\ 
HNCO			&	$237^{42}_{27}$	&	$14.1^{6.6}_{1.8}$	&	$1.9$	&	HNCO				&	$166^{22}_{19}$	&	$2.72^{0.25}_{0.22}$	&	$0.90$	\\ \hline 
\textbf{S Cavity Edge (B3)}	&			&		&		&	\textbf{Molecular Ridge (MR)}	&			&		&		 \\[-.1cm]
CH$_3$OH			&	$306^{21}_{16}$	&	$272^{47}_{30}$	&	$100$	&	CH$_3$OH				&	$216^{14}_{11}$	&	$338^{26}_{14}$	&	$100$	\\ 
CH$_3$OCH$_3$			&	$179^{30}_{32}$	&	$6.94^{2.65}_{1.57}$	&	$2.6$	&	CH$_3$OCH$_3$				&	$179^{25}_{28}$	&	$9.06^{2.47}_{1.39}$	&	$2.7$	\\ 
CH$_3$OCHO			&	$\mathit{306^{21}_{16}}$	&	$16.0^{19.6}_{8.9}$	&	$5.9$	&	CH$_3$OCHO				&	$\mathit{216^{14}_{11}}$	&	$5.28^{3.83}_{3.42}$	&	$1.6$	\\ 
CH$_3$CHO			&	$36^{13}_{5}$	&	$0.218^{0.032}_{0.036}$	&	$0.080$	&	CH$_3$CHO				&	$24^{51}_{13}$	&	$0.263^{0.027}_{0.024}$	&	$0.078$	\\ 
NH$_2$CHO			& $\mathit{135^{15}_{14}}$	&	${<}0.0102$	&	${<}0.0038$	&	NH$_2$CHO				&	$\mathit{78^{26}_{23}}$	&	${<}0.0101$	&	${<}0.0030$	\\ 
CH$_3$CN			&	$\mathit{203^{42}_{17}}$	&	$4.77^{0.821}_{0.479}$	&	$1.8$	&	CH$_3$CN				&	$\mathit{116^{13}_{11}}$	&	$2.51^{0.27}_{0.25}$	&	$0.74$	\\ 
CH$_3^{13}$CN			&	$203^{42}_{17}$	&	$0.209^{0.36}_{0.21}$	&	$0.077$	&	CH$_3^{13}$CN				&	$116^{13}_{11}$	&	$0.110^{0.012}_{0.011}$	&	$0.033$	\\ 
CH$_3$CH$_2$CN:			&	\ldots	&	$2.91^{0.26}_{0.26}$	&	$1.1$	&	CH$_3$CH$_2$CN:				&	\ldots	&	$2.65^{0.29}_{0.23}$	&	$0.79$	\\ 
\quad Primary			&	$38^{3}_{2}$	&	$2.17^{0.23}_{0.22}$	&	$0.80$	&	\quad Primary				&	$69^{11}_{7}$	&	$2.20^{0.28}_{0.22}$	&	$0.65$	\\ 
\quad Secondary			&	$254^{168}_{88}$	&	$0.74^{0.13}_{0.13}$	&	$0.27$	&	\quad Secondary				&	$169^{9}_{8}$	&	$0.45^{0.06}_{0.06}$ 	&	$0.13$	\\ 
CH$_2$CHCN			&	$67^{2}_{2}$	&	$0.973^{0.173}_{0.180}$	&	$0.36$	&	CH$_2$CHCN				&	$\mathit{69^{11}_{7}}$	&	${<}0.0138$	&	${<}0.0041$	\\ 
HNCO			&	$135^{15}_{14}$	&	$1.18^{0.14}_{0.14}$	&	$0.43$	&	HNCO				&	$78^{27}_{24}$	&	$0.160^{0.043}_{0.032}$	&	$0.047$	\\ \hline 
\enddata
\tablecomments{Uncertainties at the 1$\sigma$ level are reported. T$_{\rm{rot}}$ values in italics are those adopted from similar species, as described in the text. Column density upper limits are derived using $3\sigma$ line intensity upper limits and by assuming the median line FWHM from a chemically-related or observationally-linked species \citep[e.g., CH$_3$CH$_2$CN/CH$_2$CHCN, HNCO/NH$_2$CHO;][]{Caselli93, Allen20}}
\end{deluxetable*}

\begin{figure*}
\centering
\includegraphics[width=\linewidth]{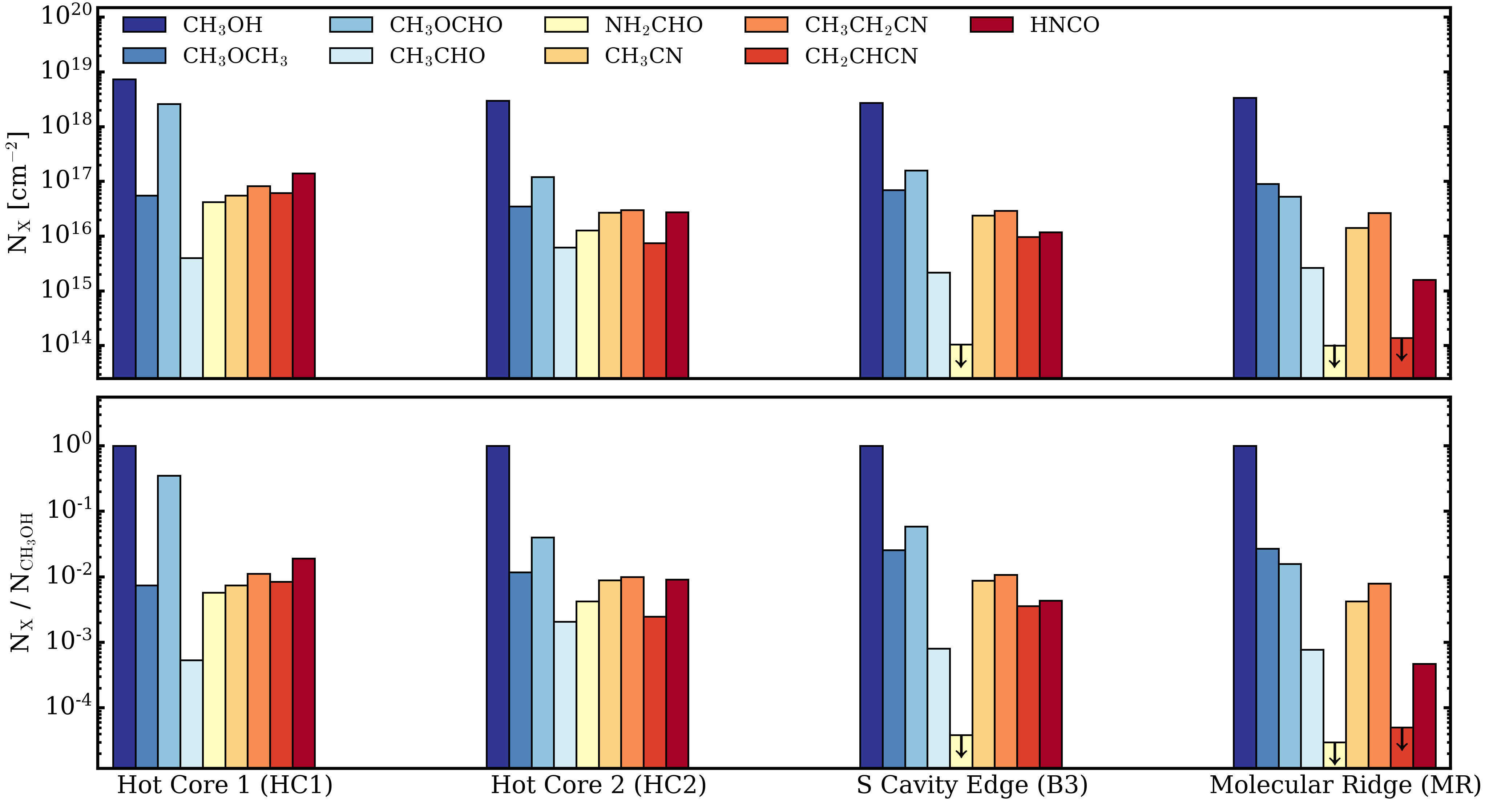}
\caption{Histogram of absolute (\textit{top}) and CH$_3$OH-normalized (\textit{bottom}) column densities toward positions of interest in G10.6. Upper limits are indicated by downward arrows.}
\label{fig:N_histogram}
\end{figure*}

In Table \ref{tab:Table3}, we provide a list of rotational temperature and column densities toward four regions of particular interest: HC1-2, B3, and MR. Figure \ref{fig:N_histogram} presents a summary of derived column densities toward these same positions. Column density variations between species span three orders of magnitude at each position, with CH$_3$OH being most abundant at all locations. HC1, HC2, and B3 appear broadly similar in their chemical compositions, as indicated by a similar distribution of COM abundances with respect to CH$_3$OH. However, HC1 exhibits at least an order of magnitude enhancement in CH$_3$OCHO relative to the other positions, while B3 is deficient in NH$_2$CHO and the MR is deficient in both NH$_2$CHO and CH$_2$CHCN.

\subsection{Spatial Column Density Correlations}

To accurately compare spatial trends of different species, we first performed 1/4-beam sampling, i.e., 4 data points extracted for each beam area, on a consistent spatial grid that was then applied uniformly to all COMs in the sample and used for all subsequent correlation analysis. We did not attempt to derive fractional abundances with respect to hydrogen due to a lack of H$_2$ gas tracers on these scales; we cannot, for example, distinguish the free-free and dust contributions to the overall continuum emission. We instead compared column densities using the cross correlation between each pair of molecules according to \citet{Guzman18} via:

\begin{equation}
\rho_{12} = \dfrac{\displaystyle \sum_{i,j} I_{1, ij} I_{2, ij} w_{ij}}{\left(\displaystyle \sum_{i,j} I^2_{1, ij} w_{ij} \displaystyle \sum_{i,j} I^2_{2, ij} w_{ij} \right)^{1/2}}.    
\end{equation}

Sums were taken over each quarter beam position with $I_{1, ij}$ and $I_{2, ij}$ corresponding to the values at each position $i, j$ and the weight $w_{ij}$ is equal to either 0 or 1 if a column density was able to be determined for that position or not. By definition, $\rho_{12} = \rho_{21}$ and $|\rho_{12}| \le 1$, and a value of $\rho_{12}=1$ implies that $I_1 = \alpha I_2$, where $\alpha$ is a positive constant. Hence, the closer the cross correlation values are to 1, the more similar the spatial distributions of column density.

\begin{figure*}
\centering
\includegraphics[width=\linewidth]{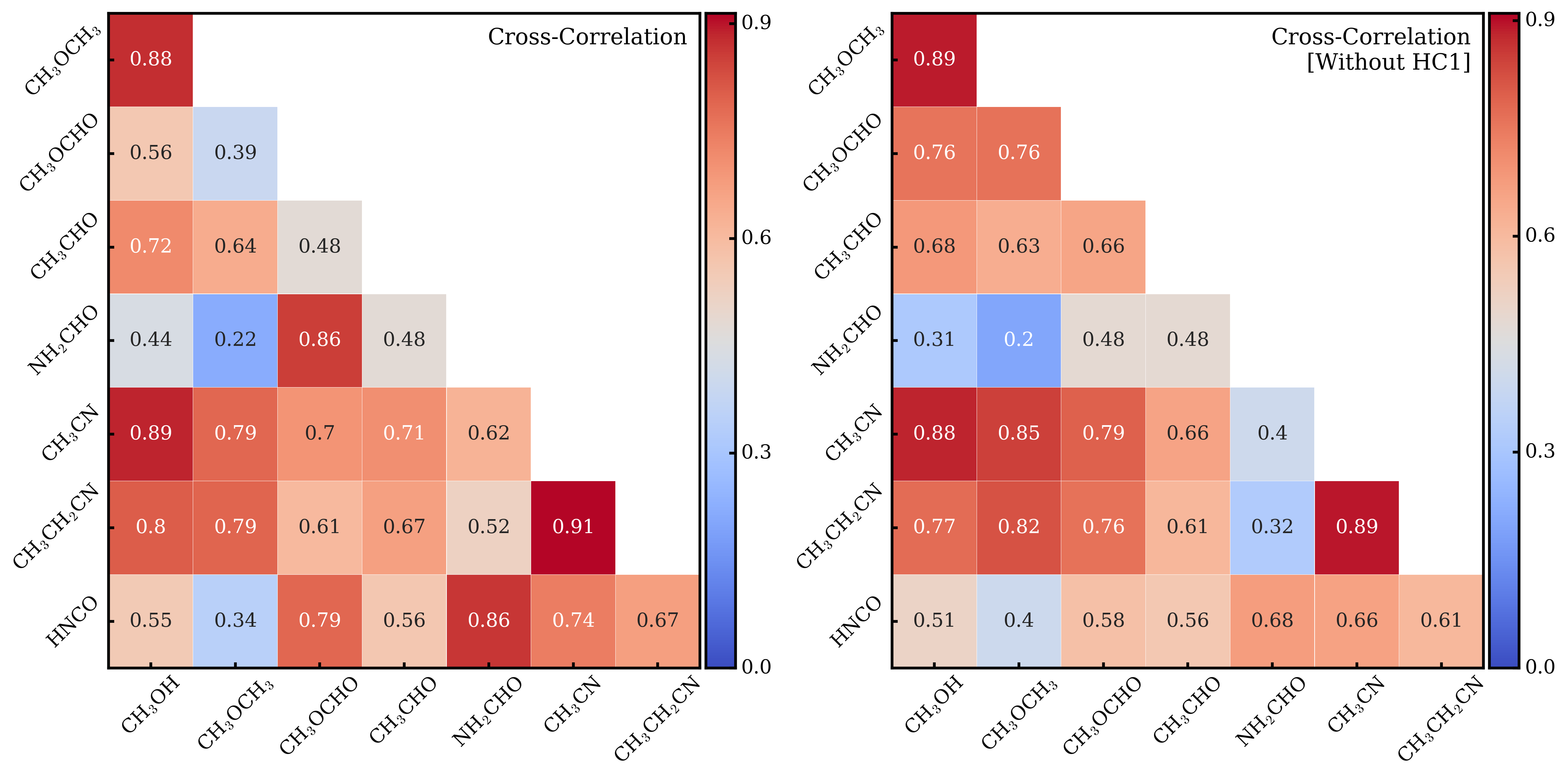}
\caption{Column density correlation matrix, showing cross correlations with (\textit{left}) and without HC1 (\textit{right}) for each pair of molecules. }
\label{fig:Fig12}
\end{figure*}

The left panel of Figure \ref{fig:Fig12} shows the calculated cross correlations for all pairs of molecules, excluding CH$_2$CHCN, which lacked a sufficient number of independent data points for meaningful comparison. Although spatially compact, the column density map of NH$_2$CHO still covers nearly 20 independent beams and thus we chose to include it in this analysis. We find strong correlations between: NH$_2$CHO/HNCO and NH$_2$CHO/CH$_3$OCHO, which are dominated by HC1 and HC2; CH$_3$CN/CH$_3$CH$_2$CN and CH$_3$OH/CH$_3$OCH$_3$, which are expected to be chemically related and trace similar column density structures; and CH$_3$OH/CH$_3$CN, which are the two most extended molecules and likely probe total gas column densities. We also note that correlations between the saturated cyanides CH$_3$CH$_2$CN and CH$_3$CN are greater than those between pairs of saturated and unsaturated N-bearing species, such as CH$_3$CN/NH$_2$CHO, CH$_3$CH$_2$CN/NH$_2$CHO, or CH$_3$CH$_2$CN/HNCO. The weakest correlations are typically found between CH$_3$OCH$_3$ and molecules such as HNCO, NH$_2$CHO, and CH$_3$OCHO, which reflects the unique column density distribution of CH$_3$OCH$_3$. 

As noted by \citet{Guzman18}, higher column density sections of an image weigh more into the calculation of $\rho_{12}$. In our case, HC1 possesses the highest column density for each species, except CH$_3$OCH$_3$. In order to avoid HC1 dominating the derived correlation coefficient and to search for additional spatial trends, we masked the pixels in a 0.16$^{\prime \prime}$ region centered on HC1. This masking region was chosen to be slightly larger than the average beam size, such that we fully excluded HC1, which is marginally-resolved in our observations. The resulting cross correlations are shown in the right panel of Figure \ref{fig:Fig12}.

As expected, correlations between NH$_2$CHO and HNCO and other species dramatically decrease, since these molecules are most prominently detected toward HC1. We do note, however, the caveat that only ${\sim}$10 independent beams are available for correlations involving NH$_2$CHO, which is a reduction of 50\% after excluding HC1. Cross correlations involving the spatially-extended molecules CH$_3$OH and CH$_3$CN are roughly the same. CH$_3$OCH$_3$, despite its unique column density distribution, remains strongly correlated with CH$_3$OH irrespective of the inclusion or exclusion of HC1. We find that CH$_3$OCHO is now well-correlated with the other O-bearing COMs CH$_3$OH and CH$_3$OCH$_3$. This suggests that outside of HC1 these molecules co-form and that the hot core environment acts to specifically enhance or destroy some COMs that are otherwise chemically-related. 

\begin{figure*}
\centering
\includegraphics[width=\linewidth]{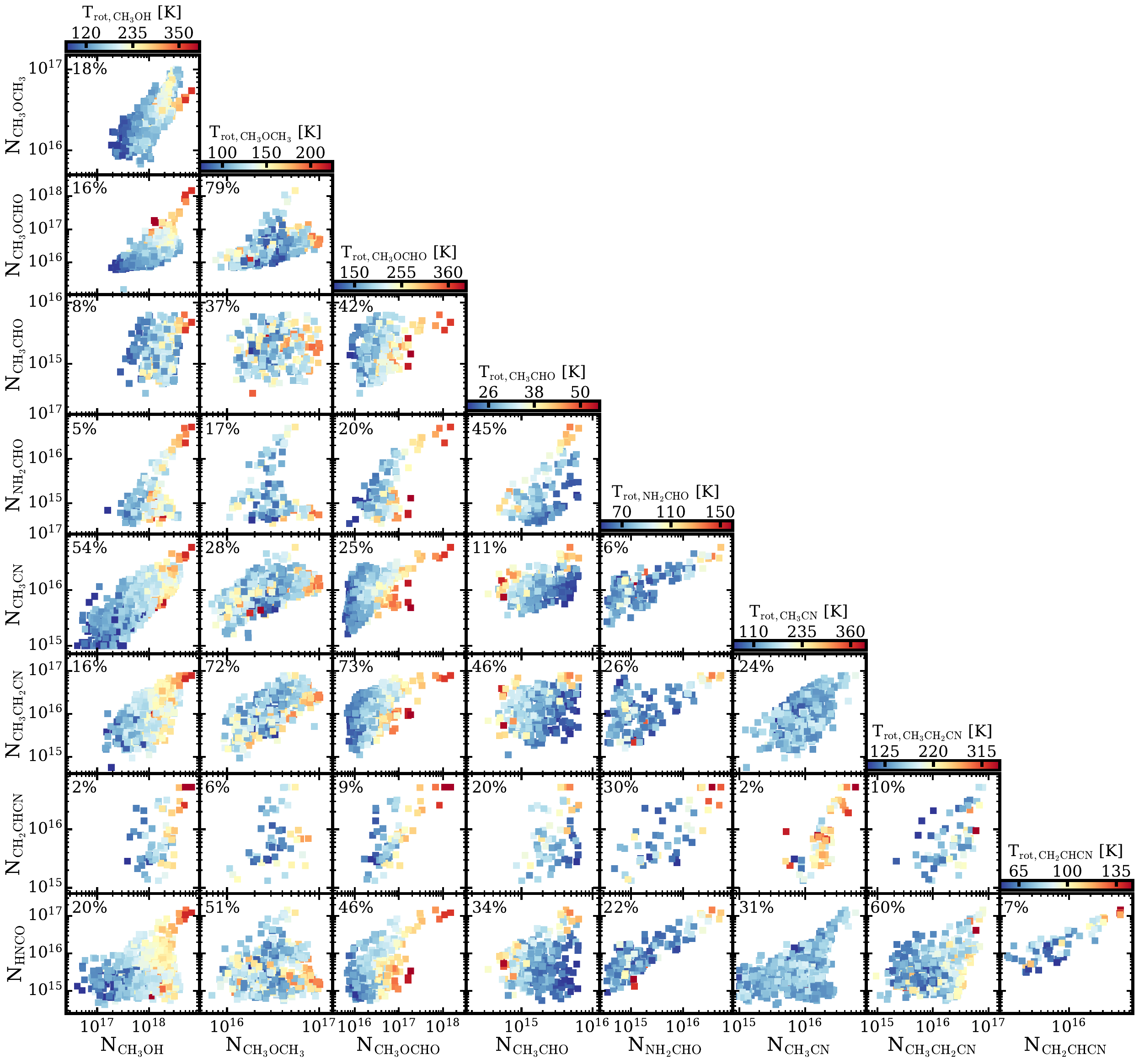}
\caption{Column densities of each molecule plotted against one another. The color-coding corresponds to the derived rotational temperature of the species on the x-axis, as indicated by the colorscale at the top of each column. The secondary temperature component of CH$_3$CH$_2$CN is shown. Percent overlap, as defined in the text, is shown in the upper left corner of each panel.}
\label{fig:corr_n}
\end{figure*}

While cross correlations are informative of general chemical relatedness, they do not capture the fact that multiple trends within single molecular pairs may exist in G10.6. To investigate the existence of such correlations directly, we show correlation plots for each pair of molecules in Figure \ref{fig:corr_n}. We labeled the percentage overlap between the two molecules being compared in the upper left corner of each panel, which helps inform the spatial extent over which these trends can be reliably interpreted. We defined this percentage overlap as the fraction of pixels from the species with a smaller spatial extent relative to the fraction of pixels from the species with a larger spatial extent. For instance, the 54\% overlap between CH$_3$OH and CH$_3$CN means that out of all the CH$_3$OH pixels with T$_{\rm{rot}}$ and N$_{\rm{col}}$ determinations, only 54\% of the corresponding pixels have T$_{\rm{rot}}$ and N$_{\rm{col}}$ derivations for CH$_3$CN.

The majority of species have positive trends, but are not well-described by a single relation and instead exhibit complex trends with multiple components. One notable exception to this trend is CH$_3$CHO, which is largely uncorrelated with all other species. To leverage the additional information provided by rotational temperatures, we also color-coded the scatter points by temperature. Two distinct hot temperature components are frequently observed: one which can be attributed to HC1, where the highest column densities and hot temperatures are typically observed, and a second from B5, which hosts the strongest H30$\alpha$ line emission, where we often see the hottest temperatures but more moderate column densities. The CH$_3$CHO/CH$_3$CN correlation plot is a good example of this phenomenon; the highest rotational temperatures are found in the upper right corner, signifying peak column densities for both species in HC1, and in a grouping of points corresponding to intermediate column densities, attributable to B5. Additional examples of these two hot temperature components are seen in the majority of CH$_3$OCHO correlations (e.g., CH$_3$OCHO/CH$_3$CN, CH$_3$OCHO/CH$_3$CH$_2$CN) and in a few correlations involving CH$_3$OH (e.g., CH$_3$OCH$_3$/CH$_3$OH). 

While strong correlations are observed for numerous species, such as NH$_2$CHO and CH$_2$CHCN, we caution broad interpretation of these trends. As reflected in the small percentage (${\lesssim}10$\%) of mutual pixels, these trends are only applicable toward the densest regions of gas, namely HC1 and HC2. A good example of this is the correlation between CH$_2$CHCN/HNCO, which is the highest correlation observed (0.95), but only has a 7\% pixel overlap. This indicates that such a correlation is applicable to a minimally-shared region between HNCO and CH$_2$CHCN, which in this case is HC1, HC2, and a few small patches of gas toward B3. However, some strongly-correlated molecules show a single component with high percentage overlap, such as CH$_3$OH and CH$_3$CN, which indicates co-formation across a wide range of environments.

\subsection{Rotational Temperature versus Column Density}

\begin{figure*}
\centering
\includegraphics[width=\linewidth]{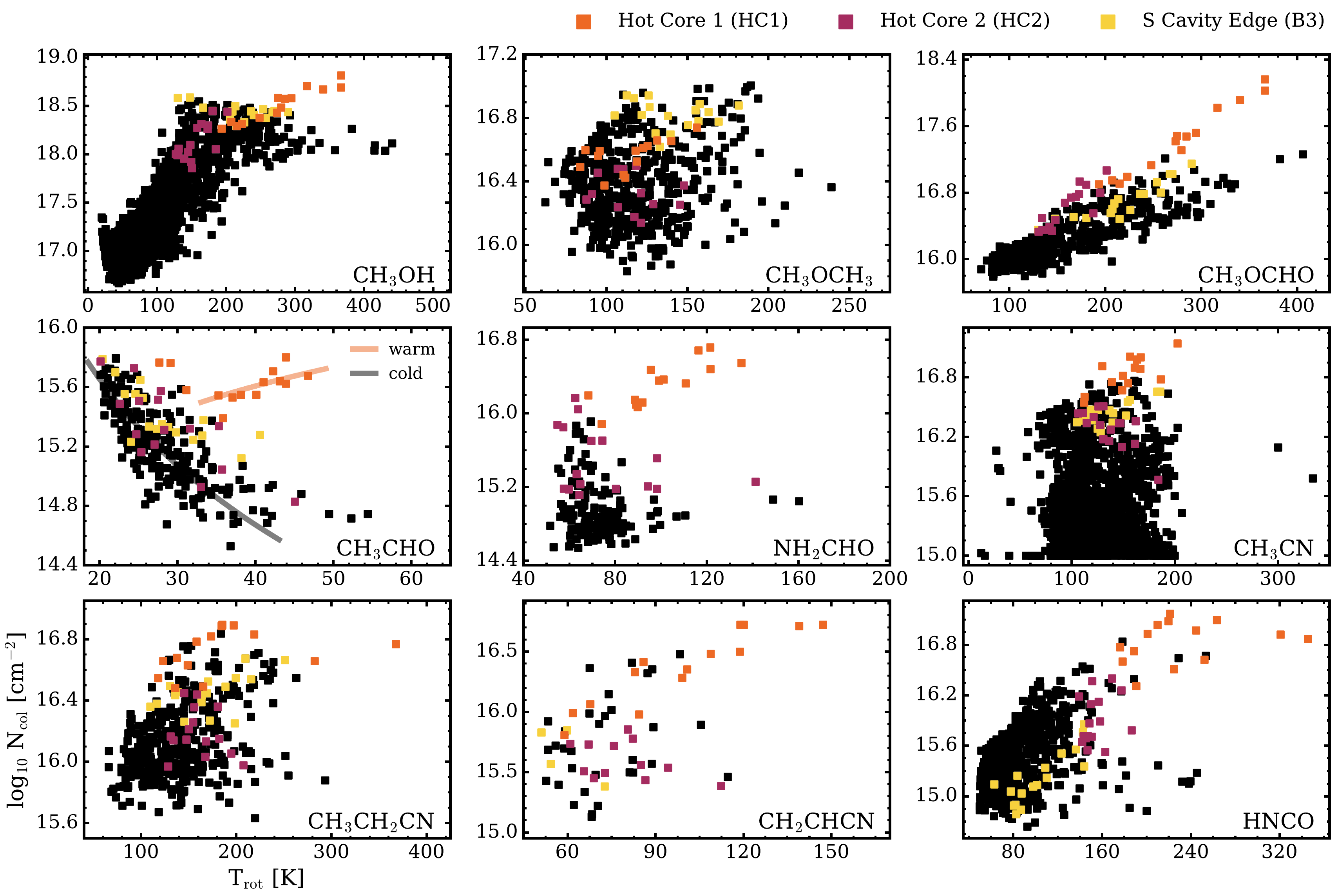}
\caption{Rotational temperature versus column density for each molecule in our sample. Regions of interest are colored according to the legend. The secondary temperature component is shown for CH$_3$CH$_2$CN. Trend lines for CH$_3$CHO illustrate the presence of a two-component temperature correlation. We note that the CH$_3$OH data closely resemble the temperature-abundance trend reported in \citet{Ginsburg17} in massive star-forming region W51~e2.}
\label{fig:N_v_T}
\end{figure*}

Next, we investigated the relationship between rotational temperature and column density in each species in Figure \ref{fig:N_v_T}. Temperature and density trends within a single species not only help characterize the gas excitation conditions present in G10.6, but can also reveal the presence of additional chemical formation and destruction pathways.

Although most species display a positive relation between T$_{\rm{rot}}$ and N$_{\rm{col}}$, there is a large range in the relative strength of this association. In general, a large fraction of the gas in N-bearing species, such as CH$_3$CN, is relatively insensitive to rotational temperature, while the column densities of O-bearing species, such as CH$_3$OH and CH$_3$OCHO, are highly correlated with gas temperature. However, this trend is less clear for CH$_3$OCH$_3$ and further illustrates its distinct temperature and column density distributions. CH$_3$CHO and, to a lesser degree, NH$_2$CHO display negative associations with temperature and are notable outliers to this overall positive T$_{\rm{rot}}$--N$_{\rm{col}}$ trend.  

To assess spatially-dependent trends, regions of interest were color-coded in Figure \ref{fig:N_v_T}. HC1 exhibits a distinct and tightly positive correlation between temperature and column density for all species, except for CH$_3$CHO and CH$_3$OCH$_3$. The diffuse correlation seen in CH$_3$CHO is due to its noisier column density map and likely does not reflect true gas conditions. For CH$_3$OCH$_3$, HC1 still exhibits a relatively tight T$_{\rm{rot}}$--N$_{\rm{col}}$ association but the largest column densities are instead found in the molecular ridge. The existence of two well-defined HNCO trends in HC1 arises from the complex, dual-peaked temperature structure seen in Figure \ref{fig:Fig9}.

HC2 and B3 span a range of different T$_{\rm{rot}}$--N$_{\rm{col}}$ associations, from extremely diffuse correlations (CH$_3$CN, HNCO) to tight positive relations (CH$_3$OH, CH$_3$OCHO) and even negative trends (CH$_3$CHO, NH$_2$CHO). No clear trends were identified among different types of molecules for either region. Although not labeled in Figure \ref{fig:N_v_T}, a diffuse component of high rotational temperature but modest column density gas is seen in several species, such as CH$_3$OH, CH$_3$OCH$_3$, CH$_3$CN, and HNCO, and corresponds to B5. 

\section{Discussion} \label{sec:discussion}

\subsection{Spatial Distribution of N- versus O-bearing Species}

Spatial separations between N- and O-bearing COMs have been observed toward several high-mass star-forming regions, such as Orion \citep{Blake87, Wright96, Beuther05, Feng15}, W3(OH)/W3(H$_2$O) \citep{Wyrowski99}, G19.61-0.23 \citep{Qin10}, and AFGL~2591 \citep{Jimenz_Serra12}. However, efforts to assess the ubiquity of such separations with single-dish surveys have so far proved inconclusive \citep{Fontani07, Suzuki18}, but rather require interferometric observations, which can resolve small-scale chemical variations to identify and explain such spatial trends \citep[e.g.,][]{Tercero18, ElAbd19}. Here, we aim to test whether chemical differentiation is seen on small scales throughout G10.6 and investigate observed COM spatial trends.

Although we find no evidence for systematic small-scale separations between N- and O-bearing COMs in G10.6, we identify several correlations within and between these molecular families. The observed spatial correlations among the O-bearing species CH$_3$OH, CH$_3$OCH$_3$, and CH$_3$OCHO probably reflect intrinsic chemical similarity, related formation pathways (see Table \ref{tab:Table-chem}), and have been previously seen in several MYSOs \citep[e.g.,][]{Bisschop07, Brouillet13}. We also find a strong spatial correlation between the nitriles CH$_3$CN and CH$_3$CH$_2$CN. Together with their high column densities, this likely implies a formation route via grain surfaces \cite[e.g.,][]{Mookerjea07}. Previous models have suggested grain surface hydrogenation and evaporation of accreted CH$_3$CN and HC$_3$N to form CH$_3$CH$_2$CN \citep{Charnley92, Caselli93}, but based on studies reporting nitrile predictions, these pathways under-produced CH$_3$CN and CH$_3$CH$_2$CN abundances \citep[][]{Caselli93, Millar97}. These small-scale observations strongly suggest that, however complex nitriles form, the formation pathways of medium and large nitriles are linked.

CH$_3$OH and CH$_3$CN, the most spatially extended O- and N-bearing species, respectively, in our sample, possess similar spatial distributions and tight column density correlations across G10.6. This is surprising as prior observations of MYSO have reported a lack of correlation between these species \citep{Bisschop07, Oberg_Faraday}, and both molecules are thought to form via different mechanisms \citep{Garrod06, Garrod08}. Indeed, in G10.6, we believe the observed correlation likely reflects that both species are tied to total gas column density, rather than intrinsic chemical similarity.

\subsection{Relationships between Specific Molecules}

During our analysis, we noticed three trends that are further explored here: the complex relationship between CH$_3$OCHO and CH$_3$OCH$_3$, tight correlation between HNCO and NH$_2$CHO, and unique temperature dependence of CH$_3$CHO.

\subsubsection{CH$_3$OCHO and CH$_3$OCH$_3$} \label{sec:CH3OCH3_CH3OCHO_disc}

Previous observations of CH$_3$OCHO and CH$_3$OCH$_3$ have revealed nearly constant abundance ratios across a wide range of sources \citep{Jaber14, Rivilla17, Coletta20}, remarkably similar spatial distributions \citep{Brouillet13}, and comparable column densities and gas temperatures \citep{Rong16, ElAbd19}. A chemical link had been previously predicted by modeling \citep{Garrod06, Garrod08} and experimental efforts \citep{Oberg09}, which indicated a common precursor hypothesis for the formation of CH$_3$OCH$_3$ and CH$_3$OCHO as either involving the CH$_3$O radical (solid phase/ice) or CH$_3$OH$_2^+$ (gas phase). Both species can also form through gas-phase chemistry at low temperatures \citep{Balucani15}, while the presence of additional warm gas-phase pathways cannot be excluded. A summary of the most commonly considered formation routes is provided in Table \ref{tab:Table-chem}. 

\begin{figure}
\centering
\includegraphics[width=\linewidth]{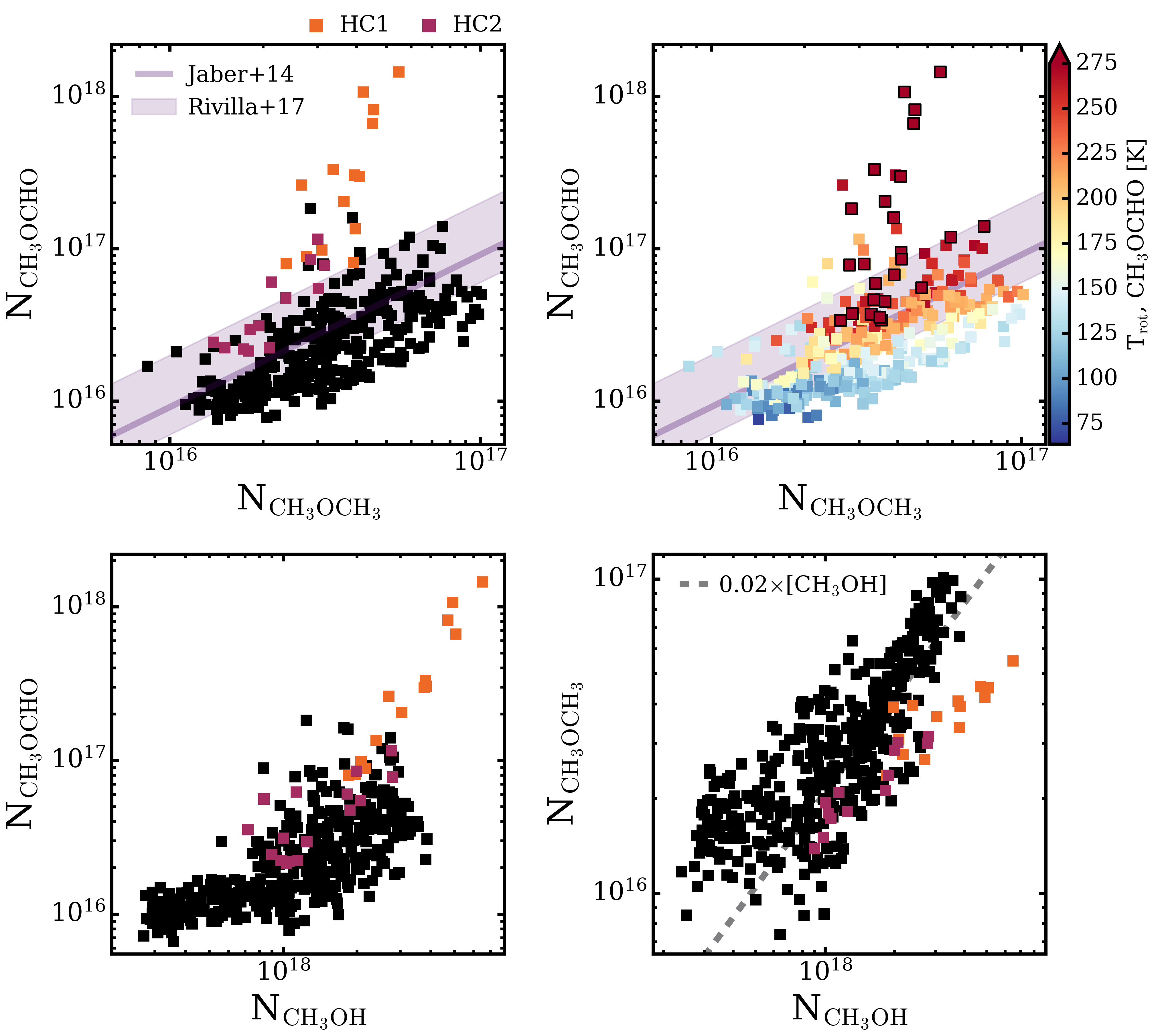}
\caption{Column density correlations between CH$_3$OCHO and CH$_3$OCH$_3$ (\textit{top panels}), and with respect to CH$_3$OH (\textit{bottom panels}). Power law relation from \citet{Jaber14} is shown as a solid purple line, while observed ratios from \citet{Rivilla17} are shaded. HC1 and HC2 are colored in orange and purple, respectively, while all other gas is shown in black. Color-coding is with respect to CH$_3$OCHO rotational temperature and all points with $\rm{T}>275$~K are shown in dark red and outlined with black borders. Dashed gray line indicates a constant ratio of CH$_3$OCH$_3$/CH$_3$OH of 2\%.}
\label{fig:CH3OCH3_fig}
\end{figure}

Since all proposed formation pathways each involve the presence of CH$_3$OH and its photodissociation products, it is unsurprising that we observe strong correlations between CH$_3$OH and both CH$_3$OCH$_3$ and CH$_3$OCHO, respectively, as shown in the bottom panels of Figure \ref{fig:CH3OCH3_fig}. For instance, we identify a nearly constant ratio of CH$_3$OCH$_3$ with respect to CH$_3$OH of 2\%, which indicates co-formation or a constant conversion of CH$_3$OH into CH$_3$OCH$_3$ at all temperatures. Constant CH$_3$OCH$_3$/CH$_3$OH ratios have been previously observed in MYSOs \citep{Oberg_Faraday} and suggest the existence of a single link between CH$_3$OH and CH$_3$OCH$_3$. The simplest explanation for this trend is co-desorption of the two ice constituents, following partial conversion of CH$_3$OH into COMs, including CH$_3$OCH$_3$. If so, this ratio serves as an indicator of overall efficiency of conversion of simple ices into more complex ones. Compared to the typical MYSOs ratios of 14\% from \citet{Oberg_Faraday}, G10.6 appears to be relatively inefficient in its CH$_3$OCH$_3$ ice conversion. In contrast to CH$_3$OCH$_3$, the CH$_3$OCHO/CH$_3$OH ratio exhibits a multi-component trend that is not well-described by a constant value and cannot be explained by ice chemistry alone. This suggests the presence of multiple formation pathways, and perhaps a mixture of ice and gas-phase chemistry.

The top panels of Figure \ref{fig:CH3OCH3_fig} show the column density correlation between CH$_3$OCHO/CH$_3$OCH$_3$. We identify a broad one-to-one correlation across G10.6, and then a second, much steeper correlation in both HC1 and HC2, where CH$_3$OCHO is over-produced compared to CH$_3$OCH$_3$. This one-to-one correlation agrees remarkably well with the power law relation derived by \citet{Jaber14} across a wide set of ISM sources over many orders of magnitude in column density and is consistent with observed ratios toward hot cores, intermediate-mass star-forming regions, and hot corinos associated with low-mass protostars \citep[][and references therein]{Rivilla17}. As this correlation is quite broad over a wide range of CH$_3$OCHO temperatures (60--275~K) and physical locations across G10.6, it is difficult to discern the relative importance of various reactions, e.g., cold versus warm gas versus ice surface chemistry, which in principle, may all contribute at some level to this trend without performing detailed chemical modeling. Regardless of their specific formation mechanisms, CH$_3$OCHO and CH$_3$OCH$_3$ seem to behave in the same way within the bulk of gas, i.e., outside of the hot cores, in G10.6 as they do in many other ISM sources.

The presence of an additional component in the CH$_3$OCHO/CH$_3$OCH$_3$ correlation indicates that this chemical similarity does not extend in the same way to the hot cores in G10.6. While this component is most distinctly associated with HC1, it may also extend to HC2, as shown in the top left panel of Figure \ref{fig:CH3OCH3_fig}. Moreover, it is not clear if this trend is being driven by gas temperature or is specific to hot core-like environments. We find that the majority, but not the entirety, of points within this steep trend exhibit temperatures ${>}$275~K. To explain the presence of this trend, we must disentangle the relative contribution of increased CH$_3$OCHO production versus more efficient CH$_3$OCH$_3$ destruction. Interestingly, we do find a steepening of the CH$_3$OCHO/CH$_3$OH correlation and a modest shallowing of the CH$_3$OCH$_3$/CH$_3$OH correlation towards higher CH$_3$OH column densities. This suggests that there is an additional CH$_3$OCHO formation pathway that kicks in within the hot cores, which would also explain the bimodal correlation between CH$_3$OCH$_3$ and CH$_3$OCHO. Further support for the idea that increased CH$_3$OCH$_3$ destruction is not driving the observed high CH$_3$OCHO/CH$_3$OCH$_3$ ratio is found in the models of \citet{Garrod08}, which predict a peak gas-phase abundance of CH$_3$OCH$_3$ at ${\sim}$200~K. Combined with the fact that CH$_3$OCH$_3$ exhibits its highest column densities in the MR, which is exposed to strong radiation fields from the nearby ionized gas in B5, it is unlikely that CH$_3$OCH$_3$ is being readily destroyed from high temperatures within the hot cores.

Observations of anti-correlations between CH$_3$OCHO and formic acid in Orion-KL \citep{NeillJPCA, Brouillet13} have suggested that the production of CH$_3$OCHO is driven by gas-phase reactions (pathway 3 in Table \ref{tab:Table-chem}) at high gas temperatures. In support of this explanation, ice chemistry pathways are not thought to be efficient at high temperatures, as the models of \citet{Garrod08} predict a peak CH$_3$OCHO abundance at relatively low temperatures (${\sim}$80~K). Thus, one possible explanation for this additional, hot core-dominated trend in G10.6 is gas-phase reactions driving extra production of CH$_3$OCHO. If CH$_3$OCH$_3$ is forming via ice conversion, which is supported by the constant, temperature-independent CH$_3$OCH$_3$/CH$_3$OH ratio across G10.6, i.e. inside and outside hot cores, then provided that the CH$_3$OCHO gas-phase reaction is sufficiently efficient, this would naturally explain the existence of the observed steeper trend.

\begin{deluxetable*}{ccllcc}[!htp]
%\tablenum{4}
\tablecaption{Chemical Pathways of Interest\label{tab:Table-chem}}
\tablehead{[-.3cm]
\colhead{COM} & \colhead{Num.} & \colhead{Pathway} & \colhead{Type} &\colhead{Ref.}  \\[-0.6cm] }
\startdata
CH$_3$OCHO      & [1] & O $+$ CH$_3$OCH$_2$ $\to$ CH$_3$OCHO $+$ H             & Cold Gas      & 1     \\
                           & [2] & CH$_3$O $+$ HCO $\to$ CH$_3$OCHO                               & Ice Surface  & 2, 3 \\
                           & [3] & CH$_3$OH$^+_2$ $+$ HCOOH $\to$ CH$_3$OCHO             & Warm Gas   &  4 \\
CH$_3$OCH$_3$ & [4] & CH$_3$O $+$ CH$_3$ $\to$ CH$_3$OCH$_3$ $+$ photon & Cold Gas     & 1      \\
                           & [5] & CH$_3$O $+$ CH$_3$ $\to$ CH$_3$OCH$_3$						 & Ice Surface & 5     \\
                           & [6] &  CH$_3$OH$^+_2$ + CH$_3$OH $\to$ CH$_3$OCH$_3$     & Warm Gas  & 6   \\
NH$_2$CHO        &  [7] & HNCO $+$ H $+$ H $\to$ NH$_2$CHO									 & Ice Surface & 7      \\
                           & [8] & NH$_2$ + HCO $\to$ NH$_2$CHO                                        & Ice Surface  & 8, 9 \\
                           & [9] & NH$_3$ + CO $\to$ NH$_2$CHO                                          & Ice Surface & 8, 10 \\
                           & [10] & H$_2$CO $+$ NH$_2$ $\to$ NH$_2$CHO $+$ H                  & Warm Gas & 11, 12  \\
CH$_3$CHO        &  [11] & CH$_3$ $+$ HCO $\to$ CH$_3$CHO                                    & Ice Surface            & 13  \\
                          & [12] & CH$_3$CH$_2$ $+$ O $\to$ CH$_3$CHO $+$ H                 & Warm Gas  & 14 \\              
\enddata
\tablecomments{References: (1) \citet{Balucani15}; (2) \citet{Garrod08}; 3) \citet{Laas11}; (4) \citet{Neill11}; (5) \citet{Oberg10}; (6) \citet{Charnley95}; (7) \citet{Charnley97}; (8) \citet{Jones11}; (9) \citet{Fedoseev16};  10 \citet{Rimola18}; (11) \citet{Kahane13}; (12) \citet{Barone15}; (13) \citet{Garrod06}; (14) \citet{Charnley04}}
\end{deluxetable*}

\subsubsection{HNCO and NH$_2$CHO}

Single dish observations, which found a constant abundance HNCO/NH$_2$CHO ratio in a range of source luminosities and masses \citep{Bisschop07, Lopez15MNRAS}, suggested that these molecules were chemically-related. Evidence of co-spatial emission in the low-mass protobinary system IRAS~16293 \citep{Coutens16}, two high-mass cores in G35.20 \citep{Allen17}, and six cores across three massive star-forming regions \citep{Allen20} further indicated a link between the two species. 

\begin{figure}
\centering
\includegraphics[width=\linewidth]{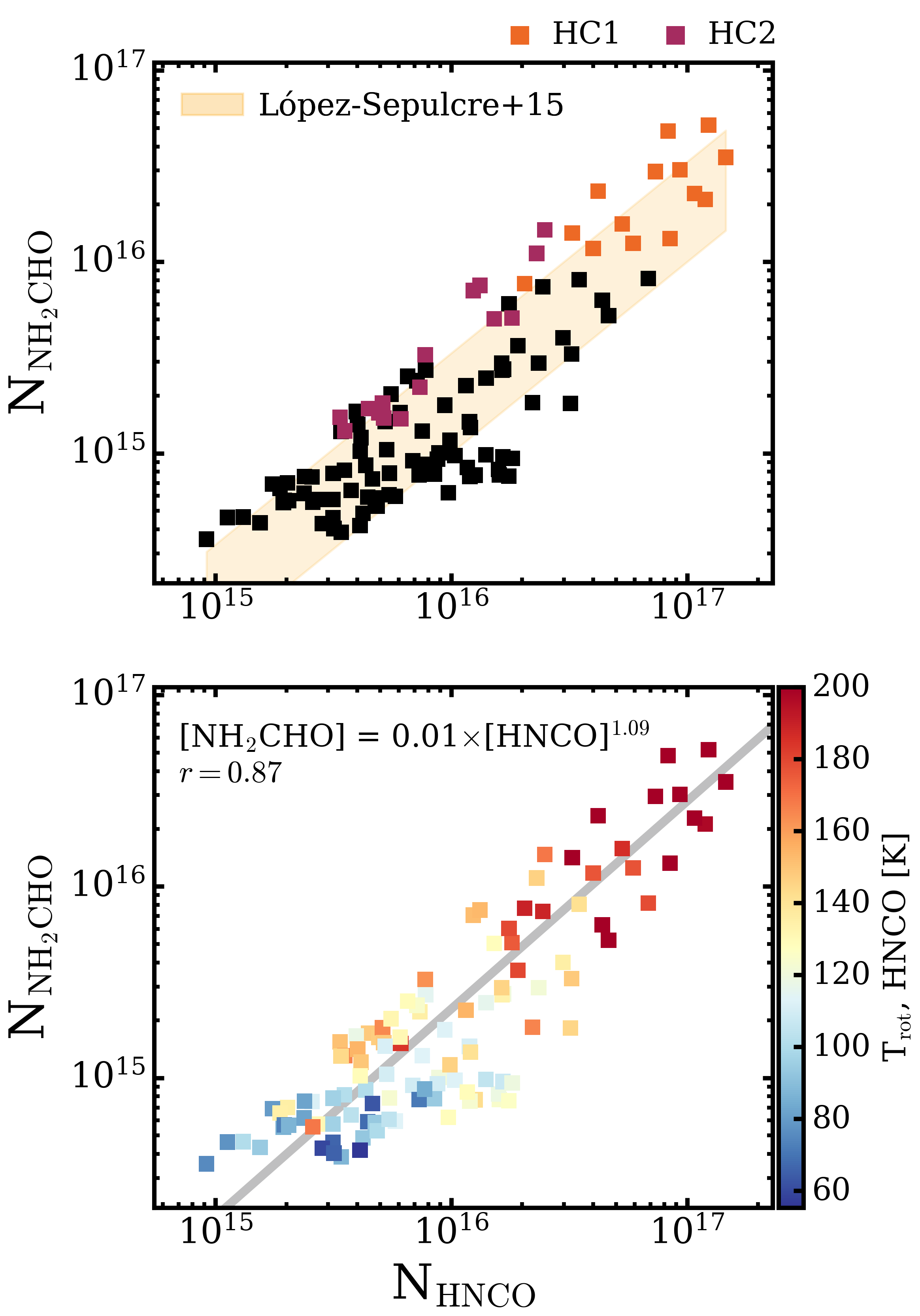}
\caption{HNCO/NH$_2$CHO column density ratios in G10.6. \textit{Top}: Observed column density ratios closely resemble those reported in \citet{Lopez15MNRAS} for a sample of low- to intermediate-mass prestellar and protostellar objects. HC1 and HC2 are colored in orange and purple, respectively. \textit{Bottom}: Column densities ratios color-coded according to HNCO rotational temperatures. Best fit power law and correlation coefficient are shown.}
\label{fig:hnco_vs_nh2cho_column_density}
\end{figure}

\begin{figure}
\centering
\includegraphics[width=\linewidth]{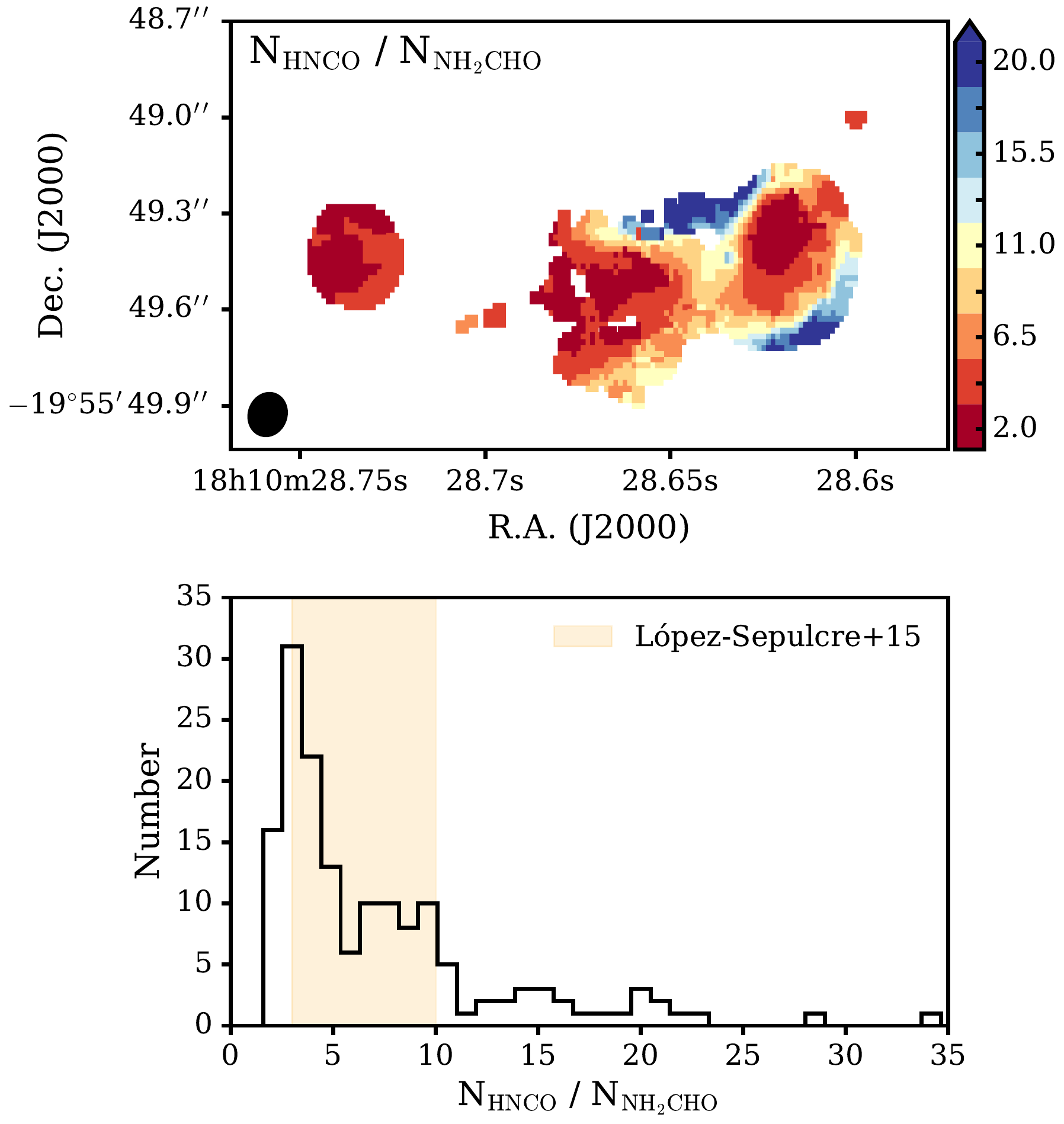}
\caption{HNCO/NH$_2$CHO column density ratios for regions of mutual overlap in G10.6 (\textit{top}) and histogram of derived ratios (\textit{bottom}), relative to those reported in \citet{Lopez15MNRAS}.}
\label{fig:hnco_nh2cho}
\end{figure}

Figure \ref{fig:hnco_vs_nh2cho_column_density} illustrates the observed HNCO/NH$_2$CHO column density correlation across G10.6. This correlation is strong with a Pearson's correlation coefficient of $r=0.87$ and extends more than two orders of magnitude in NH$_2$CHO and three orders of magnitude in HNCO. It is well-characterized by a single power law given by [N$_{\rm{NH}{_2}\rm{CHO}}] = 0.01 \times [\rm{N}_{\rm{HNCO}}]^{1.09}$. Neither HC1 nor HC2 exhibits any significant deviations from this trend, which also appears insensitive to temperature with HNCO rotational temperatures spanning approximately 200~K. In addition, NH$_2$CHO/HNCO ratios within G10.6 closely resemble those previously observed in a sample of low- and intermediate-mass prestellar and protostellar objects \citep{Lopez15MNRAS}. This consistency is further illustrated in the spatially-resolved NH$_2$CHO/HNCO ratio map shown in Figure \ref{fig:hnco_nh2cho}. The majority (60\%) of our map is consistent with the ratios derived in \citet{Lopez15MNRAS} The highest NH$_2$CHO/HNCO ratios, which show the largest deviations, are mostly located at peripheries, where NH$_2$CHO line intensities are weaker and uncertainties correspondingly elevated.

The spatial distributions of both species also provides valuable information about their formation mechanisms. As noted in Section \ref{sec:rot_temp_and_col_dense}, column density maps show that HNCO is more spatially extended relative to NH$_2$CHO. Specifically, NH$_2$CHO is never observed in a region lacking HNCO emission. Not only does NH$_2$CHO possess a spatially-limited column density map, it also displays the most compact integrated intensity map of all COMs in our sample (Figure \ref{fig:Fig_int_sum}). Particularly salient is the absence of any detectable NH$_2$CHO toward B3, which otherwise exhibits emission from all other COMs, including HNCO, in our sample. Although difficult to disentangle from the intrinsic excitation characteristics of each species, i.e., Table \ref{tab:tab2}, this nonetheless seems to suggest that the formation of NH$_2$CHO in detectable amounts depends on the presence of hot core-like physical environments and not simply on the presence of HNCO.

A variety of explanations have been put forth to explain the empirical relationship between HNCO and NH$_2$CHO. \citet{Charnley97} suggested that NH$_2$CHO forms via the hydrogenation of HNCO, but this route has been challenged by subsequent experimental work \citep{Noble15}. Simultaneously formation in ices has been explored experimentally \citep{Jones11, Fedoseev16, Ligterink18}, and gas-phase pathways \citep{Kahane13, Barone15, Codella17} have also been proposed. Other models \citep{Quenard18} argue that the observed HNCO/NH$_2$CHO correlation is instead due to both species responding in similar ways to physical parameters, such as temperature, rather than a direct chemical link between the two species. In light of this finding, the strong correlation observed here warrants a further investigation into the relationship between these molecules in G10.6.

\subsubsection{CH$_3$CHO} \label{sec:ch3och_discu}

Based on both single-dish and spatially-resolved observations of MYSOs, \citet{Oberg_Faraday} found that CH$_3$CHO displays an overall negative dependence on temperature. Specifically, at low column densities and temperatures (${<}$100~K) the CH$_3$CHO/CH$_3$OH ratio was almost constant at a few percent, which then dropped dramatically by several orders of magnitude when ${>}$100~K. However, several outliers in the form of high CH$_3$CHO/CH$_3$OH ratios even at ${>}$100~K hinted at the presence of an additional high temperature formation pathway for CH$_3$CHO. Notably, all of these outliers were derived from spatially-resolved observations, which suggests that these increased CH$_3$CHO abundances may be a common feature of hot core chemistry.

We observe the presence of two such CH$_3$CHO formation pathways explicitly in G10.6, as shown in Figure \ref{fig:N_v_T}, where superimposed trend lines are shown to illustrate the presence of these two distinct components. A positive association with temperature is only found in HC1, which indicates the activation of a lukewarm formation pathway at approximately 30~K. Outside of HC1, we see a tight negative correlation with temperature, indicating that the bulk of the CH$_3$CHO gas is forming via this cold route, and perhaps rapidly degrades in lukewarm regions. Thus, single-dish studies, or otherwise low-spatial resolution observations of MYSOs that cannot resolve hot cores, would likely be dominated by this cold component.

Although less pronounced than in CH$_3$CHO, there is evidence of a similar two component formation pathway for NH$_2$CHO. The bulk of NH$_2$CHO in G10.6 seems to form efficiently at low temperatures (50--90~K), while in HC1, an extremely tight, positive association with temperature is present. This hot core component becomes activated around 60~K and continues until nearly 150~K. Thus, there likely exists an efficient low-temperature formation pathway for these molecules, which decreases in efficiency with temperature, but then at very elevated temperatures, the species are released from the ice, along with the majority of COMs, resulting in high abundances in HC1. 

The formation route of CH$_3$CHO remains unclear, as the relative contribution of gas-phase and grain surface chemistry is not well understood. Grain surface models \citep{Garrod06, Garrod13} and laboratory experiments \citep[e.g.,][]{Bennett05, Oberg09} have predicted that CH$_3$CHO forms via the combination of radicals CH$_3$ and HCO on grain mantles. However, recent quantum chemistry computations by \citet{ER16, ER19} show that this may be inefficient, as additional pathways resulting in CH$_4$ and CO are competitive. Alternatively, gas-phase models suggest oxidation of hydrocarbons as the primary formation route. Specifically, the release of C$_2$H$_6$ from dust surfaces is expected to drive production of CH$_3$CH$_2$, which in turns reacts in the gas phase with atomic oxygen to form CH$_3$CHO \citep{Charnley04}. A summary of these proposed ice surface and gas-phase formation pathways is listed in Table \ref{tab:Table-chem}.

\citet{deSimone20} showed that a gas-phase route is responsible for CH$_3$CHO production in the outflows of NGC~1333 IRAS 4A. The rotational temperatures (${\sim}$10--30~K) in these outflows are comparable to those associated with the low temperature component in G10.6, which suggests that the bulk of CH$_3$CHO gas may also be forming via gas-phase reactions. A similar finding was reported by \citet{Codella17}, in which gas-phase formation of NH$_2$CHO was confirmed in L1157-B1. Since the temperatures (20--80~K) associated with the B1 cavities \citep[e.g.,][]{GR15} are similar to those in the cold component of NH$_2$CHO, this again may point to gas-phase formation. Although extrapolating conclusions from regions of shock chemistry is not without caveats, these results nonetheless suggest that both cold components for CH$_3$CHO and NH$_2$CHO have their origins in gas-phase reactions in G10.6.

\begin{figure*}
\centering
\includegraphics[width=\linewidth]{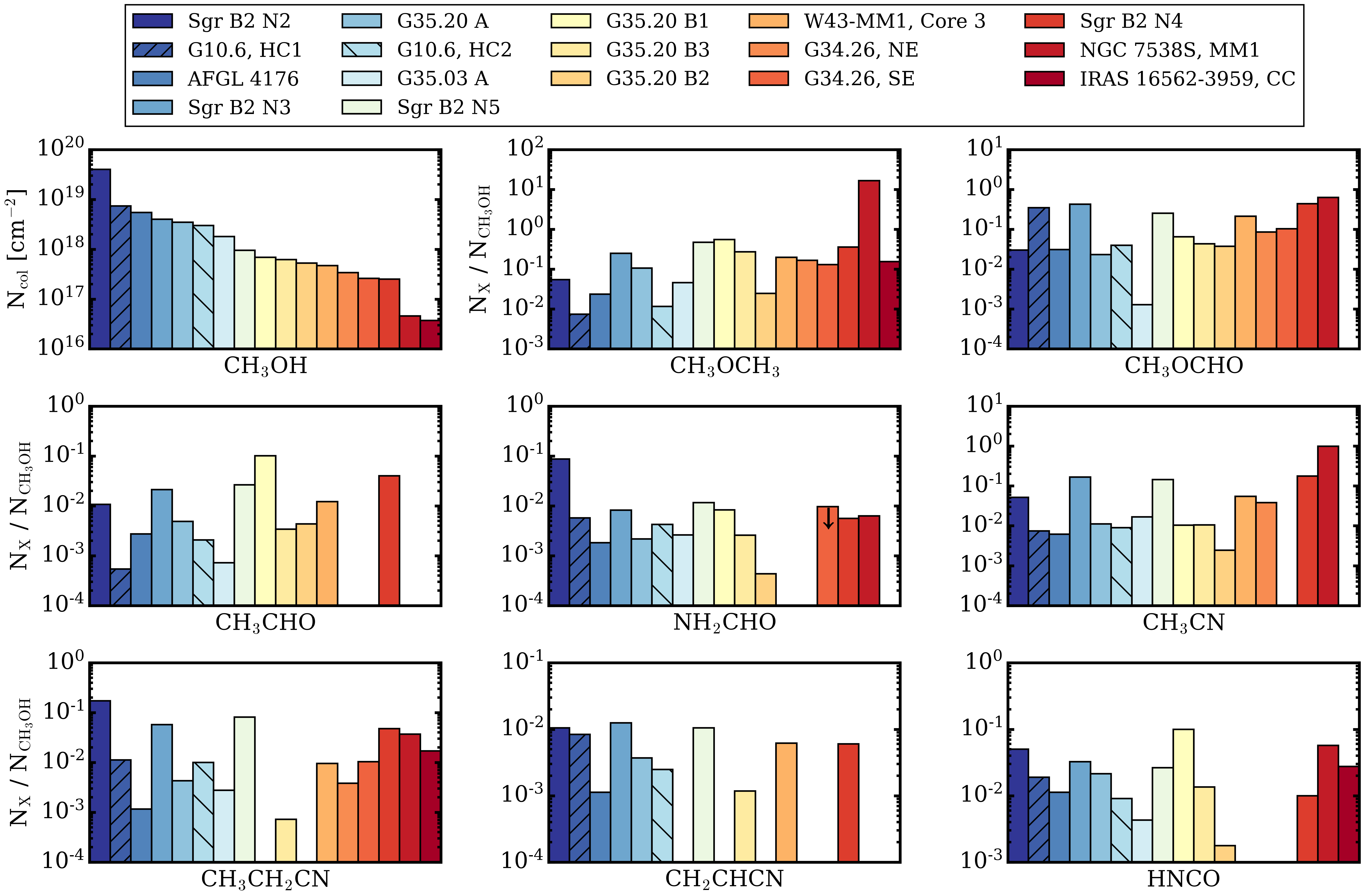}
\caption{Comparison of column densities with respect to CH$_3$OH of HC1 and HC2 in G10.6 versus a large sample of spatially-resolved MYSOs observations compiled from the literature. Sources are ordered by descending CH$_3$OH column density and upper limits are indicated by downward arrows. When the column density of a species is not reported in the literature, no bar is shown and no attempt was made to derive upper limits when not provided.}
\label{fig:MYSO_lit_sources}
\end{figure*}

\subsection{Comparisons with Other Sources}

\subsubsection{Massive Star-forming Regions}
\label{sec:mass_SF_compare}

To quantify how typical HC1 and HC2 in G10.6 are relative to other well-known MYSOs, we compared them against spatially-resolved data from: AFGL~4176 \citep{Bogelund19}; G34.26 \citep{Mookerjea07}; G35.20 A, B1--3 and G35.03 A \citep{Allen17}; IRAS 16562-3959, central compact (CC) core \citep{Guzman18}; NGC~7538S, MM1 \citep{Feng16};  Sgr B2, N2--5 \citep{Bonfand17, Bonfand19}; and Core 3, W43-MM1 \citep{Molet19}.

Figure \ref{fig:MYSO_lit_sources} shows the fractional abundances with respect to CH$_3$OH of G10.6 and the literature sample. Both G10.6 cores have relatively high CH$_3$OH column densities (${>}10^{18}$~cm$^{-2}$), which place them in the top half of the comparison sample. Both cores also exhibit the two lowest CH$_3$OCH$_3$ abundances, while HC1 possesses the lowest CH$_3$CHO abundance and HC2 is 4th lowest. However, besides these two species, both HC1 and HC2 are consistent with literature values and both appear quite typical for massive cores in their relative compositions.

\begin{figure}[h]
\centering
\includegraphics[width=\linewidth]{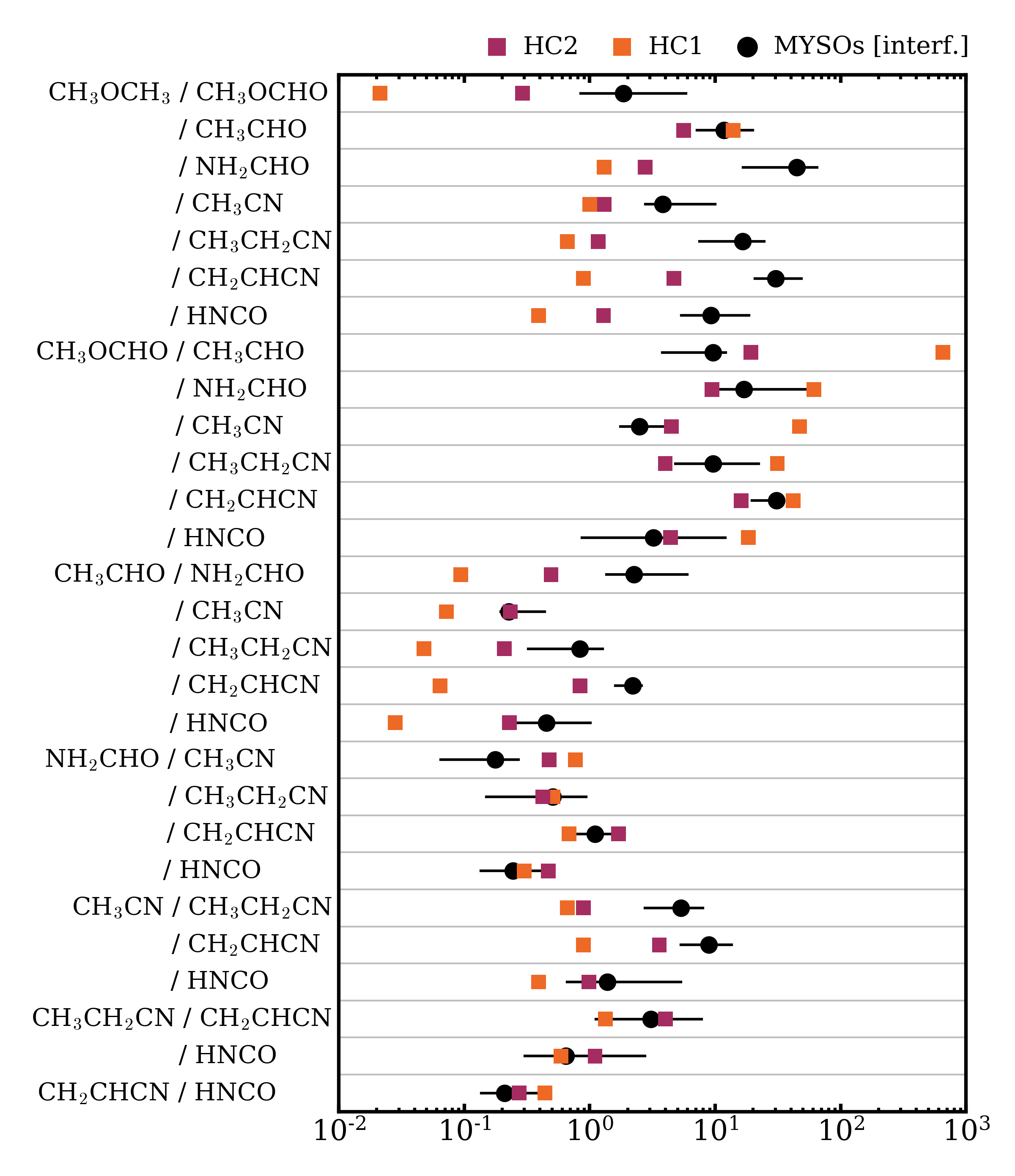}
\caption{Column density ratios in HC1 and HC2 versus those from the MYSO sample from Figure \ref{fig:MYSO_lit_sources}. HC1 and HC2 are shown as orange and purple squares, respectively. Filled black circles denote median MYSO values, while the black lines indicate the upper and lower quartiles.}
\label{fig:MYSOs_ratios}
\end{figure}

We next explore COM column density ratios, which often serve as important signposts of evolutionary state and provide constraints on the chemistry occurring within particular sources \citep[e.g.,][]{Fontani07, Suzuki18}. Figure \ref{fig:MYSOs_ratios} compares column density ratios in both hot cores with those in the literature sample of MYSOs. HC1 and HC2 consistently deviate by up to two orders of magnitude from the MYSO sample in ratios among CH$_3$CHO, CH$_3$OCH$_3$, and CH$_3$OCHO. These discrepancies are readily understood by recalling that in Figure \ref{fig:MYSO_lit_sources}, HC1 and HC2 are shown to be unusually poor in CH$_3$CHO and CH$_3$OCH$_3$, and present CH$_3$OCHO abundances in the upper range of what has been previously observed. In addition, HC1 may be unusual in its efficient CH$_3$OCHO formation, perhaps due to unique conditions favoring gas-phase formation (see Section \ref{sec:CH3OCH3_CH3OCHO_disc}).

The range of ratios exhibited by N-bearing species, such as CH$_2$CHCN/HNCO, NH$_2$CHO/HNCO, and NH$_2$CHO/CH$_2$CHCN, is substantially narrower than that of O-bearing COMs. In particular, HC1 and HC2 appear more typical relative to other MYSOs, and more consistent with each other, i.e. within factors of a few, in their overall N chemistry. The one exception to this trend is that of the complex cyanides, namely the CH$_3$CN/CH$_3$CH$_2$CN and CH$_3$CN/CH$_2$CHCN ratios, which are elevated in both HC1 and HC2.

The ratio between CH$_2$CHCN and CH$_3$CH$_2$CN is sensitive to physical parameters of hot cores \citep[][]{Caselli93} and has been used as an estimator of evolutionary age \citep[e.g.,][]{Fontani07}. We find CH$_2$CHCN/CH$_3$CH$_2$CN ratios of 0.49$\pm$0.30 and 0.23$\pm$0.04 for HC1 and HC2, respectively. These are both consistent with the MYSO comparison sample (0.3--0.9), and the values (0.3--0.5) reported in \citet{Fontani07} for a single-dish survey of well-known hot cores. Since chemical models predict a sharp decrease in the abundances of CH$_2$CHCN and CH$_3$CH$_2$CN after ${\sim}10^5$~yr \citep{Caselli93}, their mutual detection and consistent ratios in both cores suggest that they are younger than ${\lesssim}10^5$~yr with HC2 likely being a factor of a few younger than HC1. More specific conclusions require detailed chemical modeling, especially since \citet{Charnley92} and \citet{Rodgers01} have suggested that additional gas-phase reactions are important for CH$_2$CHCN production, while \citet{Allen18} demonstrated that a shorter warm-up phase can also reproduce observed cyanide abundances.

\subsubsection{Galactic Center Clouds, Comets, and Low-mass Star-forming Regions}

\begin{figure*}[th!]
\centering
\includegraphics[width=\linewidth]{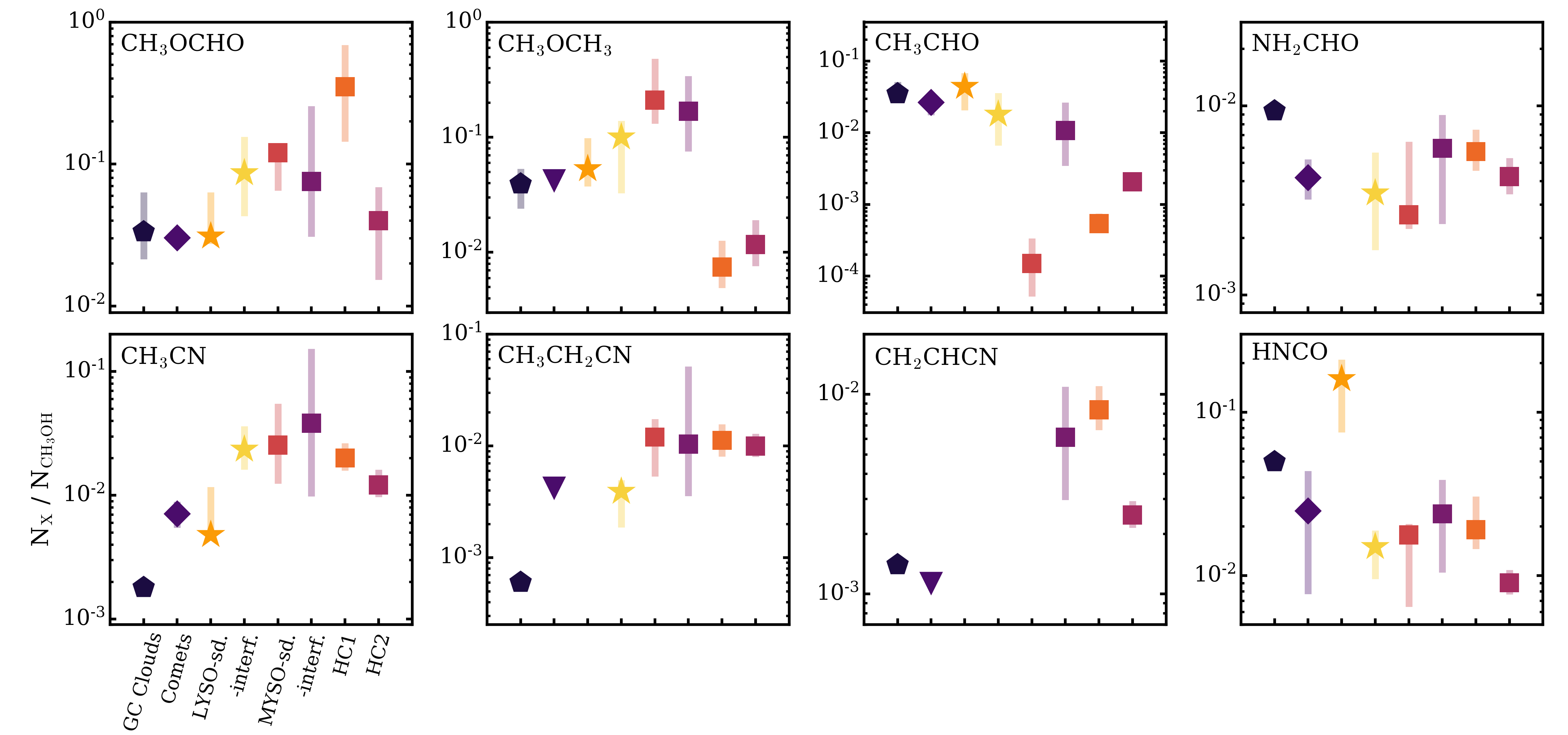}
\caption{Comparison of COM column densities with respect to CH$_3$OH. Data are taken from: GC clouds \citep[][]{Requena_Torres06, Requena_Torres08, Zeng18}; comets \citep{Biver14, Biver19}; LYSOs from single dish \citep[sd.;][]{Bergner17} and interferometric \citep[interf.;][]{Belloche20} observations; and MYSOs from single dish observations \citep[][and references therein]{Taquet15}. The MYSO sample observed with inteferometers is the same as that shown in Figures \ref{fig:MYSO_lit_sources} and \ref{fig:MYSOs_ratios}. The comet data comprise the two well-studied comets C/1995 O1 (Hale-Bopp) and C/2014 Q2 (Lovejoy). Inverted triangles indicate upper limits.}
\label{fig:LYSOs_compares}
\end{figure*}

To better understand COM trends across different physical environments, Figure \ref{fig:LYSOs_compares} compares COM column densities in HC1 and HC2 with a variety of different ISM sources, including Galactic center (GC) clouds, comets, and low-mass YSOs (LYSOs). COM abundances are remarkably similar across this wide set of sources, which suggests that the underlying chemistry is not greatly sensitive to variations in physical environments. In particular, N-bearing species, e.g., NH$_2$CHO, CH$_3$CN, CH$_3$CH$_2$CN, exhibit relatively consistent abundances across different types of sources, while O-bearing COMs display larger variations. In particular, CH$_3$OCH$_3$ is underabundant by at least an order of magnitude in both hot cores, while CH$_3$OCHO is enhanced in HC1 by a factor of a few relative to all other sources. These differences are another indicator, similar to the unusual COM ratios seen in Section \ref{sec:mass_SF_compare}, of the relative deficits or overabundances in the derived column densities of these species and further establish their relative significance.

CH$_3$CHO exhibits the largest spread across source types. \citet{Oberg_Faraday} found that spatially-resolved hot cores exhibited higher CH$_3$CHO/CH$_3$OH ratios compared to those inferred from single-dish observations. We confirm the existence of this trend by comparing observations of unresolved MYSOs versus those that are spatially resolved. The spatially-resolved values are reduced by up to two orders of magnitude in CH$_3$CHO/CH$_3$OH. We do not see a similar trend in the LYSOs, which have nearly the same CH$_3$CHO/CH$_3$OH ratios for both types of observations. Based on Figure \ref{fig:LYSOs_compares}, the majority of other COMs have consistent abundances between their resolved and unresolved samples, which confirms that CH$_3$CHO/COMs ratios in MYSOs are lower for unresolved observations.

This trend is best understood by first contrasting single dish observations of LYSOs and MYSOs. In the case of LYSOs, large beam sizes result in observations dominated by cold protostellar envelopes, where, due to their low temperatures, CH$_3$CHO is expected to be abundant. However, this is not the case for MYSOs, which exhibit higher temperatures in their envelopes and surrounding gas structures. But as seen in \citet{Oberg_Faraday} and here in G10.6, there is a rapid fall off in CH$_3$CHO abundances at warmer temperatures. Moreover, unresolved observations of MYSOs are not sensitive to high CH$_3$CHO column densities associated with hot core chemistry. The combination of these factors leads to dramatically reduced CH$_3$CHO/CH$_3$OH ratios, as is clearly seen in Figure \ref{fig:LYSOs_compares}. In this scenario, the intermediate CH$_3$CHO/CH$_3$OH ratios of HC1 and HC2, i.e., between that of the unresolved and resolved MYSO samples, is naturally explained by the fact that both hot cores are only marginally resolved in our observations of G10.6.

If this explanation is true, we would expect CH$_3$CHO / CH$_3$OH ratios in cold clouds, which do not possess hot core chemistry and contain only cold gas, to be comparable or larger to those from single dish studies of LYSOs. Recent observations of several starless and prestellar cores in Taurus by \citet{Scibelli20} provide a helpful comparison sample for this purpose. \citet{Scibelli20} found a median CH$_3$CHO/CH$_3$OH ratio of ${\sim}0.09$, which is indeed slightly larger than that associated with single dish studies of LYSOs.

\section{Conclusions}
\label{sec:conclusions}

We have presented ALMA high spatial resolution observations of the COM chemistry in massive star-forming region G10.6 at 1.3~mm. These observations reveal that while hot core chemistry appears consistent over many different environments, the chemistry of surrounding COM structures is much more varied and complex. Importantly, the discovery of extended COM structures throughout the central 20,000~au region of G10.6 illustrates that COM chemistry is not only confined to hot molecular cores, but is likely more generally associated with massive star-forming regions. The highly-structured nature of these COM features hints at the importance of interactions between hot, ionized gas and nearby surrounding dense gas.

In addition to having important implications for models of such regions (e.g., external heating), our results further reinforce the need for high spatial resolution observations and suggest that the small-scale distributions of COMs can provide crucial insight into the physical and chemical processes associated with massive star formation. These results also highlight the need for more observations at intermediate spatial resolutions, which are sensitive to larger physical scales, to properly characterize COM emission. This is best demonstrated by COMs such as CH$_3$CHO, for instance, where insufficient resolution, e.g., single dish observations, would not capture core-scale lukewarm chemistry, while very high angular resolutions would instead filter out more extended cold gas chemistry.

We identify a variety of chemical trends, e.g., CH$_3$OCH$_3$/CH$_3$OCHO, HNCO/NH$_2$CHO, previously observed in samples of diverse ISM sources, to also hold true within G10.6 in a spatially-resolved manner. Moreover, they often behave in the very same way, i.e., power law slopes, column density ratios, as they do in these other sources, suggesting that the underlying COM chemistry is tightly linked. Despite this broad agreement, we also observe several unexpected trends, including a spatial distribution of CH$_3$OCH$_3$ that is strikingly dissimilar to that of CH$_3$OCHO and an overabundance of CH$_3$OCHO within at least one hot core. These findings suggest that our current understanding of COM chemistry remains incomplete and provide valuable inputs for chemical models.

\acknowledgments

This paper makes use of the following ALMA data: ADS/JAO.ALMA {\#}2015.1.00106.S. ALMA is a partnership of ESO (representing its member states), NSF (USA) and NINS (Japan), together with NRC (Canada) and NSC and ASIAA (Taiwan), in cooperation with the Republic of Chile. The Joint ALMA Observatory is operated by ESO, AUI/NRAO and NAOJ. The National Radio Astronomy Observatory is a facility of the National Science Foundation operated under cooperative agreement by Associated Universities, Inc.

The authors thank the anonymous referees for valuable comments that improved both the content and presentation of this work. This manuscript benefited greatly from the helpful comments of David Wilner and Alyssa Goodman. C.J.L. also thanks Ryan Loomis and Rafael Mart{\'i}n-Dom{\'e}nech for discussions concerning the COM line fitting process. C.J.L. acknowledges funding from the National Science Foundation Graduate Research Fellowship under Grant DGE1745303. K.I.{\"O}. acknowledges funding from the Simons Collaboration on the Origins of Life (SCOL \#321183, K{\"O}). R.G.M. acknowledges support from UNAM-PAPIIT project IN104319. H.B.L. is supported by the Ministry of Science and Technology (MoST) of Taiwan (Grant No. 108-2112-M-001-002-MY3). P.T.P.H. is supported by Ministry of Science and Technology (MoST) of Taiwan Grant MOST 108-2112-M-001-016-MY2.


\vspace{5mm}
\facilities{ALMA}


\software{\texttt{Astropy} \citep{Astropy13}, CASA \citep{McMullin07}, \texttt{emcee} \citep{Foreman13}, \texttt{MADCUBA} \citep{Martin19}, \texttt{Matplotlib} \citep{Hunter07}, \texttt{NumPy} \citep{van_der_Walt11}}



\clearpage

\appendix

\section{Full Spectral Coverage}
\label{appendix:Extra_Spec_Appendix}

\begin{figure*}[h]
\centering
\includegraphics[width=\linewidth]{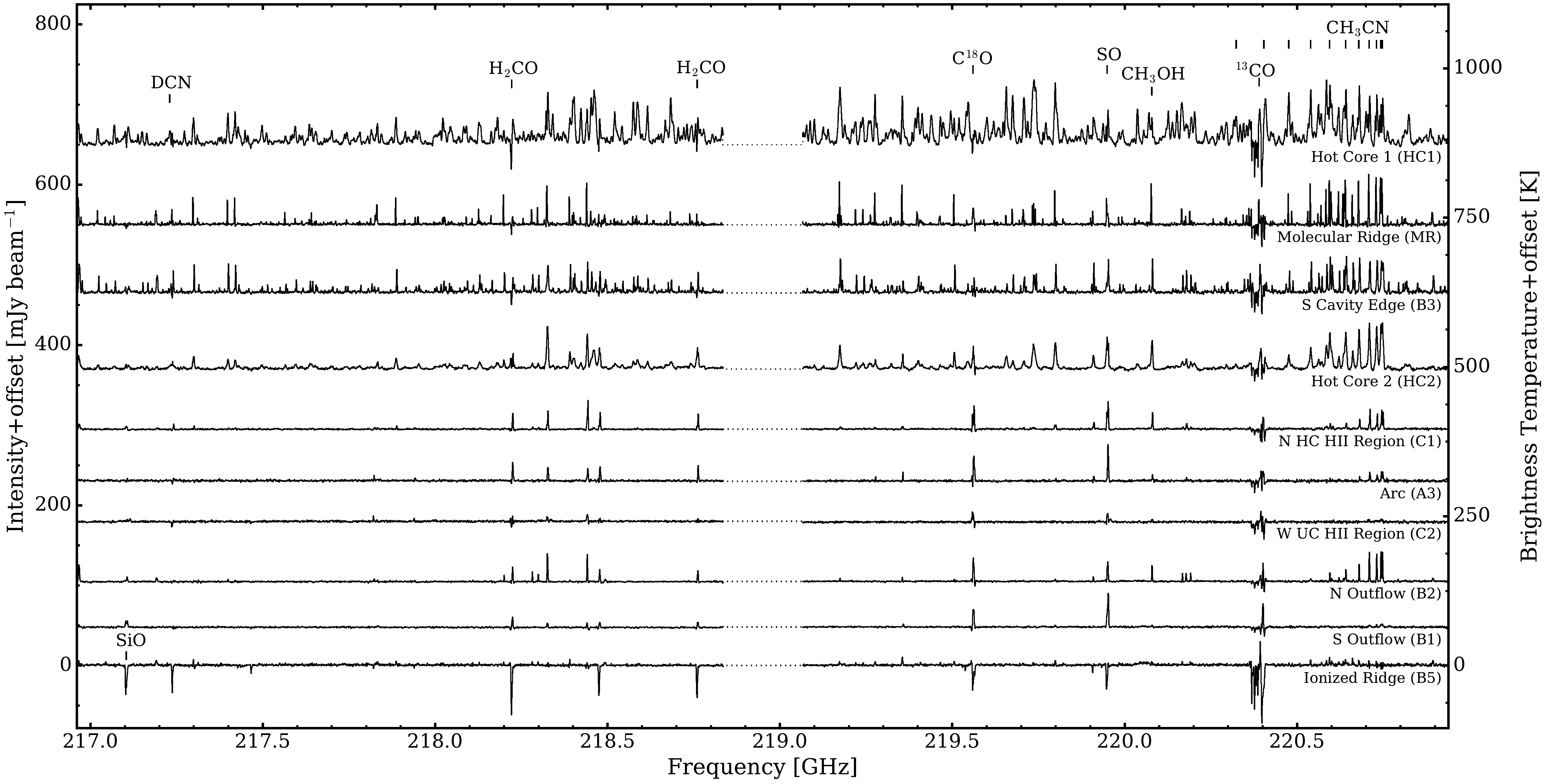}
\caption{Spectra extracted from single pixels toward the same ten positions in G10.6 as in Figure \ref{fig:Fig2}. Dashed line represent gaps in observational coverage between the spectral windows.}
\label{fig:FigA1}
\end{figure*}

Figure \ref{fig:FigA1} shows the remaining half of our spectral coverage toward the same representative positions as in Figure \ref{fig:Fig2}.

\clearpage

\section{COM Kinematics in G10.6}
\label{appendix:COM_Morph_Kine_Appendix}

\begin{figure*}[h]
\centering
\includegraphics[width=\linewidth]{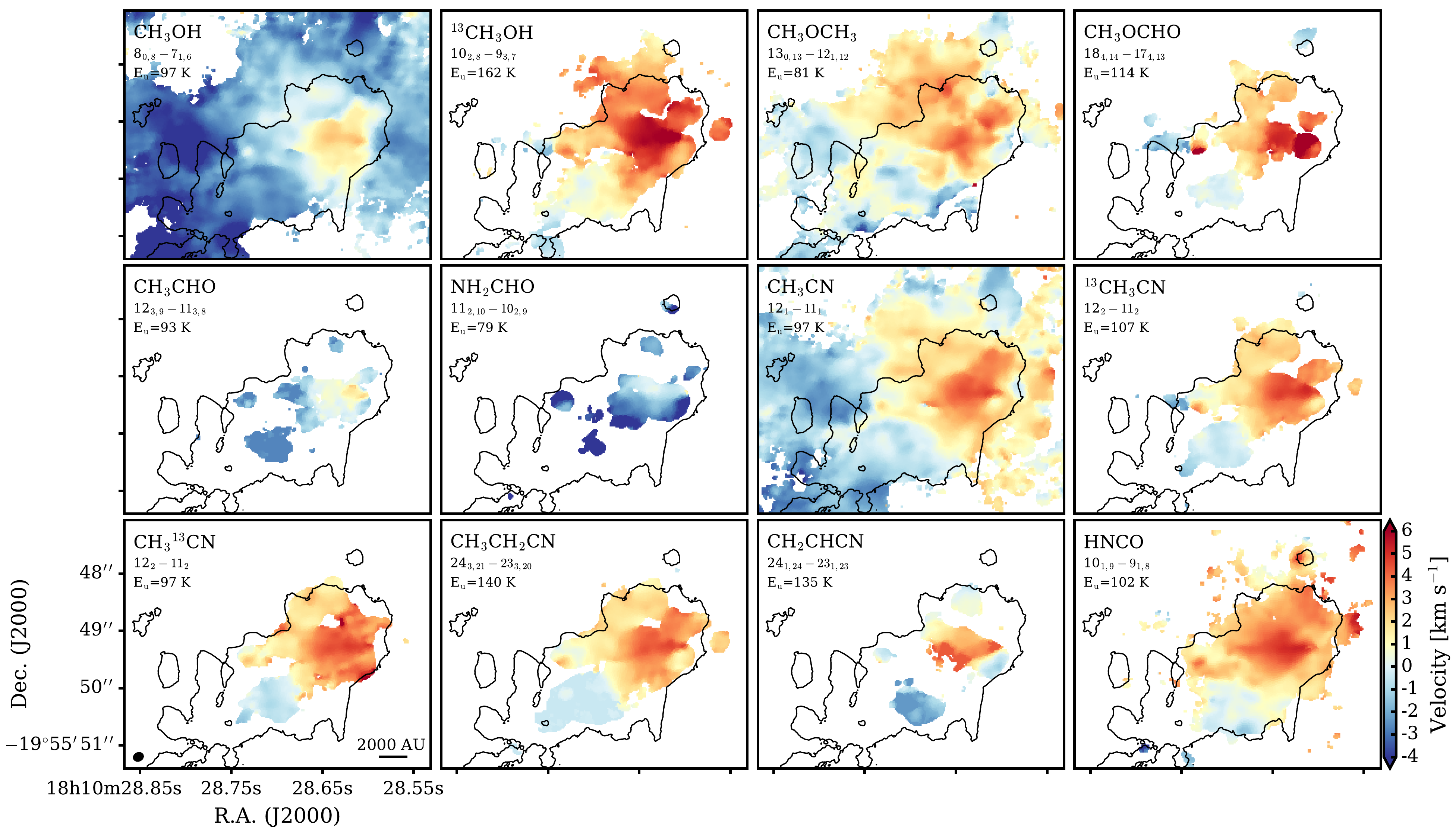}
\caption{Velocity structure for transitions with E$_{\rm{u}} \approx 100$~K for all molecules in our sample. The velocity scale in the color bar is the same for all panels. The black contours show the continuum emission at the 10\% level.}
\label{fig:FigA2}
\end{figure*}

In Figure \ref{fig:FigA2}, we examine the velocities of the same lines as in Figure \ref{fig:Fig_int_sum}. We find a coherent velocity gradient of ${\sim}$10~km~s$^{-1}$ along the NW-SE direction in both individual transitions of single molecules and across the entire sample of species. The velocities of NH$_2$CHO appear to be systematically blue shifted relative to the other COMs, which may indicate that the emission is originating from a different spatial region due to chemical layering or an expanding or contracting shell of gas. The velocities of CH$_3$CHO are also modestly blue shifted, and given that both NH$_2$CHO and CH$_3$CHO are among the molecules with the coolest temperatures in G10.6, it is possible that they are both arising from a similar region that is distinct from that of the remainder of the warmer COMs.

\section{Details of Line Data Analysis} \label{appendix:fit_details_appendix}

Below, we present a detailed description of the entire analysis process, from the initial pixel-masking to the derivation of rotational temperature and column density maps.

\subsection{Line Fitting}
Integrated intensities for each observed line were determined by fitting a single Gaussian profile to each feature. For lines with substantial wings, only the central peak was used in deriving column densities. Unresolved multiplets originating from the same species were treated as a single line by combining the degeneracies and line intensities. 

For each transition, a mask corresponding to a 5$\sigma$ level in peak intensity in a 10 km~s$^{-1}$ window centered on the transition frequency was used to define a pixel-by-pixel mask. We obtained initial constraints on line parameters for each species separately based on the most cleanly observed lines. For these initial fits, the FWHM was allowed to vary between 1.3~km~s$^{-1}$ (i.e., the velocity resolution) and 15~km~s$^{-1}$, while systemic velocity was restricted to a window centered on the expected transition frequency with a width between $\pm$3--8~km~s$^{-1}$, depending on the presence of adjacent wings from other transitions or nearby line crowding from other species. For all other lines of a single molecule, the FWHM was restricted to 0.25--1.25$\times$ the FWHM of the template line and the systemic velocity to $\pm$2 velocity channels from the template line position. We note that some variation between lines is expected, even for a single species, due to different excitation conditions of lines with different upper energy levels and Einstein A coefficients.

Since G10.6 is line-rich, blending is occasionally present, especially near regions of denser gas with large linewidths (e.g., HC1, and to a lesser extent, HC2). In general, lines that were not unambiguously free from blending throughout G10.6 are not included in the analysis. For each transition of each species, we randomly selected a set of a few hundred fits across G10.6 and confirmed via visual inspection that the fitting process was accurately capturing the observed line properties. During this process, we removed all blended transitions from the analysis. In a few rare cases, line blending was due to a nearby transition originating from the same species, e.g., the K$=0$ and K$=1$ lines of the J$=$12--11 ladder of CH$_3^{13}$CN. In such cases, we treated them as a single line by fitting a single Gaussian and combining line characteristics.

\subsection{Rotational Temperature and Column Density Derivation}
For each such pixel, we calculated column density and rotational temperature. Assuming optically thin emission, the column density of molecules in the upper state of each transition $N^{\rm{thin}}_{\rm{u}}$ is related to the emission surface brightness $I_{\nu}$ via:

\begin{equation} \label{eq:eqn1}
I_{\nu} = \dfrac{A_{ul} N^{\rm{thin}}_{\rm{u}} h c}{4\pi \Delta v},
\end{equation}

where $A_{\rm{ul}}$ is the Einstein coefficient and $\Delta v$ is the linewidth \citep[e.g.,][]{Bisschop08}. Since we are performing this analysis on individual pixels of a certain size, we need to write $I_{\nu}$ in terms of the flux density $S_{\nu}$ and the solid angle subtended by the source $\Omega$ via $I_{\nu} = S_{\nu} / \Omega$ \citep[e.g.,][]{Bergner18, Loomis18}. Rearranging Equation \ref{eq:eqn1}: 

\begin{equation} \label{eq:eqn2}
N^{\rm{thin}}_{\rm{u}} = \dfrac{4\pi S_{\nu} \Delta v}{A_{ul} \Omega hc}
\end{equation}

where $S_{\nu} \Delta v$ is the integrated flux density for each transition, and we use the pixel size ($0.018^{\prime \prime}$) to estimate the solid angle $\Omega$.

The upper state level population $N_{\rm{u}}$ is related to the total column density $N_{\rm{T}}$ by the equation:

\begin{equation} \label{eq:eqn3}
\frac{N_{\rm{u}}}{g_{\rm{u}}} = \frac{N_{\rm{T}}}{Q(T_{\rm{rot}})} e^{-E_{\rm{u}} / k_{\rm{B}} T_{\rm{rot}}},
\end{equation}

where $g_{\rm{u}}$ is the degeneracy of the upper state level, $Q$ is the partition function, and $T_{\rm{rot}}$ is the rotational temperature, and E$_{\rm{u}}$ is the upper state energy. 

The frequencies, line strengths, and upper-state energies of the observed lines of CH$_3$OH are summarized in Appendix Table \ref{tab:TableA1}, along with the complete set of unblended lines for all molecules included in our sample. In general, line characteristics were taken from the JPL\footnote{\url{https://spec.jpl.nasa.gov/}} and CDMS\footnote{\url{http://www.astro.uni-koeln.de/cdms/catalog}} catalogs, as indicated in Table \ref{tab:tab2}. We note that catalogs sometime differ in their consideration of nuclear spin in their Einstein A values and partition functions. Care therefore was taken to use matching Einstein A values and partition functions. Partition functions were linearly interpolated from catalog values for all molecules, except for CH$_3$OCHO, where the relative contributions from the vibrational and torsional states are over 40\%. In this case, we used the complete rotational-torsional-vibrational partition function from \citet{Carvajal19}.

As standard in rotational diagram analysis \citep[e.g.,][]{Goldsmith99}, taking the logarithm of Equation \ref{eq:eqn3} gives the linear equation:

\begin{equation} \label{eq:eqn4}
\ln \left( \frac{N_{\rm{u}}}{g_{\rm{u}}} \right) = \ln N_{\rm{T}} - \ln Q(T_{\rm{rot}}) - \frac{E_{\rm{u}}}{k_{\rm{B}} T_{\rm{rot}}}.
\end{equation}
  
A semi-log plot of $N_{\rm{u}}/g_{\rm{u}}$ versus upper state energies E$_{\rm{u}}$ allows for the derivation of rotational temperature and total column density from the best fit slope and intercept, respectively. The optical depth of the observed transitions is unknown \textit{a priori}, and in regions of high density such as G10.6, we may expect optically thick lines. In this case, i.e. $\tau \not \ll 1$, the optical depth correction factor $C_{\tau}$ must be applied:

\begin{equation} \label{eq:eqn5}
C_{\tau} = \frac{\tau}{1 - e^{-\tau}},
\end{equation}

which makes the true level populations:

\begin{equation}
N_{\rm{u}} = N_{\rm{u}}^{\rm{thin}} C_{\tau}.
\end{equation}

With these corrections, Equation \ref{eq:eqn4} can be rewritten as:

\begin{equation} \label{eq:eqn6}
\ln \left( \frac{N_{\rm{u}}^{\rm{thin}}}{g_{\rm{u}}} \right) + \ln C_{\tau} = \ln N_{\rm{T}} - \ln Q(T_{\rm{rot}}) - \frac{E_{\rm{u}}}{k_{\rm{B}} T_{\rm{rot}}}.
\end{equation}

The optical depths of individual transitions can then be related back to the upper state level populations via:

\begin{equation} \label{eq:eqn7}
\tau_{ul} = \frac{A_{ul}c^3}{8\pi\nu^3 \Delta v} N_{\rm{u}} \left(e^{h\nu / k_{\rm{B}} T_{\rm{rot}}} - 1 \right).
\end{equation}

\begin{figure*}[h]
\centering
\includegraphics[width=0.95\linewidth]{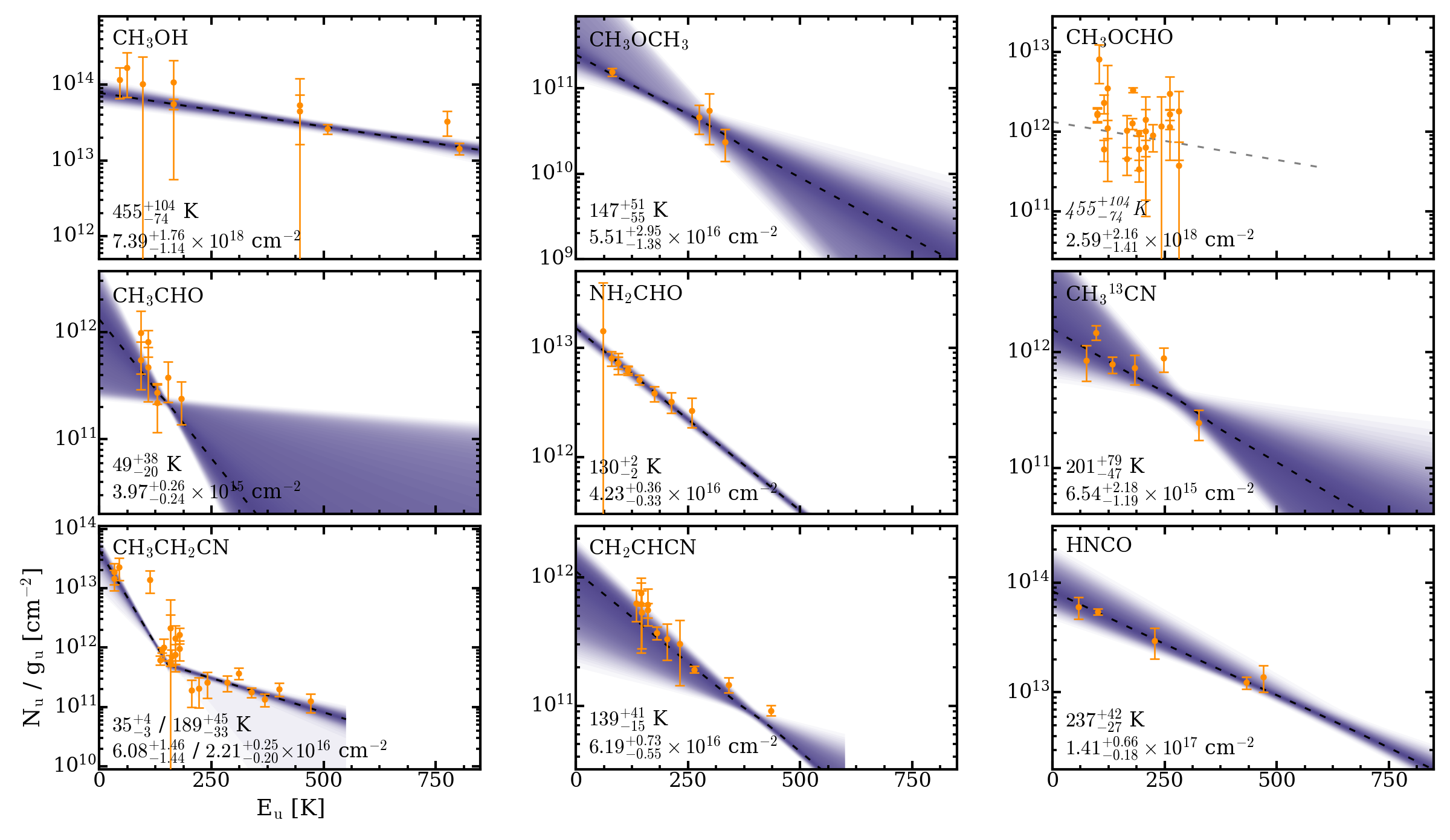}
\caption{Rotational diagrams for COMs in our sample toward HC1. Transitions are shown in orange and random draws from the fit posteriors are plotted in purple. The fit is indicated by a dashed black line and derived values are shown in the lower left corners. A gray dashed line and italicized rotational temperature indicate that this temperature was adopted from a chemically-similar species, as described in the text. (\textit{The complete figure set (4 images) showing rotational diagrams toward four positions of interest in G10.6 is available in the online journal.})}
\label{fig:FigA4}
\end{figure*}

Hence, $C_{\tau}$ can be written as a function of $N_{\rm{u}}$ and substituted back into Equation \ref{eq:eqn7} to construct a likelihood function $L (\rm{data}, N_{\rm{T}}, T_{\rm{rot}})$ to be used for $\chi^2$ minimization \citep[e.g.,][]{Loomis18}. Given this likelihood function, we then used the affine invariant MCMC code \texttt{emcee} \citep{Foreman13} to fit the data and generate posterior probability distributions for N$_{\rm{T}}$ and T$_{\rm{rot}}$, which describe the range of possible column densities and rotational temperatures consistent with the observed data. To illustrate this method, Figure \ref{fig:FigA4} shows rotational diagrams for all species in our COM sample toward HC1. Complete rotational diagrams toward HC2, B3, and MR are shown in Figure Set 1, which is available in the electronic edition of the journal.

\clearpage

\subsection{Line Blending and Example LTE Spectra}

\begin{figure*}[h]
\centering
\includegraphics[width=\linewidth]{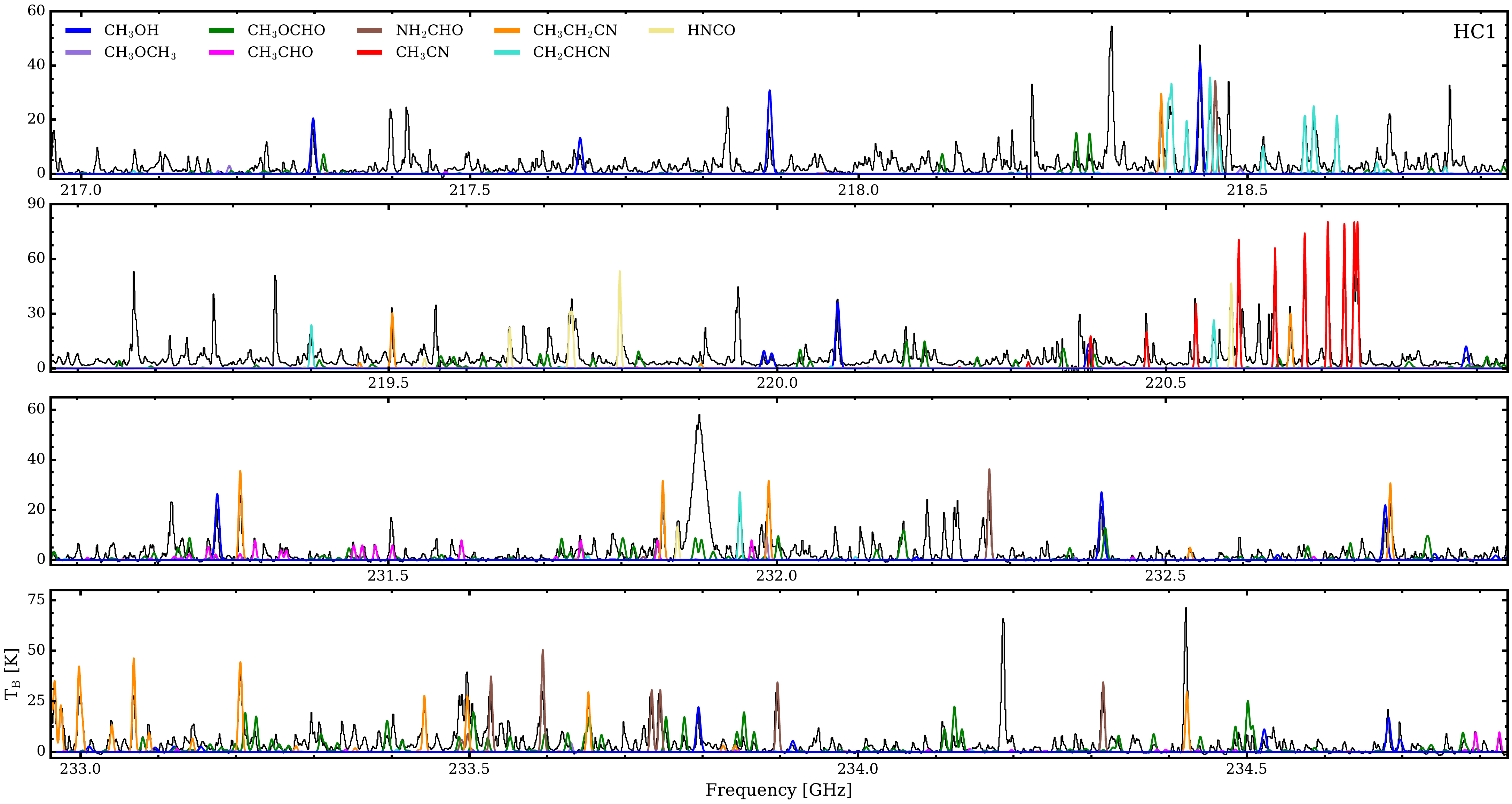}
\caption{Spectrum toward HC1 extracted from a region equal to the beam size with an overlaid synthetic LTE model spectra from \texttt{MADCUBA}. Different colors represent individual molecules according to the legend.}
\label{fig:Fig_22}
\end{figure*}

\begin{figure*}[ht!]
\centering
\includegraphics[width=\linewidth]{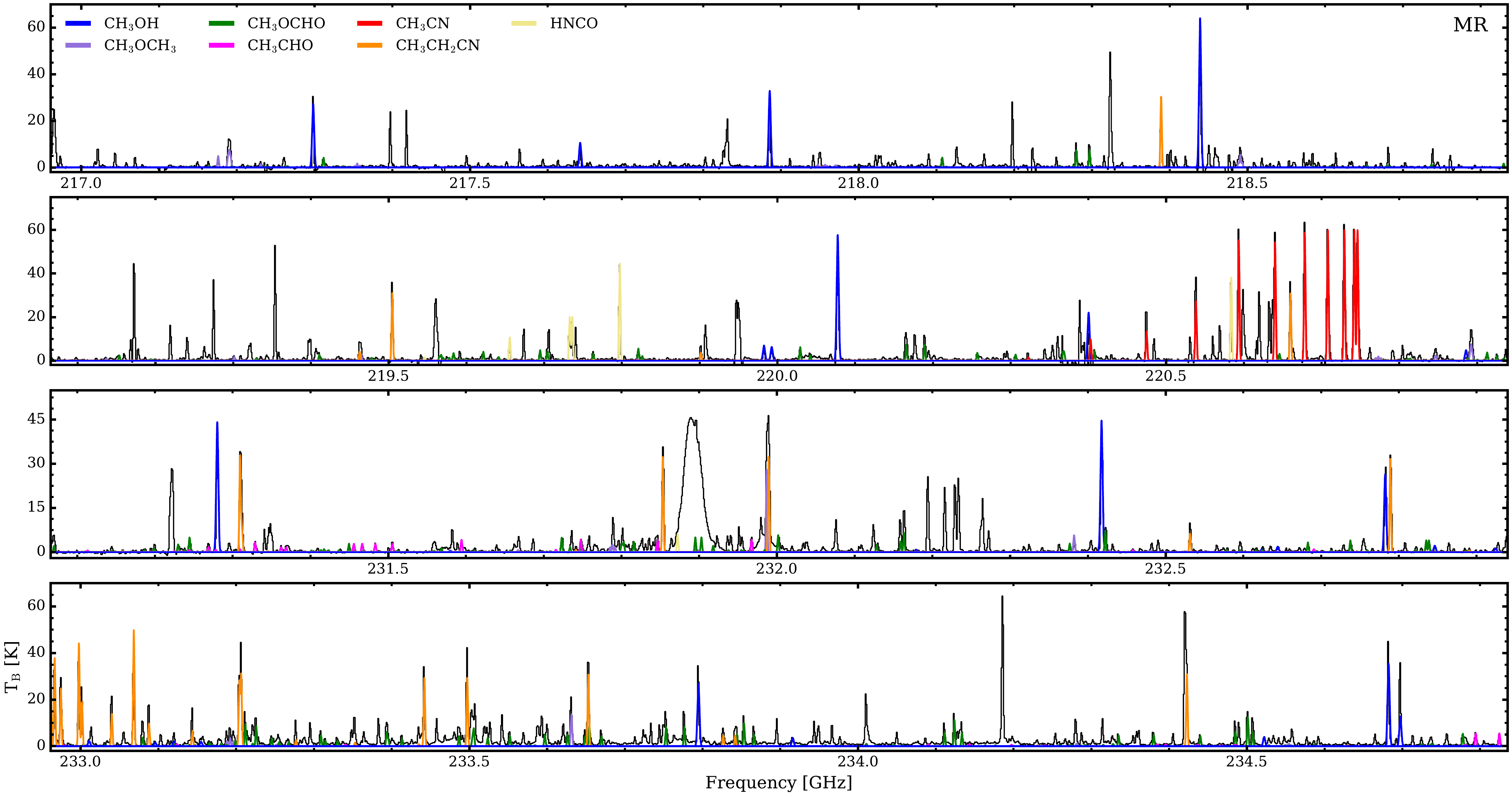}
\caption{Spectrum toward MR extracted from a region equal to the beam size with an overlaid synthetic LTE model spectra from \texttt{MADCUBA}. Different colors represent individual molecules according to the legend.}
\label{fig:Fig_23}
\end{figure*}

To assess the level of blending present, particularly toward dense regions such as HC1, we compared observed spectra with synthetic LTE model spectra generated using the \texttt{MADCUBA}\footnote{\texttt{MADCUBA} is software developed at the Center of Astrobiology of Madrid (CSIC-INTA) to visualize and analyze astronomical datacubes and spectra. \texttt{MADCUBA} is available at: \url{http://cab.inta-csic.es/madcuba/Portada.html}} (MAdrid Data CUBe Analysis) package \citep{Martin19}. We used the SLIM (Spectral Line Identification and LTE Modelling) tool within \texttt{MADCUBA}, which uses spectroscopic data from the CDMS and JPL catalogs to produce model spectra under LTE conditions for a given set of input physical parameters. For each species in our sample, we inputted the rotational temperature, column density, typical FWHM, and systemic velocity, as derived in Section \ref{sec:data_analysis}. We fixed the source size to the beam size for all molecules and extracted spectra toward HC1 and MR from a region equal in area to the beam size. Figure \ref{fig:Fig_22} shows the LTE model spectrum toward HC1, which has the densest gas and largest line widths in G10.6, and thus, we expect maximal line blending to occur toward this region. In general, a moderate degree of line blending is present for some species (e.g., CH$_3$OCHO), but in all cases, there are a sufficient number of unblended lines to allow for reliable fits and robust rotational diagrams. Next, in Figure \ref{fig:Fig_23}, we show an LTE spectrum toward the MR, which is more representative of the gas outside of HC1 and HC2 in G10.6. Here, blending is minimal, due in part to the narrower line widths.

\section{Deriving the $^{12}$C/$^{13}$C Ratio in G10.6} \label{appendix:Iso_CH3OH_Appendix}

\begin{figure*}[h]
\centering
\includegraphics[width=\linewidth]{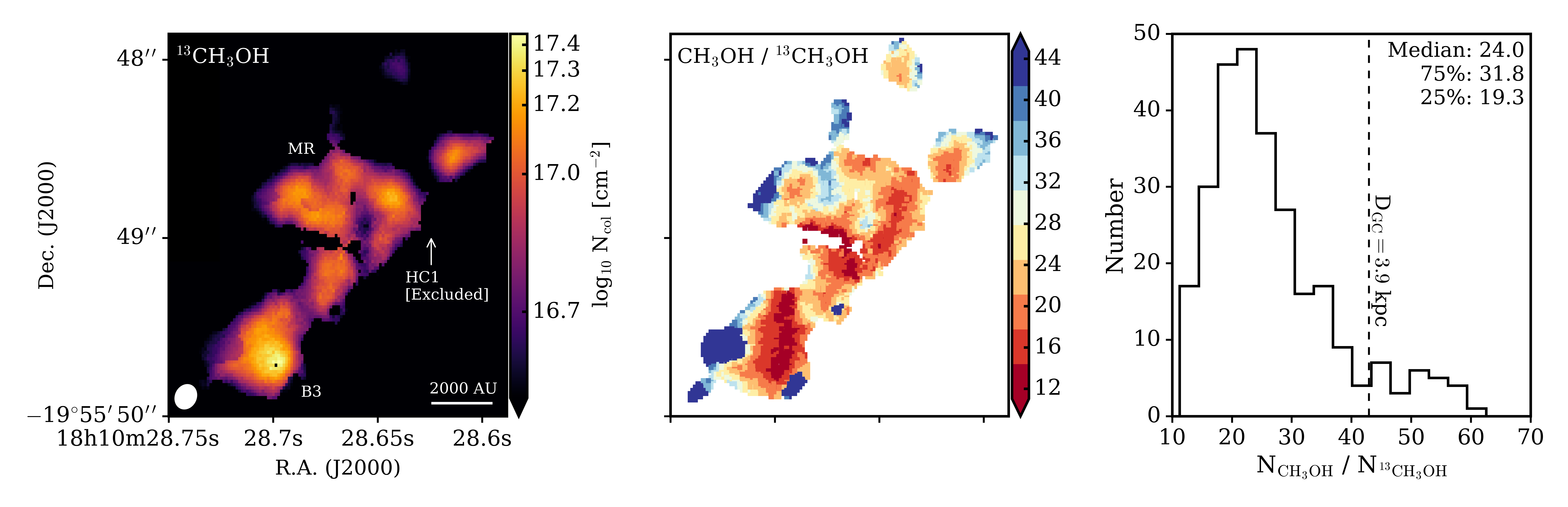}
\caption{Column density map for $^{13}$CH$_3$OH (\textit{left}), CH$_3$OH/$^{13}$CH$_3$OH column density ratio map (\textit{middle}), and histogram distribution (\textit{right}). The dashed black line shows the expected Galactic $^{12}$C/$^{13}$C ratio at the distance of G10.6 \citep{Milam05}.}
\label{fig:FigA5}
\end{figure*}

Due to the sparse number of observed $^{13}$CH$_3$OH transitions, we adopted the well-constrained rotational temperature from the main species CH$_3$OH. Toward HC1, we only detected two high E$_{\rm{u}}$ transitions (592~K and 594~K) and one low E$_{\rm{u}}$ (162~K) line. The high uncertainties associated with the integrated intensities of these lines made fitting difficult, resulting in an anomalously high column density. Thus, we purposely excluded HC1 when deriving $^{13}$CH$_3$OH column densities. Toward HC2, we only detected the E$_{\rm{u}}$ (162~K) line, which also prohibits a column density determination. The $^{13}$CH$_3$OH column density map is shown in the left panel of Figure \ref{fig:FigA5}.

In order to assess the typical $^{12}$C/$^{13}$C ratio in G10.6, we compared the distribution of relative column densities of CH$_3$OH and $^{13}$CH$_3$OH in the middle and right panels of Figure \ref{fig:FigA5}. The majority of values are clustered about the median column density ratio of 22.8, which is enhanced by nearly a factor of two relative to the expected ratio of approximately 43 \citep{Milam05} at the distance of G10.6 (D$_{\rm{GC}} = 3.9$~kpc). Ratios of ${>}40$ only occur toward the periphery of the dense gas in G10.6, where line intensities have decreased and are therefore considerably more uncertain and should be interpreted with caution. In the regions used to derive the $^{12}$C/$^{13}$C ratio, $\tau \sim $ 0.1--0.4 for the CH$_3$OH lines. While this suggests that optical depth effects are not substantially biasing the measured isotopic ratios, they cannot be definitely ruled out (e.g., unresolved internal cloud structures, inclusion of velocity ranges which are strongly affected by self absorption or high opacity but not resolved in velocity.)

\section{Full Line Lists}

The full list of COM transitions used to derive rotational temperatures and column densities is shown in Table \ref{tab:TableA1}.

\startlongtable
\begin{deluxetable*}{lccccc}
%\tablenum{4}
\tablecaption{Properties of Unblended Transitions\label{tab:TableA1}}
\tablehead{[-.2cm]\colhead{Transition} & \colhead{$\nu$ [GHz]} & \colhead{E$_{\rm{u}}$ [K]} & \colhead{A$_{\rm{ul}}\,[\log_{10} \,\rm{s}^{-1}]$} & \colhead{g$_{\rm{u}}$} & \colhead{Line List} \\[-.55cm]}
\startdata
\textbf{CH$_3$OH} & & & & & \\ 
4$_{2,3}$--3$_{1,2}$, E                     & 218.4401 & 45.5 & $-$4.32910 & 9 & CDMS \\
4$_{2,3}$--5$_{1,4}$, A                    & 234.6834 & 60.9 & $-$4.72715 & 9 & CDMS \\
8$_{0,8}$--7$_{1,6}$, E                    & 220.0786 & 96.6 & $-$4.59927 & 17 & CDMS \\
10$_{2,9}$--9$_{3,6}$, A                 & 231.2811 & 165.4 & $-$4.73685  & 21 & CDMS \\
10$_{2,8}$--9$_{3,7}$, A                 & 232.4185 & 165.4 & $-$4.72826 & 21 & CDMS \\
18$_{3,16}$--17$_{4,13}$, A           & 232.7834 & 446.5 & $-$4.66380 & 37 & CDMS \\
18$_{3,15}$--17$_{4,14}$, A          & 233.7957  & 446.6  & $-$4.65725 & 37 & CDMS \\
20$_{1,19}$--20$_{0,20}$, E           & 217.8865 & 508.4 & $-$4.47125 & 41 & CDMS \\
23$_{5,19}$--22$_{6,17}$, E           & 219.9937 & 775.9 & $-$4.75652 & 47 & CDMS \\
25$_{3,22}$--24$_{4,20}$, E           & 219.9837 & 802.2 & $-$4.69035 & 51 & CDMS \\ \hline \\[-0.2cm]
\textbf{$^{13}$CH$_3$OH} & & & & & \\ 
5$_{1,5}$--4$_{1,4}$, A        & 234.0116  & 48.3     & $-$4.27824 & 11 & CDMS \\
10$_{2,8}$--9$_{3,7}$, A      & 217.3996 & 162.4   & $-$4.81610 & 21 & CDMS \\
14$_{1,13}$--13$_{2,12}$, A  & 217.0446  & 254.3   & $-$4.62441 & 29 & CDMS \\ 
17$_{7,10}$--18$_{6,13}$, A  & 220.3218 & 592.3   & $-$4.88725 & 35 & CDMS \\
17$_{7,11}$--18$_{6,12}$, A & 220.3218 & 592.3   & $-$4.88725 & 35 & CDMS \\
22$_{1,21}$--22$_{0,22}$, E  & 231.8184 & 593.9   & $-$4.41192 & 45 & CDMS \\ \hline \\[-0.2cm]
\textbf{CH$_3$OCH$_3$} & & & & & \\
13$_{0,13}$--12$_{1,12}$, AA & 231.9879 & 80.9  & $-$4.03882 & 270 & CDMS \\
13$_{0,13}$--12$_{1,12}$, EE & 231.9879 & 80.9  & $-$4.03884 & 432 & CDMS \\
13$_{0,13}$--12$_{1,12}$, AE & 231.9879 & 80.9  & $-$4.03887 & 162 & CDMS \\
13$_{0,13}$--12$_{1,12}$, EA & 231.9879 & 80.9  & $-$4.03888 & 108 & CDMS \\
23$_{4,20}$--23$_{3,21}$, AA & 220.8950 & 274.4 & $-$4.24596 & 282 & CMDS \\
23$_{4,20}$--23$_{3,21}$, EE & 220.8934 & 274.4 & $-$4.24603 & 752 & CDMS \\
23$_{4,20}$--23$_{3,21}$, AE & 220.8918 & 274.4 & $-$4.24604 & 94  & CDMS \\
23$_{4,20}$--23$_{3,21}$, EA & 220.8918 & 274.4 & $-4$.24597 & 188 & CDMS \\
24$_{4,20}$--23$_{5,19}$, AA & 220.8465 & 297.5 & $-$4.79796 & 294 & CDMS \\
24$_{4,20}$--23$_{5,19}$, EE & 220.8476 & 297.5 & $-$4.79793 & 784 & CDMS \\
24$_{4,20}$--23$_{5,19}$, AE & 220.8487 & 297.5 & $-$4.79804 & 98  & CDMS \\
24$_{4,20}$--23$_{5,19}$, EA & 220.8488 & 297.5 & $-$4.79797 & 196 & CDMS \\
25$_{5,20}$--25$_{4,21}$, AA & 233.6329 & 331.9 & $-$4.13499 & 510 & CDMS \\
25$_{5,20}$--25$_{4,21}$, EE  & 233.6323 & 331.9 & $-$4.13491 & 816 & CDMS \\
25$_{5,20}$--25$_{4,21}$, AE & 233.6319 & 331.9 & $-$4.13494 & 306 & CDMS \\
25$_{5,20}$--25$_{4,21}$, EA & 233.6316 & 331.9 & $-$4.13495 & 204 & CDMS \\
34$_{7,27}$--33$_{8,26}$, AA & 217.9489 & 611.5 & $-$4.83593 & 414 & CDMS \\
34$_{7,27}$--33$_{8,26}$, AE & 217.9481 & 611.5 & $-$4.83601 & 138 & CDMS \\
34$_{7,27}$--33$_{8,26}$, EA & 217.9483 & 611.5 & $-$4.84024 & 276 & CDMS \\
36$_{4,32}$--36$_{3,33}$, AA & 217.1775 & 638.6 & $-$4.20439 & 438 & CDMS \\
36$_{4,32}$--36$_{3,33}$, AE & 217.1766 & 638.6 & $-$4.20447 & 146 & CMDS \\
36$_{4,32}$--36$_{3,33}$, EA & 217.1766 & 638.6 & $-$4.20440 & 292 & CDMS \\
37$_{4,33}$--37$_{3,34}$, AA & 232.3838 & 673.0 & $-$4.13634 & 750 & CDMS \\
37$_{4,33}$--37$_{3,34}$, EA & 232.3827 & 673.0 & $-$4.13630 & 300 & CDMS \\
37$_{4,33}$--37$_{3,34}$, AE & 232.3827 & 673.0 & $-$4.13629 & 450 & CDMS \\ \hline \\[-0.2cm]
\textbf{CH$_3$OCHO} & & & & & \\
17$_{3,14}$--16$_{3,13}$, E  & 218.2808 & 99.7  & $-$3.82170 & 70 & JPL \\
17$_{3,14}$--16$_{3,13}$, A  & 218.2978 & 99.7  & $-$3.82148 & 70 & JPL \\
17$_{4,13}$--16$_{4,12}$, A  & 220.1902 & 103.1 & $-$3.81694 & 70 & JPL \\
17$_{4,13}$--16$_{4,12}$, E  & 220.1669 & 103.2 & $-$3.81717 & 70 & JPL \\ 
20$_{1,20}$--19$_{1,19}$, E  & 216.9648 & 111.5 & $-$3.81496 & 82 & JPL \\
20$_{1,20}$--19$_{1,19}$, A  & 216.9660 & 111.5 & $-$3.81488 & 82 & JPL \\
20$_{0,20}$--19$_{0,19}$, E & 216.9662 & 111.5 & $-$3.81495 & 82 & JPL \\ 
20$_{0,20}$--19$_{0,19}$, A  & 216.9673 & 111.5 & $-$3.81478  & 82 & JPL \\
18$_{4,14}$--17$_{4,13}$, A   & 233.7775 & 114.4 & $-$3.73540 & 74 & JPL \\
18$_{4,14}$--17$_{4,13}$, E & 233.7540 & 114.4 & $-$3.73553 & 74 & JPL \\ 
19$_{4,16}$--18$_{4,15}$, A  & 233.2267 & 123.3 & $-$3.74021 & 78 & JPL \\
19$_{4,16}$--18$_{4,15}$, E  & 233.2128 & 123.3 & $-$3.74032 & 78 & JPL \\ 
19$_{9,11}$--18$_{9,10}$, E  & 234.5086 & 166.0 & $-$3.82053 & 78 & JPL \\
19$_{9,11}$--18$_{9,10}$, A  & 234.5022 & 166.0 & $-$3.82043 & 78 & JPL \\
19$_{9,10}$--18$_{9,9}$, A    & 234.5024 & 166.0 & $-$3.82043 & 78 & JPL \\
19$_{9,10}$--18$_{9,9}$, E     & 234.4864 & 166.0 & $-$3.82065 & 78 & JPL \\
19$_{10,9}$--18$_{10,8}$, E   & 234.1123 & 178.5 & $-$3.85309 & 78 & JPL \\
19$_{10,10}$--18$_{10,9}$, E & 234.1346 & 178.5 & $-$3.85297 & 78 & JPL \\
19$_{10,10}$--18$_{10,9}$, A & 234.1249 & 178.5 & $-$3.85298 & 78 & JPL \\
19$_{10,9}$--18$_{10,8}$, A   & 234.1249 & 178.5 & $-$3.85298 & 78 & JPL \\
19$_{11,9}$--18$_{11,8}$, E  & 233.8671 & 192.4 & $-$3.89076 & 78 & JPL \\
19$_{11,8}$--18$_{11,7}$, E  & 233.8453 & 192.4 & $-$3.89088 & 78 & JPL \\
19$_{11,9}$--18$_{11,8}$, A  & 233.8542 & 192.4 & $-$3.89077 & 78 & JPL \\
19$_{11,8}$--18$_{11,7}$, A  & 233.8542 & 192.4 & $-$3.89077 & 78 & JPL \\
19$_{12,8}$--18$_{12,7}$, E  & 233.6710 & 207.6 & $-$3.93548 & 78 & JPL \\
19$_{12,7}$--18$_{12,6}$, A  & 233.6553 & 207.6 & $-$3.93559 & 78 & JPL \\
19$_{12,8}$--18$_{12,7}$, A  & 233.6553 & 207.6 & $-$3.93559 & 78 & JPL \\
19$_{12,7}$--18$_{12,6}$, E  & 233.6499 & 207.6 & $-$3.93560 & 78 & JPL \\
19$_{13,7}$--18$_{13,6}$, E  & 233.5246 & 224.2 & $-$3.98927 & 78 & JPL \\
19$_{13,6}$--18$_{13,5}$, E  & 233.5050 & 224.2 & $-$3.98949 & 78 & JPL \\
19$_{13,6}$--18$_{13,5}$, A  & 233.5066 & 224.2 & $-$3.98938 & 78 & JPL \\
19$_{13,7}$--18$_{13,6}$, A  & 233.5066 & 224.2 & $-$3.98938 & 78 & JPL \\
19$_{14,6}$--18$_{14,5}$, E  & 233.4144 & 242.1 & $-$4.05557 & 78 & JPL \\
19$_{14,5}$--18$_{14,4}$, A  & 233.3945 & 242.1 & $-$4.05568 & 78 & JPL \\
19$_{14,6}$--18$_{14,5}$, A  & 233.3945 & 242.1 & $-$4.05568 & 78 & JPL \\
19$_{14,5}$--18$_{14,4}$, E  & 233.3967 & 242.1 & $-$4.05578 & 78 & JPL \\ 
19$_{15,5}$--18$_{15,4}$, E  & 233.3312 & 261.3 & $-$4.13989 & 78 & JPL \\
19$_{15,4}$--18$_{15,3}$, E  & 233.3158 & 261.3 & $-$4.14000 & 78 & JPL \\
19$_{15,4}$--18$_{15,3}$, A  & 233.3100 & 261.3 & $-$4.14001 & 78 & JPL \\
19$_{15,5}$--18$_{15,4}$, A  & 233.3100 & 261.3 & $-$4.14001 & 78 & JPL \\ 
18$_{16,2}$--17$_{16,1}$, A  & 220.9262 & 270.7 & $-$4.46555 & 74 & JPL \\
18$_{16,3}$--17$_{16,2}$, A  & 220.9262 & 270.7 & $-$4.46555 & 74 & JPL \\
19$_{16,4}$--18$_{16,3}$, E  & 233.2686 & 281.8 & $-$4.25236 & 78 & JPL \\
19$_{16,3}$--18$_{16,2}$, A  & 233.2466 & 281.9 & $-$4.25247 & 78 & JPL \\
19$_{16,4}$--18$_{16,3}$, A  & 233.2466 & 281.9 & $-$4.25247 & 78 & JPL \\ \hline \\[-0.2cm]
\textbf{CH$_3$CHO} & & & & & \\
12$_{2,10}$--11$_{2,9}$, E & 234.7955 & 81.9  & $-$3.35208 & 50 & JPL \\
12$_{3,9}$--11$_{3,8}$, E  & 231.8476 & 92.6  & $-$3.38760 & 50 & JPL \\
12$_{3,10}$--11$_{3,9}$, A & 231.5953 & 92.6  & $-$3.38561 & 50 & JPL \\
12$_{4,9}$--11$_{4,8}$, E  & 231.5063 & 108.3 & $-$3.40929 & 50 & JPL \\
12$_{4,8}$--11$_{4,7}$, E  & 231.4844 & 108.3 & $-$3.40927 & 50 & JPL \\
12$_{4,9}$--11$_{4,8}$, A  & 231.4567 & 108.4 & $-$3.40933 & 50 & JPL \\
12$_{5,8}$--11$_{5,7}$, E  & 231.3698 & 128.5 & $-$3.44161 & 50 & JPL \\
12$_{5,7}$--11$_{5,6}$, E  & 231.3633 & 128.5 & $-$3.44168 & 50 & JPL \\
12$_{5,8}$--11$_{5,7}$, A  & 231.3296 & 128.6 & $-$3.44164 & 50 & JPL \\
12$_{5,7}$--11$_{5,6}$, A  & 231.3296 & 128.6 & $-$3.44164 & 50 & JPL \\
12$_{6,7}$--11$_{6,6}$, A  & 231.2699 & 153.4 & $-$3.48413 & 50 & JPL \\
12$_{6,6}$--11$_{6,5}$, A  & 231.2699 & 153.4 & $-$3.48413 & 50 & JPL \\
12$_{7,5}$--11$_{7,4}$, A  & 231.2450 & 182.6 & $-$3.53997 & 50 & JPL \\
12$_{7,6}$--11$_{7,5}$, A  & 231.2450 & 182.6 & $-$3.53997 & 50 & JPL \\ \hline \\[-0.2cm]
\textbf{NH$_2$CHO}  & & & & & \\
$10_{1,9}$--$9_{1,8}$     & 218.4592 & 60.8  & $-$3.12640 & 21 & CDMS \\
$11_{2,10}$--$10_{2,9}$ & 232.2736 & 78.9  & $-$3.05471 & 23 & CDMS \\
$11_{3,9}$--$10_{3,8}$   & 233.8966 & 94.1  & $-$3.06449 & 23 & CDMS \\
$11_{3,8}$--$10_{3,7}$   & 234.3155 & 94.2  & $-$3.06216 & 23 & CDMS \\
$11_{4,8}$--$10_{4,7}$   & 233.7347 & 114.9 & $-$3.09334 & 23 & CDMS \\
$11_{4,7}$--$10_{4,6}$   & 233.7456 & 114.9 & $-$3.09332 & 23 & CDMS \\
$11_{5,6}$--$10_{5,5}$   & 233.5945 & 141.7 & $-$3.13303 & 23 & CDMS\\
$11_{5,7}$--$10_{5,6}$   & 233.5945 & 141.7 & $-$3.13303 & 23 & CDMS \\
$11_{6,5}$--$10_{6,4}$   & 	233.5278 & 174.5 & $-$3.18626 & 23 & CDMS \\
$11_{6,6}$--$10_{6,5}$   & 233.5278 & 174.5 & $-$3.18626 & 23 & CDMS \\ \hline \\[-0.2cm]
\textbf{CH$_3$CN} & & & & & \\
$12_0$--11$_0$       & 220.7473 & 68.9  & $-$3.19582 & 50  & JPL \\
$12_1$--11$_1$       & 220.7430 & 76.0  & $-$3.19899 & 50  & JPL \\
$12_2$--11$_2$       & 220.7303 & 97.4  & $-$3.20818 & 50  & JPL \\
$12_3$--11$_3$       & 220.7090 & 133.2 & $-$3.22415 & 100 & JPL \\
$12_4$--11$_4$       & 220.6793 & 183.2 & $-$3.24741 & 50  & JPL \\
$12_5$--11$_5$       & 220.6411 & 247.4 & $-$3.27937 & 50  & JPL \\
$12_6$--11$_6$       & 220.5944 & 325.9 & $-$3.32175 & 100 & JPL \\
$12_7$--11$_7$       & 220.5393 & 418.6 & $-$3.37778 & 50  & JPL \\
$12_8$--11$_8$       & 220.4758 & 525.6 & $-$3.45279 & 50  & JPL \\
$12_{10}$--11$_{10}$ & 220.3236 & 782.0 & $-$3.71328 & 50  & JPL \\
$12_{11}$--11$_{11}$ & 220.2350 & 931.4 & $-$3.99556 & 50  & JPL \\ \hline \\[-0.2cm]
\textbf{$^{13}$CH$_3$CN} & & & & & \\
13$_0$--12$_0$       & 232.2342 & 78.0  & $-$2.96674 & 54  & JPL \\
13$_1$--12$_1$       & 232.2298 & 85.2  & $-$2.96929 & 54  & JPL \\
13$_2$--12$_2$       & 232.2167 & 106.7 & $-$2.97723 & 54  & JPL \\
13$_3$--12$_3$       & 232.1949 & 142.4 & $-$2.99071 & 108 & JPL \\
13$_4$--12$_4$       & 232.1644 & 192.5 & $-$3.01034 & 54  & JPL \\
13$_5$--12$_5$       & 232.1251 & 256.9 & $-$3.03683 & 54  & JPL \\
13$_6$--12$_6$       & 232.0772 & 335.5 & $-$3.07160 & 108 & JPL \\
13$_7$--12$_7$       & 232.0206 & 428.4 & $-$3.11660 & 54  & JPL \\
13$_8$--12$_8$       & 231.9554 & 535.5 & $-$3.17503 & 54  & JPL \\
13$_{11}$--12$_{11}$ & 231.7081 & 942.1 & $-$3.51635 & 54  & JPL \\ \hline \\[-0.2cm]
\textbf{CH$_3^{13}$CN}  & & & & & \\
12$_0$--11$_0$ & 220.6381& 68.8  & $-$3.03477 & 50  & JPL \\
12$_1$--11$_1$ & 220.6338 & 76.0  & $-$3.03783 & 50  & JPL \\
12$_2$--11$_2$ & 220.6211 & 97.4  & $-$3.04713 & 50  & JPL \\
12$_3$--11$_3$ & 220.6000 & 133.1 & $-$3.06309 & 100 & JPL \\
12$_4$--11$_4$ & 220.5704 & 183.1 & $-$3.08635 & 50  & JPL \\
12$_5$--11$_5$ & 220.5323 & 247.4 & $-$3.11821 & 50  & JPL \\
12$_6$--11$_6$ & 220.4859 & 325.9 & $-$3.16068 & 100 & JPL \\
12$_7$--11$_7$ & 220.4310 & 418.6 & $-$3.21671 & 50  & JPL \\ \hline \\[-0.2cm]
\textbf{CH$_3$CH$_2$CN} & & & & & \\
8$_{4,5}$--7$_{3,4}$           & 233.8275 & 33.3  & $-$4.27063 & 17 & CDMS \\
8$_{4,4}$--7$_{3,5}$           & 233.8428 & 33.3  & $-$4.27050 & 17 & CDMS \\
12$_{3,10}$--11$_{2,9}$     & 219.9025 & 43.6  & $-$4.51661 & 25 & CDMS \\
22$_{2,21}$--21$_{1,20}$   & 219.4636 & 112.5 & $-$4.36804 & 45 & CDMS \\
24$_{2,22}$--23$_{2,21}$   & 219.5056 & 135.6 & $-$3.05185 & 49 & CDMS \\
24$_{3,21}$--23$_{3,20}$   & 218.3900 & 139.9 & $-$3.06221 & 49 & CDMS \\
25$_{2,24}$--24$_{2,23}$   & 220.6609 & 143.0 & $-$3.04508 & 51 & CDMS \\
26$_{1,25}$--25$_{1,24}$   & 231.3104 & 153.4 & $-$2.98243 & 53 & CDMS \\
27$_{0,27}$--26$_{0,26}$   & 231.9904 & 157.7 & $-$2.97677 & 55 & CDMS \\
27$_{0,27}$--26$_{1,26}$   & 231.3123 & 157.7 & $-$4.03702 & 55 & CDMS \\
27$_{1,27}$--26$_{1,26}$   & 231.8542 & 157.7 & $-$2.97758 & 55 & CDMS \\
27$_{1,27}$--26$_{0,26}$   & 232.5323 & 157.7 & $-$4.03004 & 55 & CDMS \\
26$_{3,24}$--25$_{3,23}$   & 232.7900 & 161.0 & $-$2.97762 & 53 & CDMS \\
26$_{4,23}$--25$_{4,22}$   & 233.6540 & 169.0 & $-$2.97716 & 53 & CDMS \\
26$_{4,22}$--25$_{4,21}$   & 234.4240 & 169.1 & $-$2.97285 & 53 & CDMS \\
26$_{5,22}$--25$_{5,21}$   & 233.4431 & 178.8 & $-$2.98431 & 53 & CDMS \\
26$_{5,21}$--25$_{5,20}$   & 233.4983 & 178.9 & $-$2.98400 & 53 & CDMS \\
26$_{7,19}$--25$_{7,18}$   & 233.0694 & 205.4 & $-$3.00266 & 53 & CDMS \\
26$_{7,20}$--25$_{7,19}$   & 233.0693 & 205.4 & $-$3.00266 & 53 & CDMS \\
26$_{8,18}$--25$_{8,17}$   & 232.9987 & 222.0 & $-$3.01361 & 53 & CDMS \\
26$_{8,19}$--25$_{8,18}$   & 232.9987 & 222.0 & $-$3.01361 & 53 & CDMS \\
26$_{9,17}$--25$_{9,16}$   & 232.9676 & 240.9 & $-$3.02597 & 53 & CDMS \\
26$_{9,18}$--25$_{9,17}$   & 232.9676 & 240.9 & $-$3.02597 & 53 & CDMS \\
26$_{11,15}$--25$_{11,14}$ & 232.9755 & 285.2 & $-$3.05617 & 53 & CDMS \\
26$_{11,16}$--25$_{11,15}$ & 232.9755 & 285.2 & $-$3.05617 & 53 & CDMS \\
26$_{12,14}$--25$_{12,13}$ & 233.0027 & 310.7 & $-$3.07436 & 53 & CDMS \\
26$_{12,15}$--25$_{12,14}$ & 233.0027 & 310.7 & $-$3.07436 & 53 & CDMS \\
26$_{13,13}$--25$_{13,12}$ & 233.0411 & 338.3 & $-$3.09506 & 53 & CDMS \\
26$_{13,14}$--25$_{13,13}$ & 233.0411 & 338.3 & $-$3.09506 & 53 & CDMS \\
26$_{14,12}$--25$_{14,11}$ & 233.0889 & 368.2 & $-$3.11859 & 53 & CDMS \\
26$_{14,13}$--25$_{14,12}$ & 233.0889 & 368.2 & $-$3.11859 & 53 & CDMS \\
26$_{15,11}$--25$_{15,10}$ & 233.1448 & 400.2 & $-$3.14537 & 53 & CDMS \\
26$_{15,12}$--25$_{15,11}$ & 233.1448 & 400.2 & $-$3.14537 & 53 & CDMS \\
26$_{17,9}$--25$_{17,8}$    & 233.2779 & 470.6 & $-$3.21101 & 53 & CDMS \\
26$_{17,10}$--25$_{17,9}$  & 233.2779 & 470.6 & $-$3.21101 & 53 & CDMS \\ 
26$_{18,8}$--25$_{18,7}$   & 233.3540 & 509.1 & $-$3.25180 & 53 & CDMS \\
26$_{18,9}$--25$_{18,8}$   & 233.3540 & 509.1 & $-$3.25180 & 53 & CDMS \\
26$_{19,7}$--25$_{19,6}$   & 233.4361 & 549.7 & $-$3.29960 & 53 & CDMS \\
26$_{19,8}$--25$_{19,7}$   & 233.4361 & 549.7 & $-$3.29960 & 53 & CDMS \\
49$_{5,45}$--49$_{4,46}$   & 233.2363 & 556.0 & $-$4.22129 & 99 & CDMS \\
50$_{3,47}$--50$_{3,48}$   & 233.2366 & 565.2 & $-$3.16959 & 1  & CDMS \\
26$_{22,4}$--25$_{22,3}$   & 	233.7144 & 684.1 & $-$3.51299 & 53 & CDMS \\
26$_{22,5}$--25$_{22,4}$   & 233.7144 & 684.1 & $-$3.51299 & 53 & CDMS \\
26$_{23,3}$--25$_{23,2}$   & 233.8174 & 733.1 & $-$3.62847 & 53 & CDMS \\
26$_{23,4}$--25$_{23,3}$   & 233.8174 & 733.1 & $-$3.62847 & 53 & CDMS \\
26$_{25,1}$--25$_{25,0}$   & 234.0380 & 837.3 & $-$4.08692 & 53 & CDMS \\
26$_{25,2}$--25$_{25,1}$   & 234.0380 & 837.3 & $-$4.08692 & 53 & CDMS \\ \hline \\[-0.2cm]
\textbf{CH$_2$CHCN} & & & & & \\
24$_{1,24}$--23$_{1,23}$   & 220.5614 & 134.9 & $-$2.55618 & 47 & CDMS \\
23$_{3,21}$--22$_{3,20}$   & 218.5851 & 145.3 & $-$2.53367 & 41 & CDMS \\
23$_{3,20}$--22$_{3,19}$   & 219.4006 & 145.5 & $-$2.52873 & 41 & CDMS \\
24$_{2,22}$--23$_{2,21}$   & 231.9523 & 146.8 & $-$2.49262 & 47 & CDMS \\
23$_{4,20}$--22$_{4,19}$   & 218.5736 & 160.4 & $-$2.53953 & 41 & CDMS \\
23$_{4,19}$--22$_{4,18}$   & 218.6151 & 160.4 & $-$2.53934 & 41 & CDMS \\
23$_{5,18}$--22$_{5,17}$   & 218.4524 & 179.8 & $-$2.54797 & 41 & CDMS \\
23$_{5,19}$--22$_{5,18}$   & 218.4513 & 179.8 & $-$2.54797 & 41 & CDMS \\
23$_{6,17}$--22$_{6,16}$   & 218.4025 & 203.5 & $-$2.55783 & 41 & CDMS \\
23$_{6,18}$--22$_{6,17}$   & 218.4024 & 203.5 & $-$2.55783 & 41 & CDMS \\
23$_{7,16}$--22$_{7,15}$   & 218.3986 & 231.5 & $-$2.56948 & 41 & CDMS \\
23$_{7,17}$--22$_{7,16}$   & 218.3986 & 231.5 & $-$2.56948 & 41 & CDMS \\
23$_{8,15}$--22$_{8,14}$   & 218.4218 & 263.8 & $-$2.58311 & 41 & CDMS \\
23$_{8,16}$--22$_{8,15}$   & 218.4218 & 263.8 & $-$2.58311 & 41 & CDMS \\
23$_{10,13}$--22$_{10,12}$ & 218.5200 & 341.1 & $-$2.61750 & 41 & CDMS \\
23$_{10,14}$--22$_{10,13}$ & 218.5200 & 341.1 & $-$2.61750 & 41 & CDMS \\
23$_{12,11}$--22$_{12,10}$ & 218.6665 & 435.0 & $-$2.66366 & 41 & CDMS \\
23$_{12,12}$--22$_{12,11}$ & 218.6665 & 435.0 & $-$2.66366 & 41 & CDMS \\ \hline \\[-0.2cm]
\textbf{HNCO} & & & & & \\
10$_{0,10}$--9$_{0,9}$     & 219.7983 & 58.0 & $-$3.83290   & 21 & CDMS \\
10$_{1,9}$--9$_{1,8}$       & 220.5848 & 101.5 & $-$3.83753  & 21 & CDMS \\
10$_{2,8}$--9$_{2,7}$       & 219.7372 & 228.3 & $-$3.87103  & 21 & CDMS \\
10$_{2,9}$--9$_{2,8}$       & 219.7339 & 228.3 & $-$3.87104  & 21 & CDMS \\
10$_{3,7}$--9$_{3,6}$       & 219.6568 & 433.0 & $-$3.92039  & 21 & CDMS \\
10$_{3,8}$--9$_{3,7}$       & 219.6568 & 433.0 & $-$3.92039  & 21 & CDMS \\
28$_{1,28}$--29$_{0,29}$ & 231.8733 & 469.9 & $-$4.17514  & 57 & CDMS \\
10$_{5,5}$--9$_{5,4}$       & 219.3924 & 1049.5 & $-$4.09372 & 21 & CDMS \\
10$_{5,6}$--9$_{5,5}$       & 219.3924 & 1049.5 & $-$4.09372 & 21 & CDMS\\ 
\enddata
\end{deluxetable*}




%\bibliographystyle{apj}
\bibliography{G10_COMs.bib} % if your bibtex file is called example.bib



\end{document}